\documentclass[article]{aa}
\usepackage{epstopdf}
\usepackage{graphicx}
\usepackage[varg]{txfonts}
\usepackage{amsmath}
\usepackage{aas_macros}
\usepackage{float}
\usepackage{caption}
\usepackage{esdiff}
\usepackage{marginnote}
\usepackage{color}

\graphicspath{{figures.res/}}

\bibpunct{(}{)}{;}{a}{}{,} 
\newcommand{\specialcell}[2][c]{%
  \begin{tabular}[#1]{@{}c@{}}#2\end{tabular}}

\begin{document}

\author{Hans J.T. Buist 
\and Amina Helmi}

\institute{Kapteyn Astronomical Institute, University of Groningen, P.O. Box 800, 9700 AV Groningen, The Netherlands\\ \email{buist@astro.rug.nl}}

\title{On the behaviour of streams in angle and frequency spaces in different potentials}

\date{}
\abstract{We have studied the behaviour of stellar streams in the Aquarius fully
cosmological N-body simulations of the formation of Milky Way halos.
In particular, we have characterised the streams in angle and
frequency spaces derived using an approximate but generally
well-fitting spherical potential. We have also run several
test-particle simulations to understand and guide our interpretation
of the different features we see in the Aquarius streams. Our goal is
both to establish which deviations of the expected action-angle
behaviour of streams exist because of the approximations made on the
potential, but also to derive to what degree we can use these
coordinates to model streams reliably. \\
We have found that many of the Aquarius streams wrap in angle space
along relatively straight lines, and distribute themselves along
linear structures also in frequency space. On the other hand, from our
controlled simulations we have been able to establish that deviations
from spherical symmetry, the use of incorrect potentials and the
inclusion of self-gravity lead to streams in angle space to still be
along relatively straight lines but also to depict wiggly behaviour
whose amplitude increases as the approximation to the true potential
becomes worse. In frequency space streams typically become thicker and
somewhat distorted. Therefore, our analysis explains most of the features
seen in the approximate angle and frequency spaces for the Aquarius
streams with the exception of their somewhat `noisy' and `patchy'
morphologies. These are likely due to the interactions with the large
number of dark matter subhalos present in the cosmological
simulations. Since the measured angle-frequency misalignments of
the Aquarius streams can largely be attributed to using the wrong
(spherical) potential, the determination of the mass growth history of
these halos will only be feasible once (and if) the true potential has
been determined robustly.}


\maketitle

\section{Introduction}

In the last two decades much progress has been made on the discovery
and characterisation of tidal streams around our Milky Way and in
other nearby galaxies \citep[see e.g.][]{Koposov2012,
  Martin2014}. Tidal streams consist of stars stripped from satellites
(dwarf galaxies and globular clusters) that move on nearby almost parallel
orbits. As such they constitute extremely sensitive probes of the mass
distribution in the host system \citep{Johnston1999, Ibata2001,
  Johnston2004, Law2005, LawMajewski2010, Koposov2010,
  VeraCiroHelmi2013, Sanders2013a, Sanderson2014}. This is one of the
drivers of the observational and theoretical studies of streams, as
one of the ultimate goals is to establish the mass distribution in
the dark halo of galaxies like the Milky Way, which in turn will lead
to a better understanding of the nature of dark matter \citep[see
e.g.][]{Strigari2013}.

The recently launched {\it Gaia} satellite \citep{Perryman2001} will
provide the phase-space coordinates of a vast sample of stars in the
Milky Way in the next decade. Together with spectroscopic follow-up
surveys of the fainter {\it Gaia} stars such as {\it 4MOST}
\citep{DeJong2012} and {\it WEAVE} \citep{Dalton2012}, these datasets
will provide an unprecedented detailed view of our Galaxy. Our
understanding of the dynamics of the halo and its streams needs to be
sharpened to maximally exploit the wealth of data that will soon
become available \citep[see e.g.][]{Johnston1999, LawMajewski2010,
  Eyre2011, Bonaca2014}.

Action-angle coordinates provide an excellent tool to describe the
evolution of streams \citep{Tremaine1999, Helmi1999}. For example, the
evolution of stars in angle space is linear with time for a static
potential, while the actions are adiabatic invariants. The difficulty
lies in finding the necessary coordinate transformations to map
observables into action-angle space. Only for specific types of potentials, including spherical and those of Staeckel form, 
can we directly compute the angles and actions because the
Hamilton-Jacobi equation is separable \citep{Goldstein1950,
  DeZeeuw1985, BinneyTremaine2008}. However, in recent years several
approximate schemes have been developed to overcome this problem. For
example, an appropriate toy potential can be used to compute the true
action-angles, based on the work of \citet{McGill1990} and
\citet{Kaasalainen1994} \citep[see e.g.][]{McMillan2008, Fox2012,
  Sanders2014a, Bovy2014}. Another method is to approximate the
potential locally by a Staeckel potential (axisymmetric or triaxial),
a procedure known as the `Staeckel fudge' \citep{Binney2012,
  Sanders2014b}.

\begin{figure*}[!htbp]
\centering
\noindent\makebox[\textwidth]{
\includegraphics[width=0.7\textwidth]{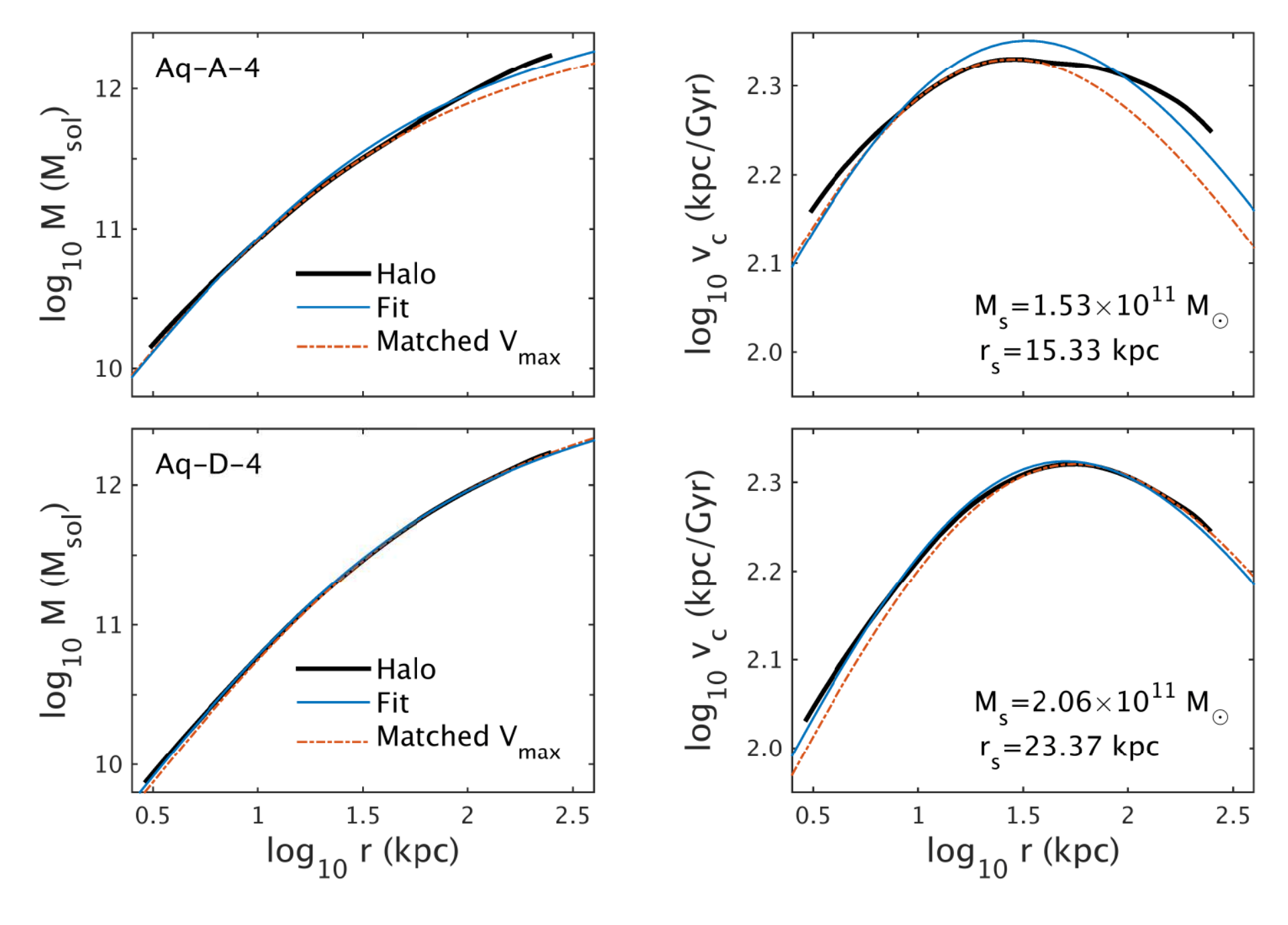}}
\caption{\small The enclosed mass (left) and circular velocity (right)
  for halo Aq-A (top panels) and halo Aq-D (bottom panels) with NFW
  fits to the circular velocity (blue lines). The red dashed curve
  shows the profile when we match the peak velocity. For an NFW the
  position of this maximum depends only on $r_\textrm{s}$ and its
  magnitude only on $M_\textrm{s}$. The radial range plotted starts at the
  convergence radius \citep{Power2003, Navarro2010} and extends up to $r_{200}$
  \citep[see][]{Springel2008}.}
\label{fig4:fitAqAD}
\end{figure*}

The availability of full phase-space information for a large number of
stars in the {\it Gaia} dataset will assist greatly in exploiting the power
of action-angle coordinates for streams \citep{McMillan2008, Gomez2010b}. Actions
(and integrals of motion) may be used as well to derive the accretion
history of the halo of the Milky Way, because even when a stream is
fully phase mixed it is still clumped in this
space (\citealt{HelmiDeZeeuw2000}; also see \citealt{Gomez2010a}). We do require the potential to evolve
adiabatically slowly for the actions to remain clumped. This clumpiness in action space can
also be employed to determine the gravitational field in which the
stars have evolved because the largest degree of clustering occurs
when the actions are computed in the true potential
\citep{Sanderson2014, Magorrian2014, Penarrubia2012}. The angles and frequencies can
also be used to this end, because streams should lie along straight
lines that have the same slope in angle and in frequency space for the
correct gravitational potential under the condition that it is
static \citep{Sanders2013b, Sanders2013a}. In \citet{BuistHelmi2015}
we argued that an adiabatically growing potential will cause a small
difference in these slopes, or an ``angle-frequency misalignment''. We
also pointed out that in angle and frequency space there are several other indicators 
that the potential used in the computation may be incorrect. This will allow the
determination of the characteristic parameters of the true potential to be separated from its
time-evolution.

We continue along that line of study in this Paper, where we explore
the behaviour of streams evolved in fully cosmological N-body
simulations. In particular, we characterise the streams in the angle and
frequency spaces obtained using an approximate spherical potential. We
also run several test-particle simulations to understand and guide our
interpretation of the different features we see in the cosmological
simulations. Our goal is both to establish which perturbations of the
action-angle behaviour of streams exist because of the approximations
made on the potential, but also to derive to which degree we can use
these coordinates to study cosmologically evolved streams.

The structure of this Paper is as follows. In
Sec.~\ref{sec4:streamsinaquarius} we describe the cosmological N-body
simulations and the stream catalogue used. In
Sec.~\ref{sec4:aabehaviouraquarius} we discuss their behaviour in 
action-angle coordinates, computed using an appropriately chosen
spherical potential. To guide our understanding of the behaviour of these streams
we present in Sec.~\ref{sec4:testparticlesaxisymmetric} a set of test-particle streams evolved in an axisymmetric potential, loosely based on the cosmological simulations. In Sec.~\ref{sec4:aabehavioursimulations} we gain further insight into the generic behaviour
of streams by determining the impact of
computing the action-angles of the test-particle simulations of
Sec.~\ref{sec4:testparticlesaxisymmetric} in several incorrect
potentials, and discuss the effect of self-gravity. We end in
Sec.~\ref{sec4:discussionconclusion} with a discussion and conclusions.

\section{Streams in cosmological simulations}
\label{sec4:streamsinaquarius}

\subsection{Description of the Aquarius project and its stellar halos}
The Aquarius project \citep{Springel2008, Navarro2010} consists of a
set of six re-simulations of Milky-Way mass \mbox{($\sim10^{12} \textrm{
  M}_\odot$)} dark matter halos extracted from a larger cosmological
parent simulation \citep{Boylan2009}. 
They were selected to have no close
massive neighbours at $z=0$, and form late-type galaxies when evolved
using semi-analytic galaxy formation models. Halos Aq-A to Aq-E are 
believed to be representative of the Milky Way, while Aq-F experiences 
a recent major merger and hence is less suitable \citep{Wang2011}. 
We focus our study of streams to two of the halos, Aq-A and Aq-D to avoid
flooding this article with examples  while providing a flavour of the variations found in different systems. 

We extract streams from the accreted component of stellar halos modelled
using the Durham semi-analytic model GALFORM. \cite{Cooper2010} have
associated stellar populations with dark matter particles in the
simulations via a `tagging' scheme. \cite{Lowing2014} took this a step
further and generated individual stars from these populations by
re-sampling the dark matter particles and using stellar population
synthesis modelling. Another difference between these works is that Cooper et al.\ used
the \cite{Bower2006} version of GALFORM, while Lowing et al.\ used the
\cite{Font2011} version which has improved physics on dwarf galaxy
scales that makes model satellites more similar to those observed
around the Milky Way. Here we use the public catalogue that Lowing et al.\ offer
online\footnote{http://galaxy-catalogue.dur.ac.uk:8080/StellarHalo},
which has all stars with magnitude $M_g < 7$.

The Lowing dataset re-samples the dark matter particle positions and
velocities in a way that aims to preserve their distribution in
phase-space. To observe streams in the halo of the Milky Way it is
common to select the Red Giant Branch (RGB) stars because they are
intrinsically bright. We noticed that some of the thinner simulated
streams look quite clumpy when only using the RGB stars, most likely
because of the re-sampling of the dark matter particles. Since our
interest is only in the dynamical features of the streams, we instead
decided to use the source (`tagged') dark matter particles. To this
end we have matched the dark matter ID's to those in the outputs of
the Aquarius simulations and found the positions and velocities of the
source particles from the Lowing dataset. Whether we would have used
the RGB stars or the dark matter particles does not matter much for
the number statistics as both sets have similar sizes.

We aligned the coordinate system to that of the parent dark halo by
using the principal axes determined at 100 kpc from the centre by
\cite{VeraCiro2011}. These authors used the reduced
inertia tensor method \citep{Allgood2006} which closely follows
isodensity contours. The $z$-direction is set along the major axis of these halos.
Further details on the shapes are discussed in
Sec.~\ref{sec4:potentialsetup}. 

\subsection{Mass distribution of the Aquarius halos}
\label{sec4:massdistributionaquarius}

\begin{figure}[!htbp]
\centering
\includegraphics[width=0.5\textwidth]{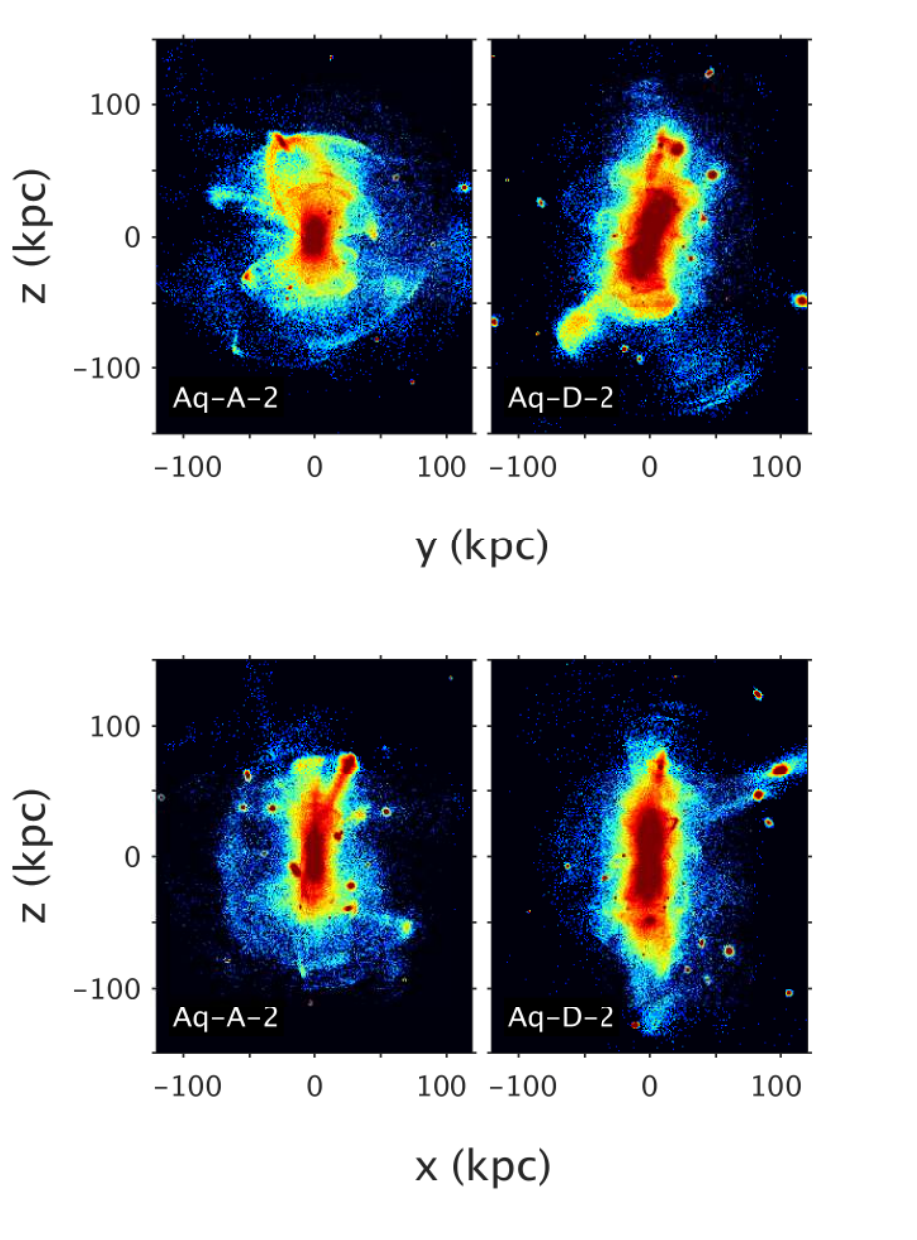}
\caption{\small Density maps of all the dark matter particles tagged with
  stellar populations originating in objects in halos Aq-A and Aq-D with more than 500
  dark matter particles in their progenitor. The coordinate system
  is aligned with the principal axes of the halo at 100 kpc. The minor
  and intermediate axes are in the $x$-$y$ plane and the major axis is
  along the $z$ direction.}
\label{fig4:aquarius_streams}
\end{figure}

Although intrinsically triaxial, with triaxiality parameters in the range
  2/3 to 1 (see Sec.~\ref{sec4:potentialsetup}), the Aquarius halos
can be fit reasonably well with a spherical mass profile. Here we use
Navarro-Frenk-White functional form \citep[hereafter
NFW]{NFW1996,NFW1997}, which provides a relatively good description in
the regions where we study streams \citep[$r \sim 50-100$ kpc,
see][]{Springel2008, Navarro2010}, and we prefer it because of its
simplicity and computational efficiency compared to the slightly
better fitting Einasto profile \citep{Einasto1965}.

In Fig.~\ref{fig4:fitAqAD} we show the results obtained when fitting
the spherically averaged circular velocity profile with two free parameters, the scale mass $M_\textrm{s}$ and the scale radius $r_\textrm{s}$, for halo Aq-A and halo Aq-D
\citep[see also][]{Navarro2010}. The scale mass is defined as $M_\textrm{s} = M(r_\textrm{s})$ and is related to the virial mass as
\begin{equation}
	M_\textrm{s} = \frac{M_\textrm{vir}}{f(c)} ;\ \ f(x) = \frac{\log(1+x)-x/(1+x)}{\log(2) - 1/2},
\end{equation}
where the concentration $c \equiv r_\textrm{vir} / r_\textrm{s}$. Note that halo Aq-D is much
better fitted by an NFW profile than halo Aq-A, which appears to have a bump in its circular velocity profile. Aq-A is also more triaxial than Aq-D \citep{VeraCiro2011}. 

\subsection{Selection of streams}

Not all objects in the Aquarius stellar halos are apparent as
stream-like structures, for example this may be the case if the object
is too small or if it has not been significantly
disrupted. Fig.~\ref{fig4:aquarius_streams} shows that large streamy
structures exist out to large radii (see also \citealt{Cooper2010}). 
Each of the 2 Aquarius stellar halos have of the order of 100-200 individual
progenitors, of which about 20\% have produced discernible stream-like
features by $z=0$, for example, a visible piece of a loop. We selected these by eye in physical space and verified this
selection by checking that they also appeared streamy in their
approximate angle coordinates, as discussed later. For comparison we also included several massive and seemingly phase mixed objects that most likely represent debris on quite radial orbits.
We also impose a
lower limit of at least 500 `tagged' dark matter particles for each
progenitor to exclude the really small streams. This corresponds to a
`tagged' dark matter mass of $\sim7\times10^6 \textrm{ M}_\odot$, and
yields a total of 35 streams for halo Aq-A and Aq-D together. In the
main part of this Paper we focus on 10 streams from each halo to
illustrate their behaviour, a selection which includes a few of the
highly phase-mixed shell-like structures that appear near the long
axis (here the $z$-axis) and which move on more radial orbits. The remaining 25 streams are shown in the Appendix.

\subsection{The morphology of selected streams}
\label{sec4:streamsmorphologyandorbits}
\begin{figure*}[!htbp]
\centering
\noindent\makebox[\textwidth]{
\includegraphics[width=0.95\textwidth]{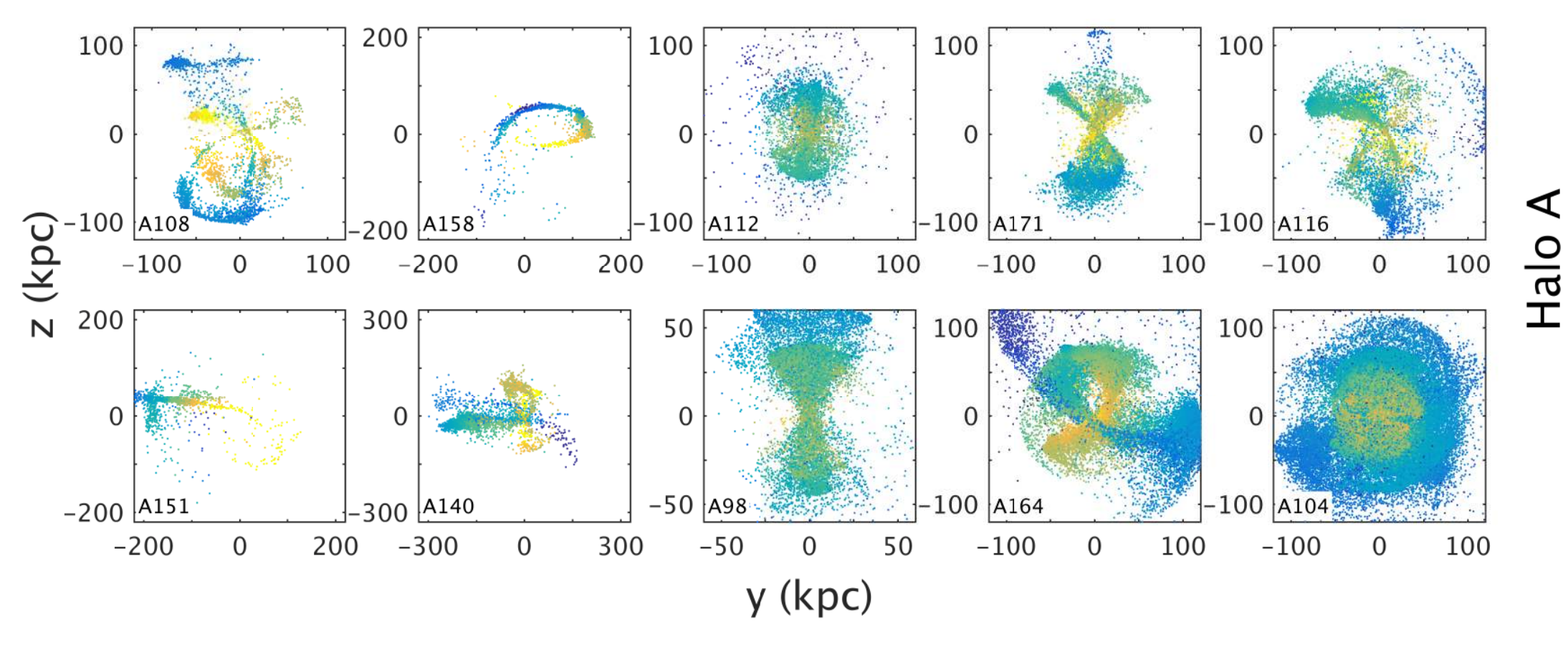}}
\noindent\makebox[\textwidth]{
\includegraphics[width=0.95\textwidth]{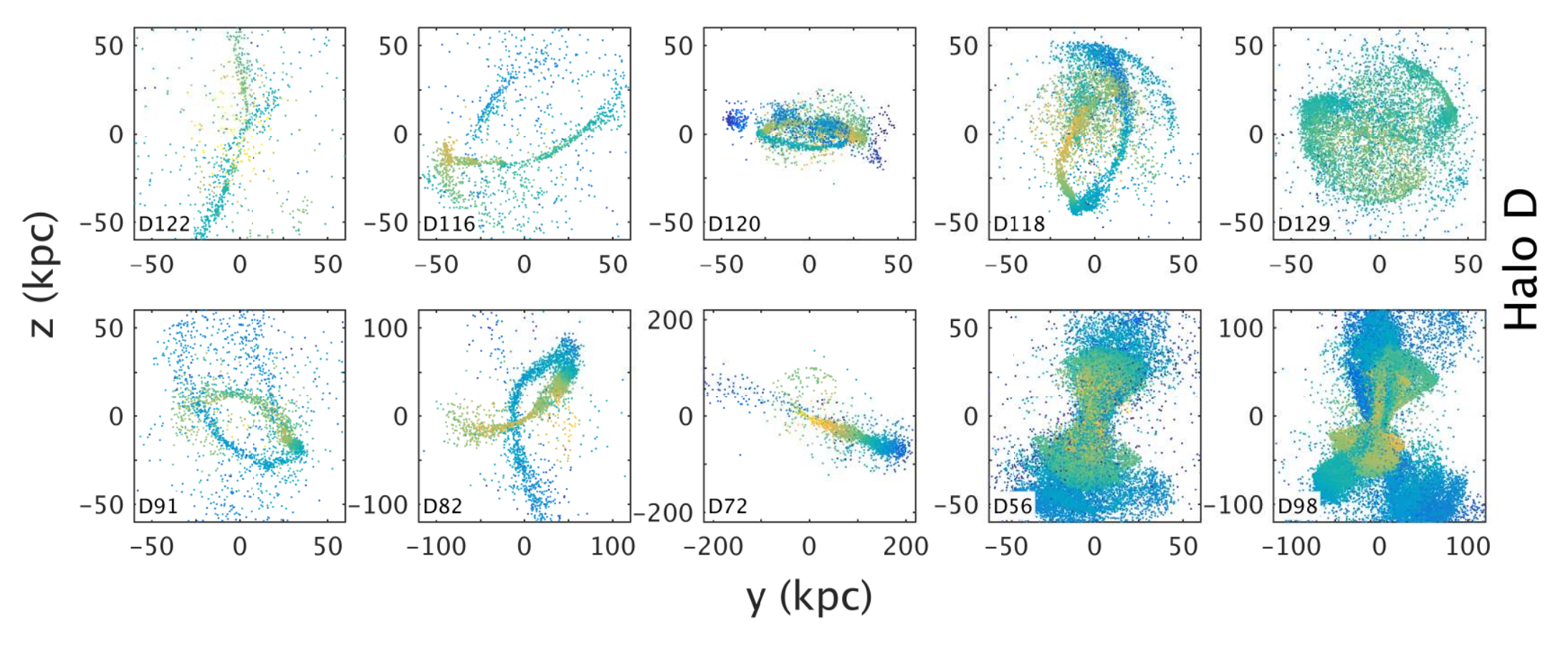}}
\vspace{-0.8cm}
\caption{\small Stream-like objects in Aq-A (top) and Aq-D (bottom) 
in the same projection as in the top panel of Fig.~\ref{fig4:aquarius_streams}. Shown here are the `tagged' dark matter particles while the labels indicate the object IDs used throughout this Paper. The ranges are different in each panel to show the streams in maximum detail. The colours represent the energy gradient computed using the best-fitting spherical NFW potential, with yellow the most bound particles, and blue those least bound. The streams have been sorted by dark matter mass, with the lightest stream on the top-left and the most massive on the bottom-right.}
\label{fig4:aquarius_AqAD}
\end{figure*}

Streams consist of groups of stars that have similar orbits, and the
relatively small variance in their orbits creates structures whose
trajectory follows closely the orbit of the progenitor
\citep{JinLyndenBell2007, Binney2008}, although not exactly \citep{Choi2007,
  Eyre2009, Sanders2013a}. We can therefore analyse the streams in
terms of the orbits permitted by the potential.

Our selection of individual streams for halo Aq-A and Aq-D 
are shown in Fig.~\ref{fig4:aquarius_AqAD}. To some extent, the consideration of structures in these two different haloes help us 
gauge the variety and similarities in the characteristics of their streams. The IDs of
the streams given in this figure correspond one-to-one with tree-IDs in the Lowing catalogue.
 We provide the progenitor's masses of these streams in Appendix~\ref{sec4:Appendix4C}.
 The
colours in the figure represent the binding energy computed in the
best fitting spherical NFW potential (yellow is the most bound, blue
is the least bound). 
 Most of the streams depict at least one clear stream-like feature or
loop (almost by construction, and for as far as a 2-D projection allows us to show this), except 
for the debris for A98, A104, D56 and D98, which we selected to
demonstrate massive and seemingly very phase mixed objects on radial orbits.

Typically the streams consist of one or more petals, but they are not
the clean rosette-like figures seen in the case of spherical
potentials \citep[see e.g.][]{BuistHelmi2015}, because in a triaxial
potential many more orbital families exist. In a triaxial potential
orbits are typically box or tube orbits around the major, intermediate
and minor axes. Box orbits get arbitrarily close to the centre of the
potential, a property they share with purely radial orbits in a
spherical potential, and do not have a sense of rotation. Tube orbits
circulate about one of the axes of the potential and never get to the
centre of the potential, which they have in common with loop orbits in
a spherical potential. For example, stream D56 and stream A98 are
distributed on a structure that appears similar to that defined by a box orbit,
while stream A108 resembles a `fish'-orbit \citep[3:2
resonance, see e.g.][]{MiraldaEscude1989, Merritt1999,
  BinneyTremaine2008}. We also notice that the streams in halo D also seem to
  be a bit better defined (i.e. smoother and with longer loops) than those in halo A.

\begin{figure*}[!htbp]
\centering
\noindent\makebox[\textwidth]{
\includegraphics[width=0.98\textwidth]{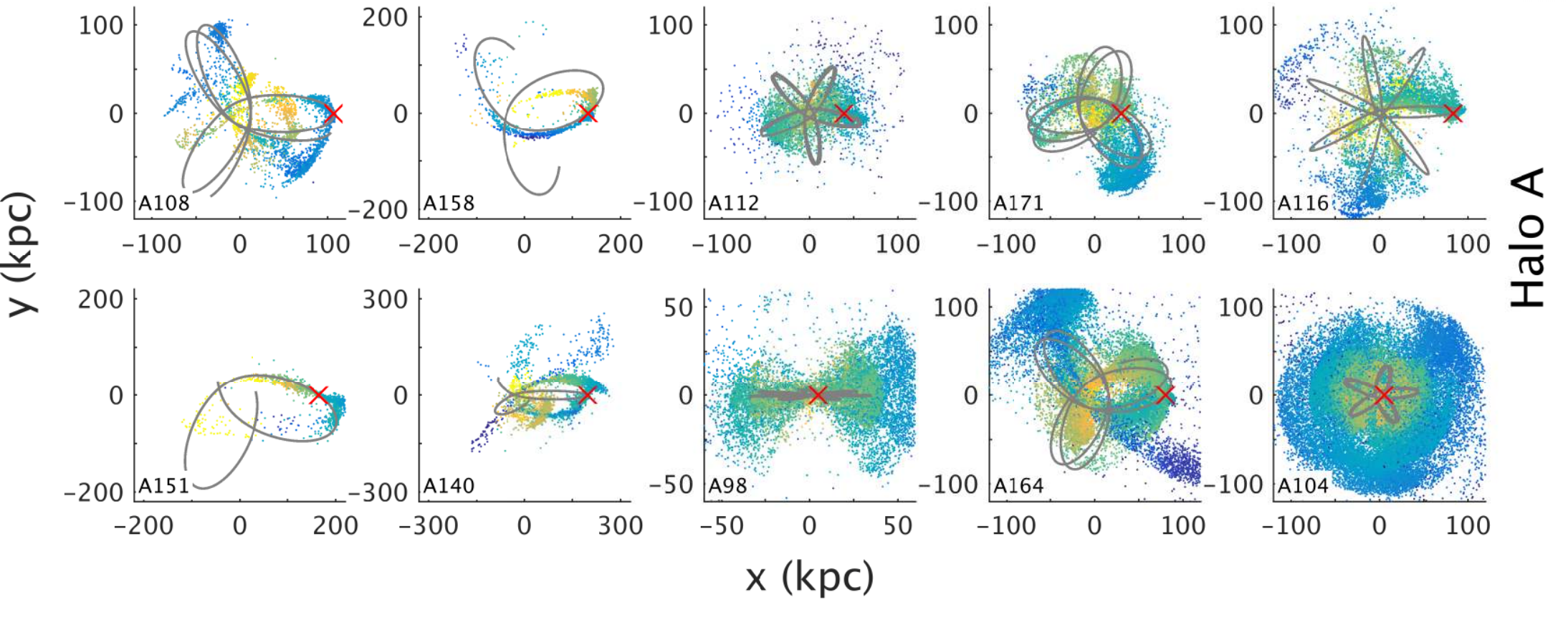}}
\noindent\makebox[\textwidth]{
\includegraphics[width=0.98\textwidth]{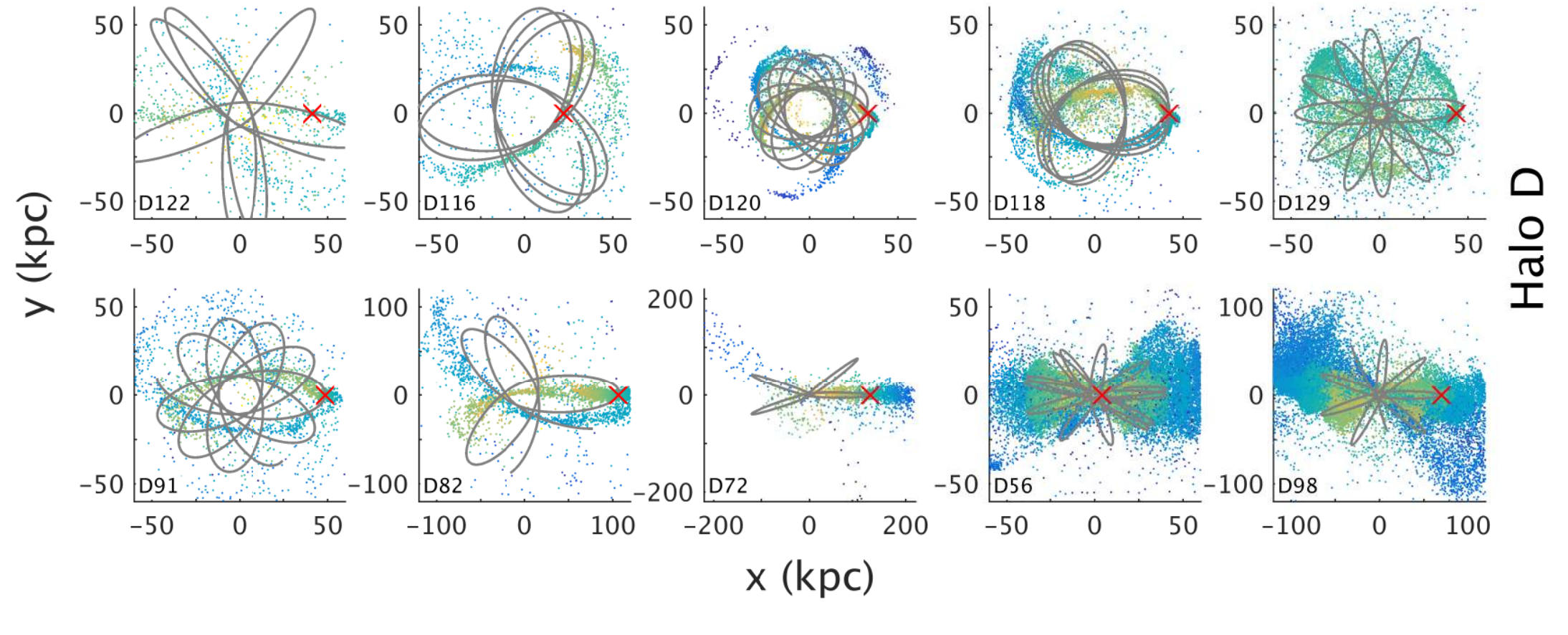}}
\caption{\small The same as Fig.~\ref{fig4:aquarius_AqAD} but now
  overplotting orbits integrated in the best fitting spherical
  potential to halo Aq-A (top panels) and halo Aq-D (bottom
  panels). The orientation of the streams has been changed such that
  the $z$-direction corresponds with the mean angular momentum of the
  particles. The plotted orbit is that of a particle (indicated with a
  red cross) near or in the bound part of the progenitor and has been
  evolved 4 Gyr forward and backwards in time assuming a static
  spherical potential.}
\label{fig4:sphericalOrbitsAqADxy}
\end{figure*}

\begin{figure*}[!htbp]
\centering
\noindent\makebox[\textwidth]{
\includegraphics[width=0.98\textwidth]{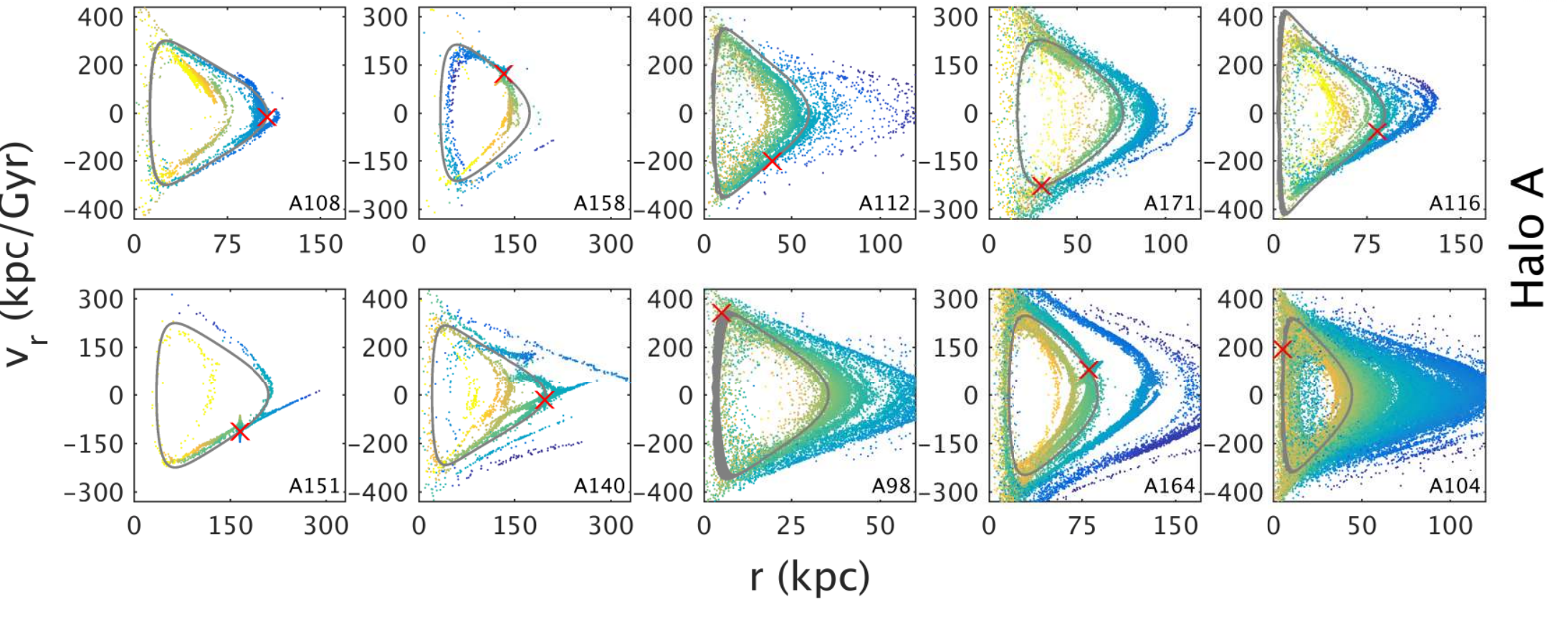}}
\noindent\makebox[\textwidth]{
\includegraphics[width=0.98\textwidth]{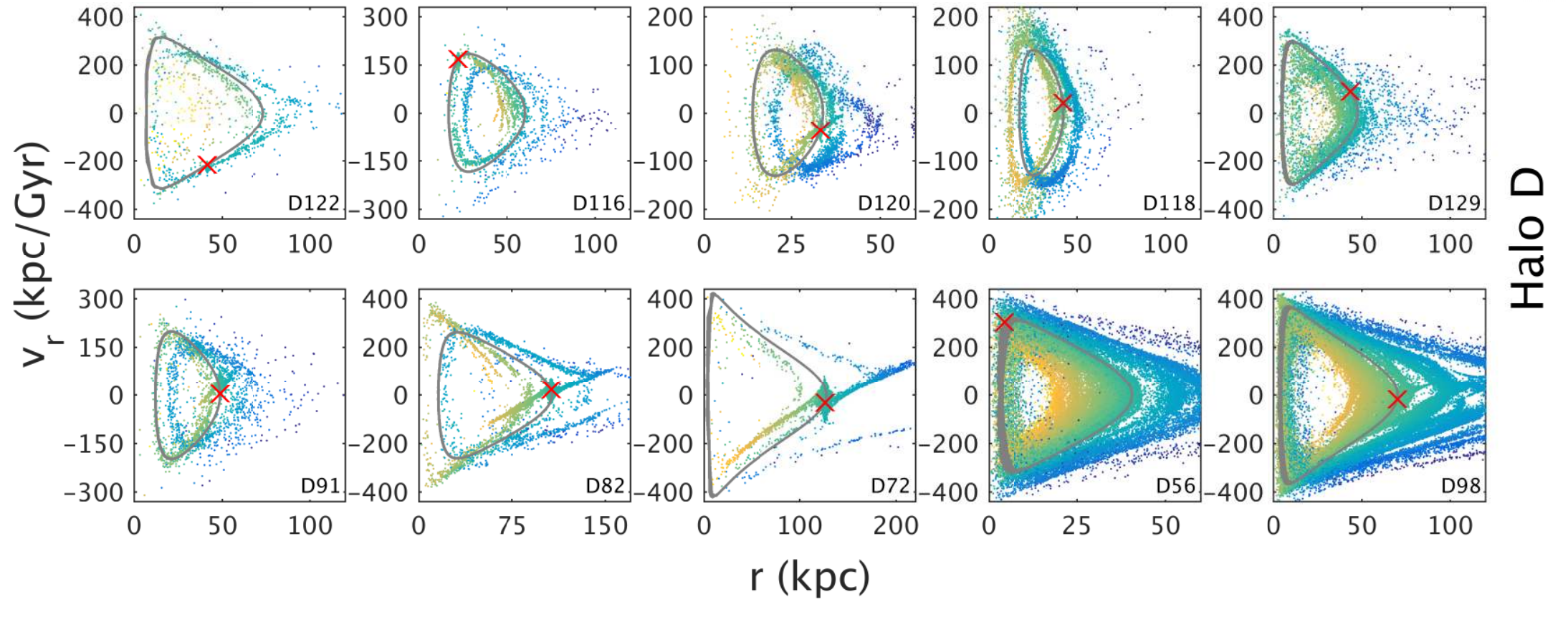}}
\caption{\small The same as Fig.~\ref{fig4:sphericalOrbitsAqADxy} but now showing the $r$-$v_r$ projection for halo Aq-A (top panels) and halo Aq-D (bottom panels).}
\label{fig4:sphericalOrbitsAqADrvr}
\end{figure*}

As a first step in our characterisation of streams, and for simplicity, we explore how orbits integrated in the best fitting spherical potential follow the trajectories delineated by the streams. For the relatively massive progenitors we naturally expect worse agreement, as discussed below.
The initial conditions of the orbit were taken from a
particle located in the highest density portion of the stream (typically a bound particle in the progenitor). We
integrate this particle forward and backwards for 4 Gyr in our best
fitting NFW potential to approximately match the streams' length.

In Fig.~\ref{fig4:sphericalOrbitsAqADxy} we show the resulting orbits on the streams, but in an
orientation where the $z$-axis is in the direction of the mean angular
momentum of the stream. An overview of the apocentre and pericentre distances of the spherical orbits is given in Appendix~\ref{sec4:Appendix4C}. For thinner streams, with only one or two wraps,
the orbits seem to follow the stream, such as for
A158 and A151, but also the much thicker A164 has at least one loop
reasonably matched. Many of the heavier streams that have more wraps
show a big difference in radial extent when compared to the orbits,
with stream A104 giving one of the least satisfactory results. This is
not too surprising given that the spherical potential only supports a
very limited range of orbit families. Also, a single orbit cannot fit
the stream's radial extent because this depends on the range of
energies of the particles, and this is particularly large for some of
our objects.

In Fig.~\ref{fig4:sphericalOrbitsAqADrvr} we show the streams in the
$r$-$v_r$ plane with the corresponding orbits overlaid. For all objects individual stream wraps are discernible, even for those objects that are very massive and well phase-mixed. For these it is clear that the radial extent is not well matched by the orbits, as discussed earlier.  Part of the complex structure seen in this phase-space projection may also be attributed to
the triaxiality of the true potential. Nonetheless, in all cases the spherically integrated orbit seems to match reasonably
at least a single wrap. Often the still bound part of the progenitor appears as a vertical
diamond shaped object in the stream. Its location coincides quite well
with the location of the highest density of the stream, which was
chosen to define the initial conditions for the orbital integration
and is indicated by the red cross. Near the progenitors of D82 and D72 we see very long arms that at the ends seem to dissolve in
the stream. Most likely these particles are not yet completely unbound
from the progenitor \citep[see e.g.][]{Gibbons2014}

We conclude that our orbits do reproduce some of the wraps of the
streams, but typically they lack the complexity that is seen in the
Aquarius simulations' streams. This is partially because a single orbit
cannot represent these streams stream well, but for a major part because the spherical
potential does not support the exact same types of orbits.

\section[Action-angle behaviour of Aquarius streams in spherical potentials]{Action-angle behaviour of Aquarius streams in\\ spherical potentials}
\label{sec4:aabehaviouraquarius}

\subsection{General behaviour in a static potential}
We now investigate the properties of our streams in action-angle
space, where their behaviour is expected to be
particularly simple. For an individual particle in a static potential, angles evolve linearly with time as 
\begin{equation}\theta_i(t) = \Omega_i t + \theta_i(0),
\end{equation}
here $\Omega_i$ are the orbital frequencies, which only depend on the adiabatically invariant actions and hence are constant. For an ensemble this implies the spread in angles evolves as 
\begin{equation}
\Delta \theta_i(t) = \Delta \Omega_i t + \Delta \theta_i(0).
\end{equation} 
For streams that have evolved long enough ($t \gg \Omega_i^{-1}$) we can
ignore the initial angle spread $\Delta \theta_i(0)$, and therefore
the appearance of the stream in angle and frequency is rather similar,
with the stream in angle space being stretched out in time.

We can understand the shape in frequency space more quantitatively using a Taylor expansion of the orbital frequencies $\Omega_i$ with respect to the actions $J_j$ \citep{Helmi1999}
\begin{equation}
  \Delta \Omega_i \approx H_{ij} \Delta J_j,
\label{eq4:delta_omega_ch4}
\end{equation}
with the spreads measured with respect to the centre of mass of the progenitor, and the Hessian of the Hamiltonian $H_{ij}$ also evaluated at this point. If we diagonalise the Hessian (and assuming the Einstein notation convention)
\begin{equation}
  H_{ij} \Delta J_j = V_{ik} D_{kl} (V^T)_{lj} \Delta J_j = V_{il} \left( \lambda_l \Delta\tilde{J}_l \right),
\label{eq4:hessian_exp_ch4}
\end{equation}
with $V$ the matrix of eigenvectors $\vec{e}_i$ of $H_{ij}$ (as
its columns), $D_{kl}$ the diagonal matrix with the eigenvalues
$\lambda_i$, and $\Delta\tilde{J}_l = V_{lj}^T \Delta J_j$, the action
spreads in the eigenspace. In vector notation this expression takes the simpler form
\begin{equation}
	\vec{\Delta\Omega} = \sum_i \lambda_i\, \Delta \tilde{J}_i \, \vec{e}_i.
\end{equation}
Generally one of the eigenvalues of the Hessian is much larger than
the others\footnote{For the experiments explored in this paper, the largest eigenvalue of the Hessian matrix is approximately 100
times bigger than the other two eigenvalues in the spherical case, while the ratio is typically 10:1 for the axisymmetric case.} , and this is responsible for the stream being a thin 1-D
structure in frequency space(and therefore also in angle space), that
is elongated mostly in the direction associated to the corresponding
eigenvector \citep{Tremaine1999}. Note here that also the relative 
magnitude of the action spreads is relevant for the direction in which the frequency distribution is elongated.

The action-angle coordinates of course depend on the underlying potential. For
example, in a spherical potential there are two independent
frequencies $\Omega_\phi$ and $\Omega_r$ (and 
$\Omega_\phi = \Omega_\vartheta$, apart from a possible sign
difference), and therefore for sufficiently long times $\Delta
\theta_r(t) = \Delta \Omega_r t$ and $\Delta \theta_\phi(t) = \Delta
\Omega_\phi t$. This implies that in this regime in the space of angles, streams
follow a straight line with slope $S(\Delta \theta) =
\Delta\theta_\phi/ \Delta \theta_r$ which is the same slope as in frequency
space $S(\Delta \Omega) = \Delta \Omega_\phi /\Delta \Omega_r$.

\begin{figure*}[!htbp]
\centering
\vspace{-0.2cm}
\noindent\makebox[\textwidth]{
\includegraphics[width=0.95\textwidth]{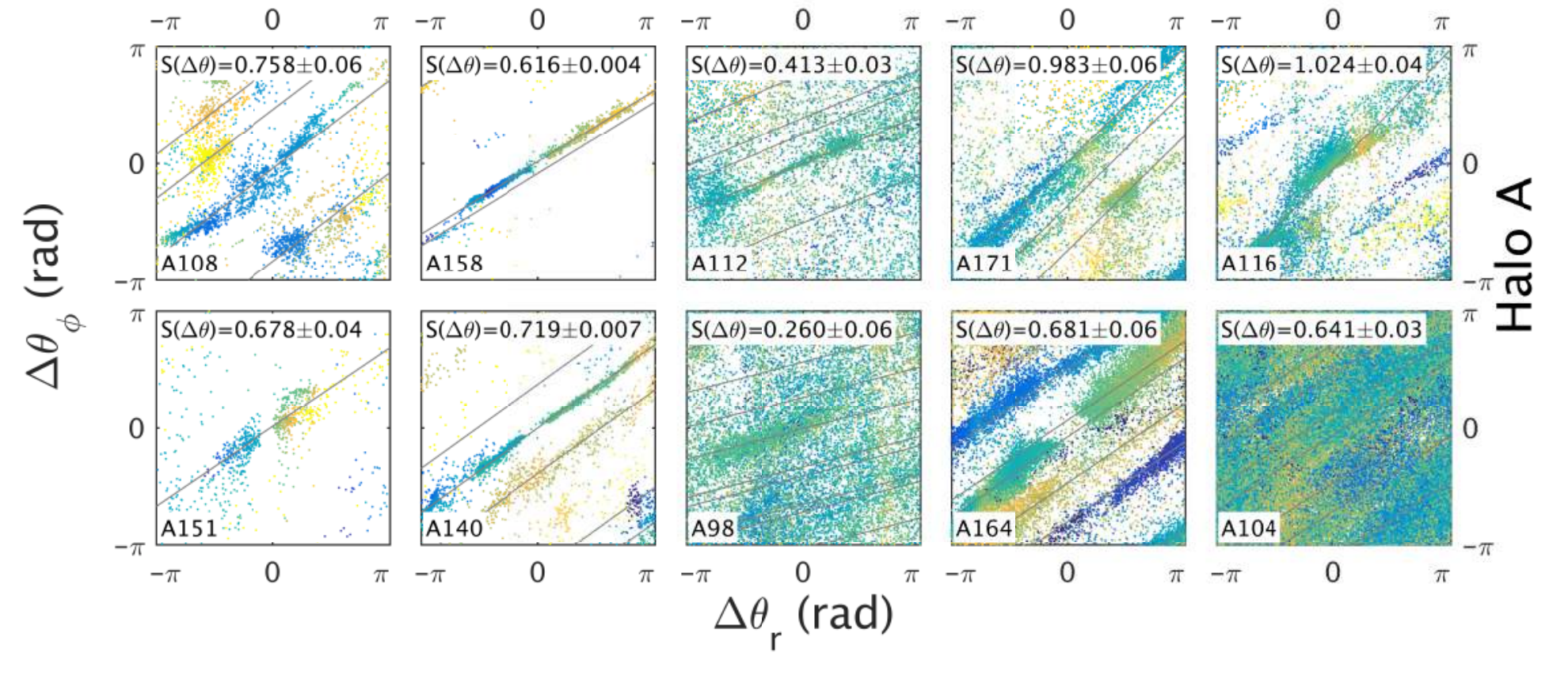}}
\noindent\makebox[\textwidth]{
\includegraphics[width=0.95\textwidth]{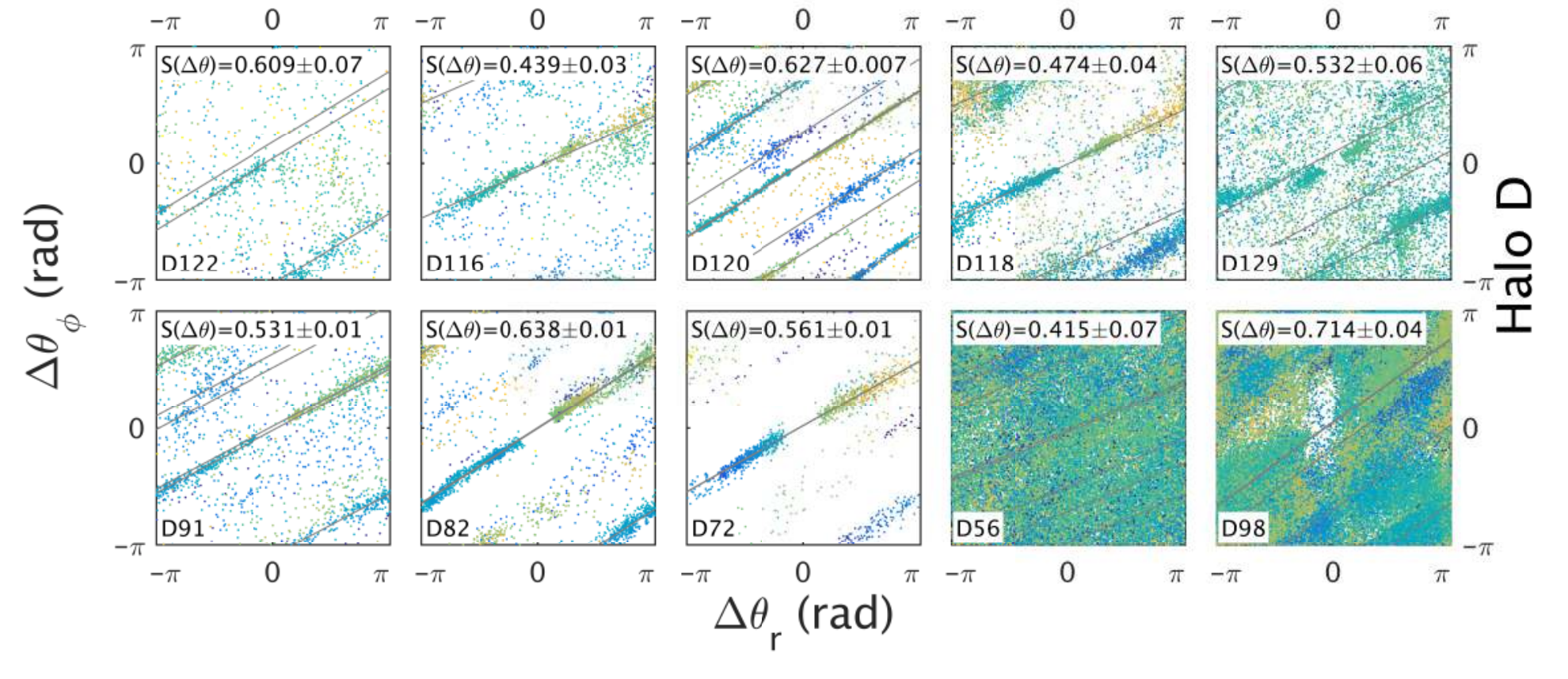}}
\vspace{-0.8cm}
\caption{\small The $\theta_r$-$\theta_\phi$ space for the streams selected from halo Aq-A (top panels) and Aq-D (bottom panels) shown in previous figures. The angles have been computed using the best fitting spherical NFW potentials, and are centred around the most bound particle in the progenitor, or around a particle closest to the highest density in $\theta_r$-$\theta_\phi$ space. The colours indicate the energy gradient assuming the spherical NFW potential. The distributions were fitted with straight lines after removing generously particles still bound to the progenitor. The insets show the fitted slope and its error as estimated from bootstrapping 200 times.}
\label{fig4:aquarius_anglesAD}
\end{figure*}

\begin{figure*}[!htbp]
\centering
\vspace{-0.2cm}
\noindent\makebox[\textwidth]{
\includegraphics[width=0.95\textwidth]{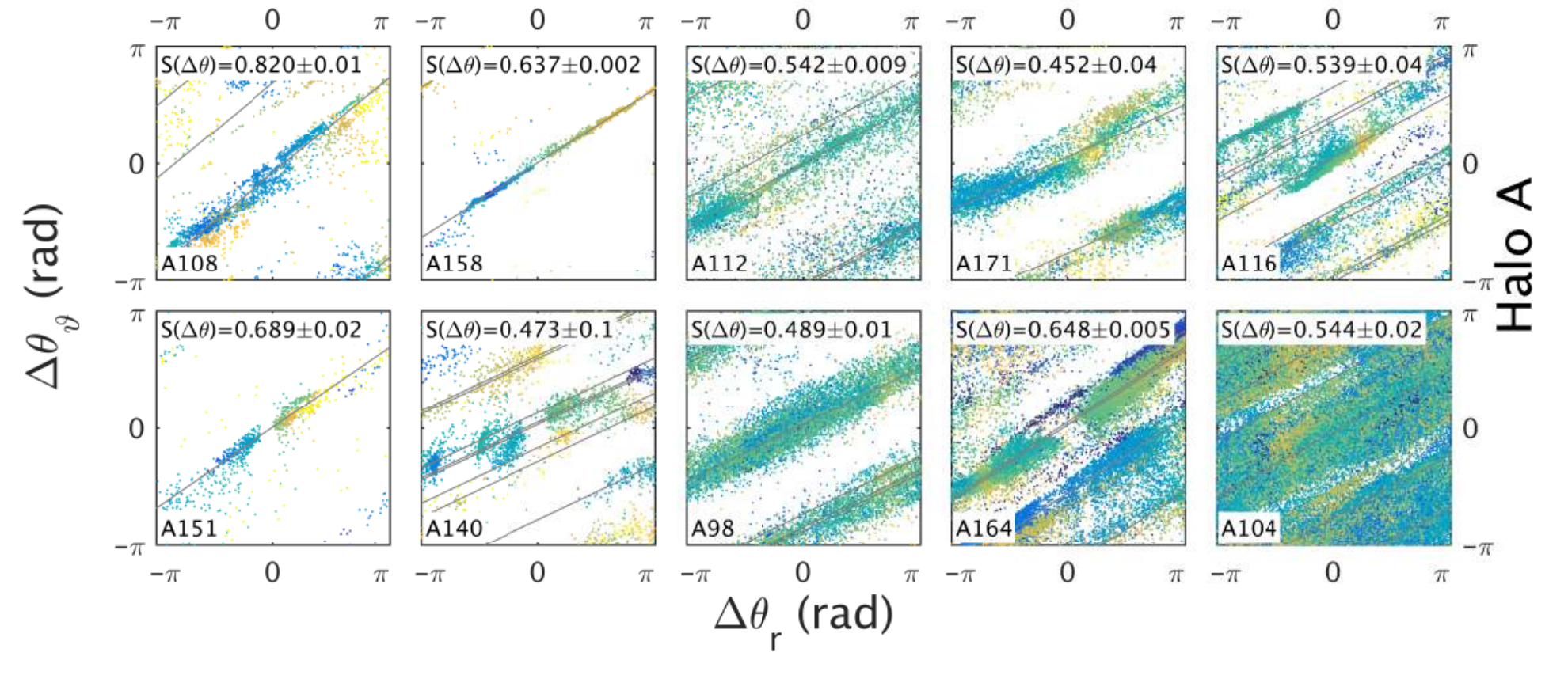}}
\noindent\makebox[\textwidth]{
\includegraphics[width=0.95\textwidth]{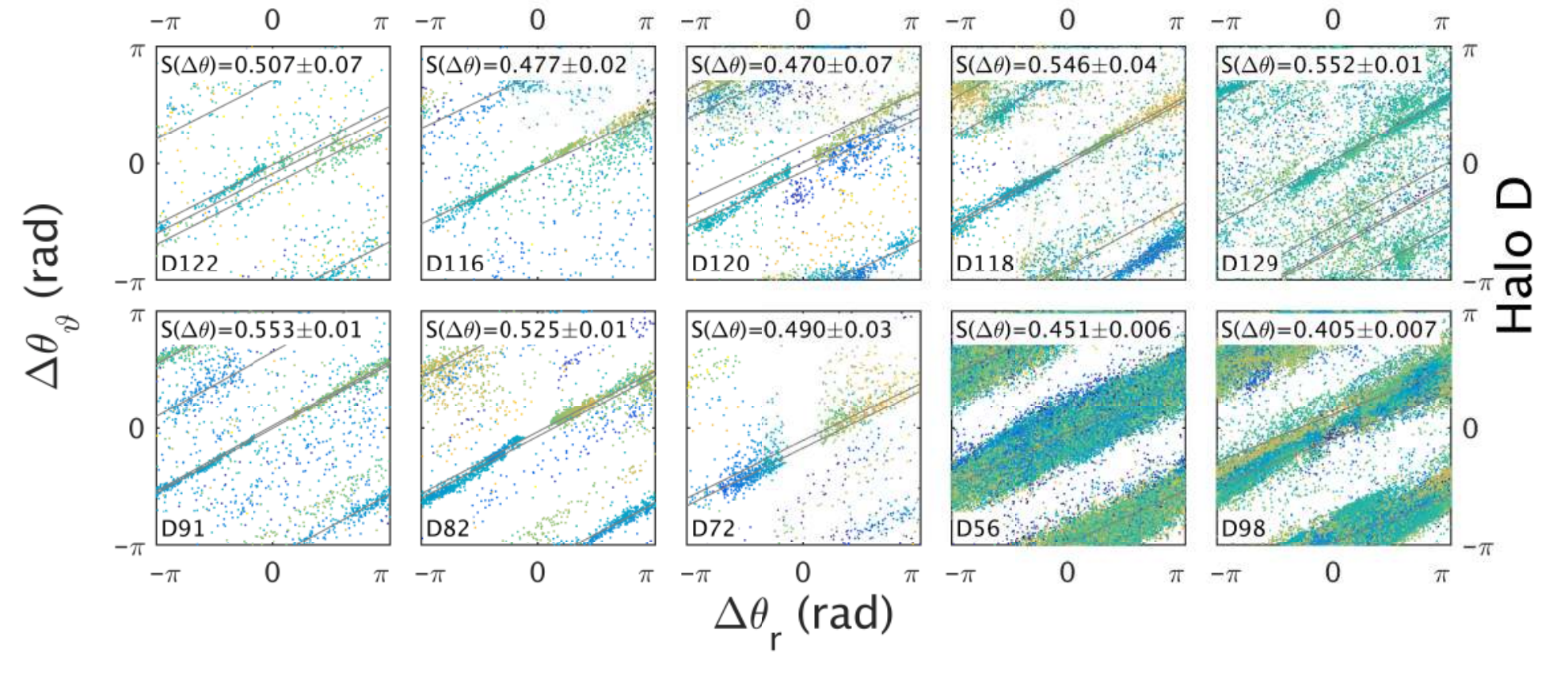}}
\vspace{-0.8cm}
\caption{\small As Fig.~\ref{fig4:aquarius_anglesAD}, but now the $\theta_r$-$\theta_\vartheta$ angle space for the streams selected from halo Aq-A (top panels) and Aq-D (bottom panels).}
\label{fig4:aquarius_anglesAD_2}
\end{figure*}

\subsection{Computing the angles and frequencies}
In this section, and as a first approximation, we assume the
mass distribution of the Aquarius halos is represented by our
best fitting spherical NFW potential. For this potential we compute the
actions, angles and frequencies \citep[see][for the general procedure]{Goldstein1950, BinneyTremaine2008}. As noted above, in
a spherical potential, $\Omega_\phi$ and $\Omega_\vartheta$ are equal
by definition, however the corresponding
angles $\theta_\vartheta$ and $\theta_\phi$ will be different for the
particles in our streams. This is because they depend on the current
positions and velocities of the particles which are the result of the
evolution in a different potential than used to compute the
angles\footnote{When applied to our streams, we lose a small fraction
  of the particles in this procedure, for example because some of the
  numerical integrals do not converge well if the particles are almost
  unbound in the approximated spherical potential.}.
  
In Figs.~\ref{fig4:aquarius_anglesAD} and
\ref{fig4:aquarius_anglesAD_2} we show the distribution of particles
in the $\theta_r$-$\theta_\phi$ and
$\theta_r$-$\theta_\vartheta$ spaces respectively. The most striking feature
in these figures is that each of the streams is distributed along more
or less straight lines \citep{Sanders2013b,BuistHelmi2015}, even
though the host potential is really not spherical. For all streams,
the behaviour in $\theta_r$-$\theta_\vartheta$ space is generally
cleaner and the individual streams are seen more clearly than in
$\theta_r$-$\theta_\phi$ space. That such a difference exists is of course a reflection of the halos being non-spherical.

\begin{table*}[!htbp]
\small
\centering
\caption{\small Overview of the fitted slopes obtained assuming a spherical NFW potential for Aquarius haloes Aq-A and Aq-D from Figs.~\ref{fig4:aquarius_anglesAD}, \ref{fig4:aquarius_anglesAD_2}, and \ref{fig4:aquarius_frequenciesAD}.} 
\def\arraystretch{1.2}
\begin{tabular*}{0.9\textwidth}{rlll|rlll}
\hline
\hline
\specialcell[t]{Stream} & \specialcell[t]{$S(\Delta\theta_{r,\phi})$\ \ \ } & \specialcell[t]{$S(\Delta\theta_{r,\vartheta})$\ \ \ } & \specialcell[t]{$S(\Delta\Omega)$\ \ \ } &  \specialcell[t]{Stream} & \specialcell[t]{$S(\Delta\theta_{r,\phi})$\ \ \ } & \specialcell[t]{$S(\Delta\theta_{r,\vartheta})$\ \ \ } & \specialcell[t]{$S(\Delta\Omega)$\ \ \ } \\
\hline
A98 & $0.260\pm0.06$ & $0.489\pm0.01$ & $0.566\pm0.001$ & D56 & $0.415\pm0.07$ & $0.451\pm0.006$ & $0.561\pm0.0004$\\
A104 & $0.641\pm0.03$ & $0.544\pm0.02$ & $0.568\pm0.0005$ & D72 & $0.561\pm0.01$ & $0.490\pm0.03$ & $0.585\pm0.002$\\
A108 & $0.758\pm0.06$ & $0.820\pm0.01$ & $0.525\pm0.002$ & D82 & $0.638\pm0.01$ & $0.525\pm0.01$ & $0.544\pm0.001$\\
A112 & $0.413\pm0.03$ & $0.542\pm0.009$ & $0.577\pm0.002$ & D91 & $0.531\pm0.01$ & $0.553\pm0.01$ & $0.571\pm0.002$\\
A116 & $1.024\pm0.04$ & $0.539\pm0.04$ & $0.574\pm0.003$ & D98 & $0.714\pm0.04$ & $0.405\pm0.007$ & $0.548\pm0.0003$\\
A140 & $0.719\pm0.007$ & $0.473\pm0.1$ & $0.618\pm0.002$ & D116 & $0.439\pm0.03$ & $0.477\pm0.02$ & $0.605\pm0.002$\\
A151 & $0.678\pm0.04$ & $0.689\pm0.02$ & $0.710\pm0.005$ & D118 & $0.474\pm0.04$ & $0.546\pm0.04$ & $0.566\pm0.002$\\
A158 & $0.616\pm0.004$ & $0.637\pm0.002$ & $0.652\pm0.003$ & D120 & $0.627\pm0.007$ & $0.470\pm0.07$ & $0.579\pm0.001$\\
A164 & $0.681\pm0.06$ & $0.648\pm0.005$ & $0.632\pm0.0008$ & D122 & $0.609\pm0.07$ & $0.507\pm0.07$ & $0.588\pm0.004$\\
A171 & $0.983\pm0.06$ & $0.452\pm0.04$ & $0.572\pm0.002$ & D129 & $0.532\pm0.06$ & $0.552\pm0.01$ & $0.575\pm0.002$\\ 
\hline\\
\end{tabular*}
\label{tab:aquarius_slopes}
\end{table*}

We fit straight lines to the streams, in analogy to what we did in
\cite{BuistHelmi2015}, and the results are given also in the figures and in Table~\ref{tab:aquarius_slopes}. Our method to fit lines to the distributions of
streams in angle space is discussed in Appendix
\ref{sec4:Appendix4A}. We fit the distributions in
$\theta_r$-$\theta_\phi$ and $\theta_r$-$\theta_\vartheta$ separately because as we just saw,
the behaviour in these spaces is different. Particles that are still
bound to the progenitor are removed from the fitting procedure because they are
not expected to follow the general action-angle behaviour of the stream. The result of this removal is clearly visible for example in stream D72 and A164 where we see a gap in the middle of the distribution in angle space.

The fitting procedure puts into evidence several noticeable
distortions in angle space because the different wraps of a
stream are not always on parallel lines, such as for stream A108 and A116. Much more subtle is the deviation seen in the bottom-left of stream A158. These distortions clearly pose a challenge to the determination of the
slope. For this reason we have to take the slopes determined with our
method generally with some care (especially in the case of A116) as
the quoted errors do not account for such systematic uncertainties. A
quick visual inspection usually helps to evaluate the outcome and
the reliability of the fit.  The slopes in $\theta_r$-$\theta_\phi$ space are much more varied than those
in $\theta_r$-$\theta_\vartheta$ space, which especially for
halo Aq-D all cluster around $0.5$. Generally, we also notice that the streams in halo Aq-D seem less distorted than those in halo Aq-A. 

In our selection of streams we also included several objects on very
radial orbits, and whose debris is distributed in an hour-glass shape
in configuration space, such as A98, D56 and D98. In
$\theta_r$-$\theta_\phi$ space they seem very mixed and do not show
very distinct structures, yet their behaviour in
$\theta_r$-$\theta_\phi$ is remarkable with streams along straight
lines being clearly apparent (see e.g. D56 and D98 in
Fig.~\ref{fig4:aquarius_anglesAD_2}).  These objects are typically
more massive and have deposited debris close to the centre of the halo
and therefore are more phase mixed.

\begin{figure*}[!htbp]
\centering
\vspace{-0.2cm}
\noindent\makebox[\textwidth]{
\includegraphics[width=0.95\textwidth]{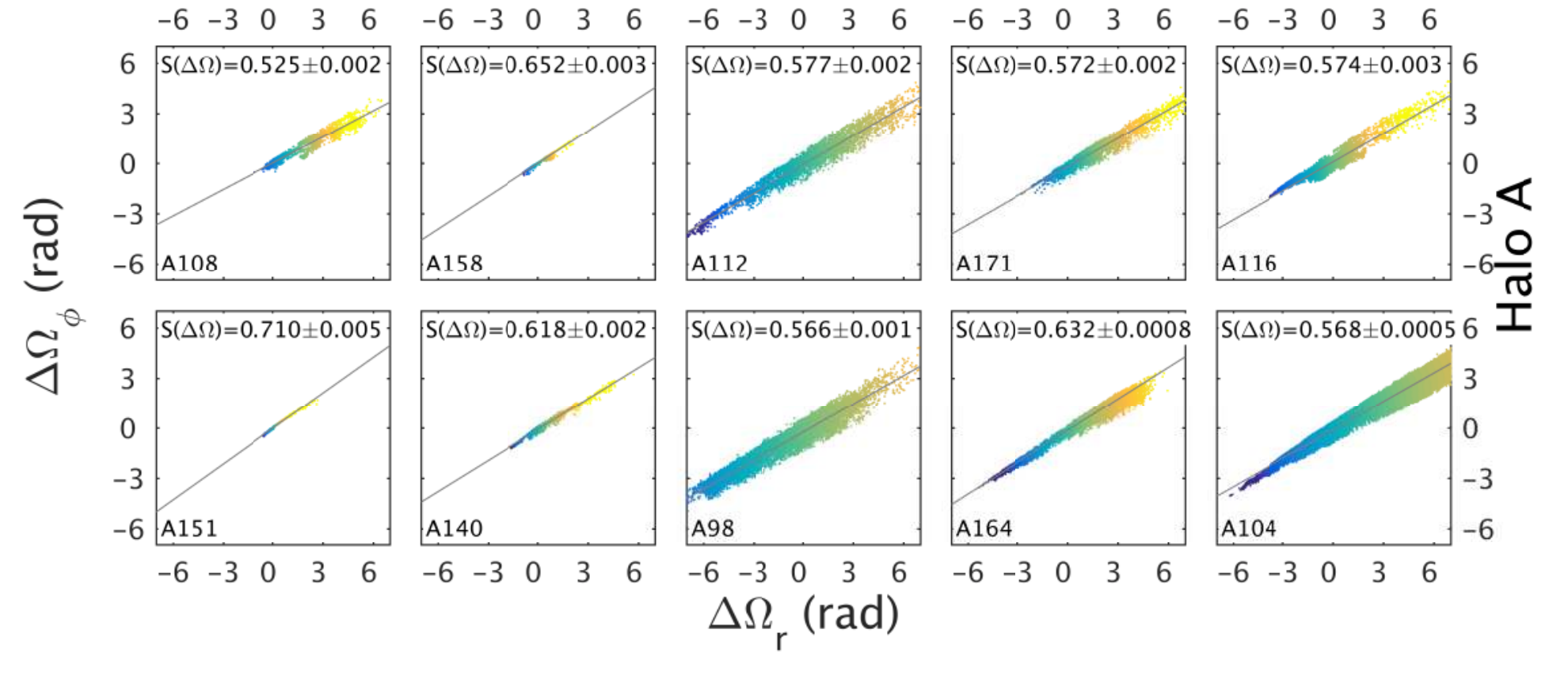}}
\noindent\makebox[\textwidth]{
\includegraphics[width=0.95\textwidth]{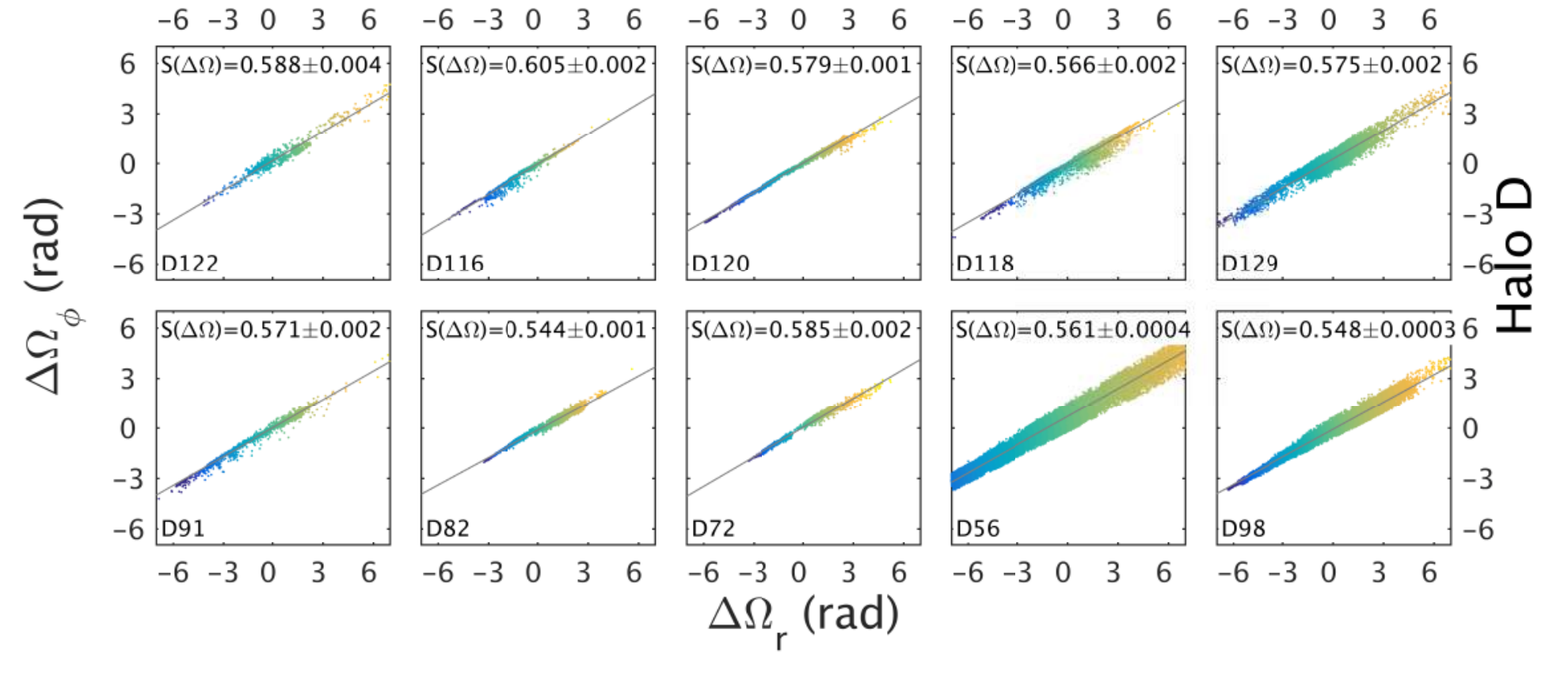}}
\vspace{-0.8cm}
\caption{\small Frequency distributions computed using the best fitting spherical potentials for our selection of streams from halo Aq-A (top panels) and Aq-D (bottom panels) shown in previous figures. The panels are centred around the same particle as in Figs.~\ref{fig4:aquarius_anglesAD} and \ref{fig4:aquarius_anglesAD_2}. The distributions were fitted with straight lines after removing generously the bound particles from the progenitor. The colours indicate the energy gradient for the spherical potential. The insets show the fitted slope and estimated errors.}
\label{fig4:aquarius_frequenciesAD}
\end{figure*}

\subsection{Comparing the slopes in frequency and angle space}

In Fig.~\ref{fig4:aquarius_frequenciesAD} we show the frequency
distributions of the streams. These follow closely a straight
line, although sometimes they are quite thick. For A108
and A116 the width in frequency space is not everywhere the
same. This is one of the signatures expected when using the wrong
potential to compute the frequencies. The determination of the fitted slopes is more robust in frequency
space, and these show a considerably smaller range than the slopes in
angle space. 

The slopes in angle space and frequency space
differ, especially when comparing to the $\theta_r$-$\theta_\phi$
space to frequency space. They are expected to be equal in the true
(static) potential \citep{Sanders2013b,Sanders2013a}. In
\cite{BuistHelmi2015} we found that in a time-dependent potential, the
slope in angle space is steeper than in frequency space
(i.e. $S(\Delta\theta) > S(\Delta\Omega)$). Therefore, since the potentials
in the Aquarius N-body simulations have grown in time we might expect the angle
space slope to be larger. For some of the streams in our sample this
is indeed the case, but there are quite a few streams for which this
does not hold, such as D116 and A112, for which the
magnitude of the angle-frequency difference is far greater than what
can be expected from (adiabatic) evolution of the halo. Also in
$\theta_r$-$\theta_\vartheta$ space many streams are not in line with
our expectations. Overall we find that in $\theta_r$-$\theta_\vartheta$,
28 of the 35 streams have a larger slope in frequency than in angle
space, i.e.\ they do not follow the expected behaviour, and while for
$\theta_r$-$\theta_\phi$, this is the case for 17 out of the 35 streams.

The simulations of \cite{BuistHelmi2015} that follow the
  evolution of streams in a spherical potential, show that differences between
  the slopes can be obtained when the wrong potential is assumed. For
  example, $S(\Delta\theta) - S(\Delta\Omega) > 0$ when the enclosed
  mass is too high, and $S(\Delta\theta) - S(\Delta\Omega) < 0$ when
  the enclosed mass is too low.  We therefore may attribute the
differences in slope to having assumed an incorrect potential (mostly in shape), and not to time-dependence.

\subsection{Influence of the potential}

We also expect an energy gradient to be present along a stream
\citep{BuistHelmi2015}, and although this is visible in frequency
space it is less clear in angle space. This is not unexpected as 
the frequencies depend on energy (via the actions) and more energetic particles move faster,  
while, on the other hand, the angles depend also on other phase-space coordinates. Even small (of order $10$\%)
differences in the characteristic parameters of the potential can lead
to the energy gradient being lost in angle space, even when the stream itself has a
normal appearance \citep{BuistHelmi2015}. An exception is A164 in
$\theta_r$-$\theta_\phi$-space, which does seem to have a continuous
energy gradient when following the stream along its various wraps. 

To see if the behaviour in angle space and the
energy gradient could be improved we experimented by changing one of the characteristic parameters, namely the enclosed mass for the particular case of A158. However we were unable to remove the bend seen 
at the bottom-left in $\theta_r$-$\theta_\phi$ space, a behaviour that
in the spherical case is known to be indicative of wrong values of the
characteristic parameters of the potential. It seems clear from the
analysis presented in this section that it is the shape of the potential that is
wrong rather than the value of the enclosed mass.

We conclude that streams can still look rather regular in angle and
frequency space when making an incorrect assumption about the shape of
the potential, even if they evolved in a potential that grew via
accretion and merging. In later sections of this Paper we will
investigate the conditions for streams to be distributed along
straight lines such as those seen in
Figs.~\ref{fig4:aquarius_anglesAD}, \ref{fig4:aquarius_anglesAD_2},
and \ref{fig4:aquarius_frequenciesAD}.

\section{Test-particle simulations of streams in axisymmetric potentials}
\label{sec4:testparticlesaxisymmetric}

The Aquarius halos are not spherically symmetric, and it is important
to understand which of the deviations seen in the streams' behaviour
in angle space arise from having assumed a spherical
potential. Rather than attempting to study their dynamics in a
  full triaxial potential, we add only one degree of complexity and
  now analyse a set of example streams evolved in the
  axisymmetric Kuzmin-Kutuzov Staeckel potential \citep{Dejonghe1988}
 with similar density axis ratios as halo Aq-D of the Aquarius
  simulations.\footnote{We focus on Aq-D because it has a less complex mass distribution than Aq-A, as can be seen from Fig.~\ref{fig4:fitAqAD}}.  For the streams we have used as initial orbital conditions those
extracted from the present-day positions and velocities of particles
in halo Aq-D (as described in
Sec.~\ref{sec4:streamsmorphologyandorbits}). An axisymmetric Staeckel potential allows for a
  relatively straightforward computation of the actions and angles.  This implies that we can directly compare their
behaviour using true action-angles to those computed assuming a
spherical approximation to this potential.

\subsection{Potential set-up}
\label{sec4:potentialsetup}

To motivate our choice of the characteristic parameters of our
axisymmetric potential, we use the Aquarius halos. These halos are
triaxial, and may be described with two axis ratios: $q = c/a$ (long
to short axis ratio) and $s=b/a$ (intermediate to long axis ratio,
i.e. $a \geq b \geq c$). A useful quantity is the triaxiality
parameter \citep{Franx1991}
\begin{equation}
	T = \frac{a^2-b^2}{a^2-c^2} = \frac{1-s^2}{1-q^2},
	\label{eq4:triaxiality}
\end{equation}
which is zero for an oblate halo and equals unity for a prolate halo.
Halos Aq-A and Aq-D have triaxiality parameters between $T=2/3$ and
$1$ at a radius of $50$ kpc, making them somewhat more prolate, with
halo Aq-A being more triaxial than halo Aq-D. Therefore these halos are not in the range of being extremely 
triaxial ($T=1/3$ to $2/3$, see also
\citealt{Warren1992}), although at larger radii their triaxiality increases \citep{VeraCiro2011}.

The functional form of the Kuzmin-Kutuzov potential is given by \citep{Dejonghe1988}:
\begin{equation}
	\Phi_\textrm{K}(R, z) = -\frac{G M_\textrm{K}}{\sqrt{a_\textrm{K}^2 + c_\textrm{K}^2 + R^2 + z^2 + 2\sqrt{a_\textrm{K}^2 c_\textrm{K}^2 + R^2 c_\textrm{K}^2 + a_\textrm{K}^2 z^2}}},
\end{equation}
where $M_{\rm K}, a_{\rm K}, c_{\rm K}$ are characteristic mass and scale parameters, respectively. 
A coordinate transformation to prolate ellipsoidal coordinates ($\lambda$, $\nu$, $\phi$) leads to a simple expression
\begin{equation}
	\Phi_\textrm{K}(\lambda, \nu) = -\frac{G M_\textrm{K} }{\sqrt{\lambda} + \sqrt{\nu}},
\label{eq:KK_lambda-nu}
\end{equation}
where the relations between $\lambda$, $\nu$ and $R$, $z$ can be expressed as
\begin{equation}
\begin{split}
\lambda\nu = c_\textrm{K}^2 R^2 + a_\textrm{K}^2 z^2 + a_\textrm{K}^2 c_\textrm{K}^2, \\
\lambda+\nu = R^2 + z^2 + a_\textrm{K}^2 + c_\textrm{K}^2.
\end{split}
\label{eq4:ellipsoidalcoords}
\end{equation}
We take $\lambda > \nu$. Depending on the ratio $c_\textrm{K}/a_\textrm{K}$, this coordinate system can be used to represent a prolate ($q_\textrm{K}>1$) or an oblate ($q_\textrm{K} < 1$) mass distribution (see also the discussion in \citealt{Dejonghe1988}). The potential in Eq.~(\ref{eq:KK_lambda-nu}) reduces in the spherical limit ($a_\textrm{K} =
c_\textrm{K}$) to Henon's isochrone potential \citep{Dejonghe1988,
  Henon1959}. 
For convenience we link the parameters $M_\textrm{K}$,
$a_\textrm{K}$ and $c_\textrm{K}$ to the isochrone scale radius
$r_\textrm{iso}$ and scale mass $M_\textrm{iso}$, and introduce the
flattening parameter $q_\textrm{K} \equiv c_\textrm{K} /
a_\textrm{K}$. We define the isochrone scale mass as
$M_\textrm{iso} = M_\textrm{isochrone}(r_\textrm{iso})$, such that
\begin{equation}
\begin{split}
	M_\textrm{K} &= M_\textrm{iso} \left(\frac{3}{\sqrt{2}} - 2\right)^{-1}, \\
	a_\textrm{K} &= r_\textrm{iso}\frac{2}{1+q_\textrm{K}}, \\
    c_\textrm{K} &= r_\textrm{iso}\frac{2 q_\textrm{K}}{1+q_\textrm{K}}.
\end{split}
\label{eq4:isochrone_KuzminKutuzov}
\end{equation}
These relations satisfy $a_\textrm{K} + c_\textrm{K} = 2 r_\textrm{iso}$, i.e.\ the isochrone scale radius is the average of the Kuzmin-Kutuzov axis lengths. 
For the determination of the numerical values of the parameters we refer to Appendix \ref{sec4:potentialparametersnumerical}, where we set $q_\textrm{K} = 1.87$, $M_\textrm{iso} = 1.58 \times 10^{11}$ M${}_\odot$ and $r_\textrm{iso} = 22.84$ kpc.

\begin{figure*}[!htbp]
\centering
\noindent\makebox[\textwidth]{
\includegraphics[width=0.95\textwidth]{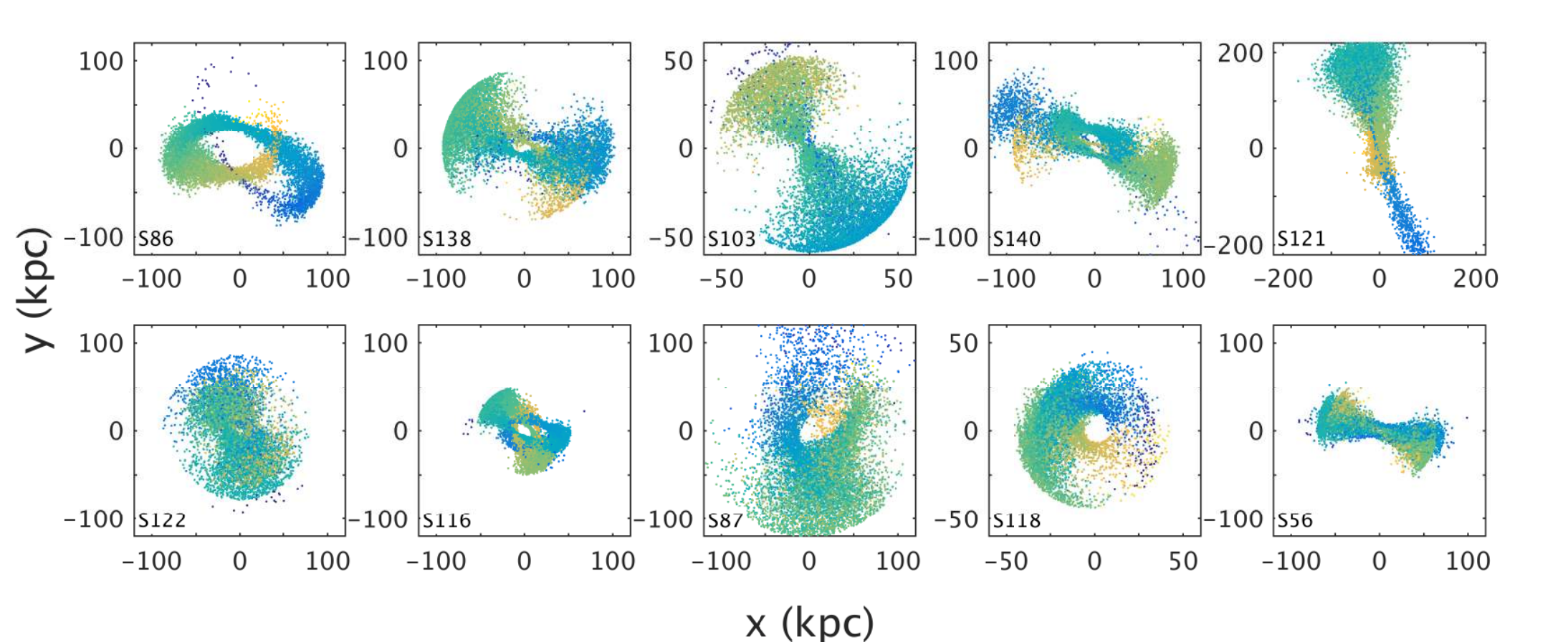}}
\noindent\makebox[\textwidth]{
\includegraphics[width=0.95\textwidth]{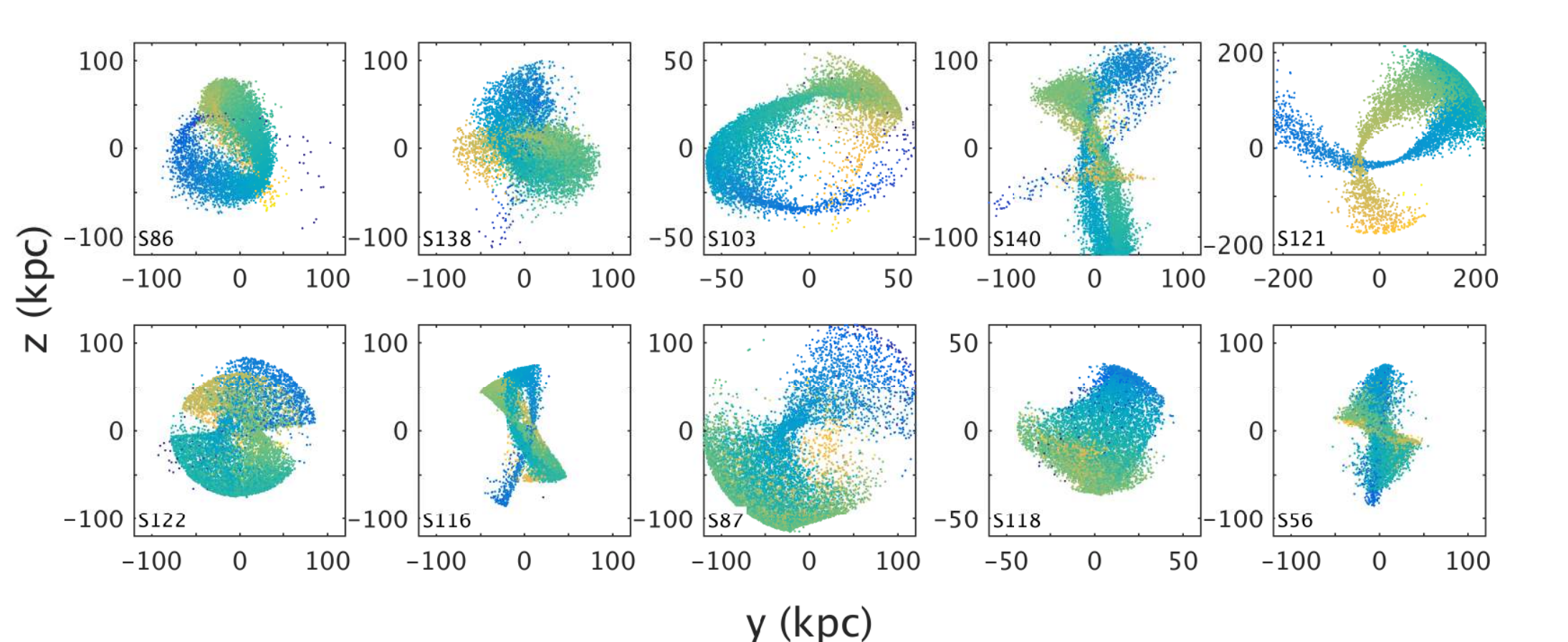}}
\caption{\small Spatial distribution of our selection of test-particle streams after 10 Gyr of evolution in the axisymmetric Kuzmin-Kutuzov potential. The colours represent the energy gradient with the most bound particles in yellow and those least bound in blue. The $z$-axis is aligned with the major axis of the potential.}
\label{fig4:simulation_XY}
\end{figure*}

\subsection{Streams set-up}

As mentioned earlier we use a subset of the streams' positions and velocities extracted from the Aquarius simulations to evolve our test-particle simulations. The stream progenitor is composed of 10,000 particles that follow an isotropic Gaussian distribution in position and velocity, characterised by dispersions $\sigma_\textrm{pos} = 0.3$~kpc and $\sigma_\textrm{vel} = 10$~kpc/Gyr respectively. This can be roughly translated into a progenitor mass $M$ using that $\sigma_\textrm{vel}^2 \sim GM/R$ and where we take $R = \sigma_\textrm{pos}$, which results in  $M \approx 7.4 \times 10^7$ M${}_\odot$. The orbits are integrated for $10$ Gyr in the Kuzmin-Kutuzov potential with the parameters derived from Aquarius halo D. 

Fig.~\ref{fig4:simulation_XY} shows our streams, where the inset labels are the IDs of the corresponding Aq-D streams (e.g. S140 is based on the position and velocity of one particle in D140). Recall that it is not our goal to reproduce the original streams from
the Aquarius halos in the axisymmetric limit, but to understand the behaviour of streams in the non-spherical regime.

\subsection{Action-angles in the true potential}
\label{sec4:true_angles}
\begin{figure*}[!htbp]
\centering
\noindent\makebox[\textwidth]{
\includegraphics[width=0.95\textwidth]{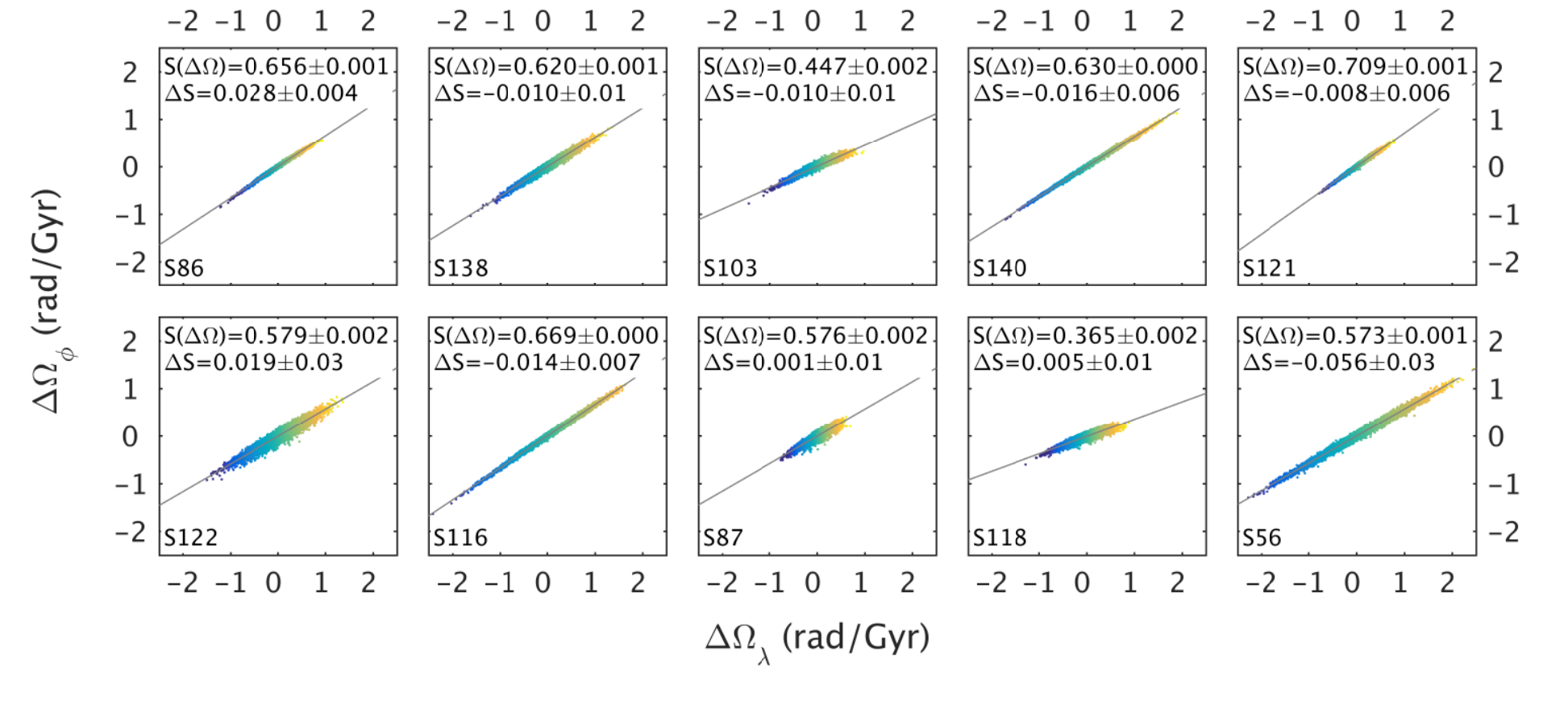}}
\noindent\makebox[\textwidth]{
\includegraphics[width=0.95\textwidth]{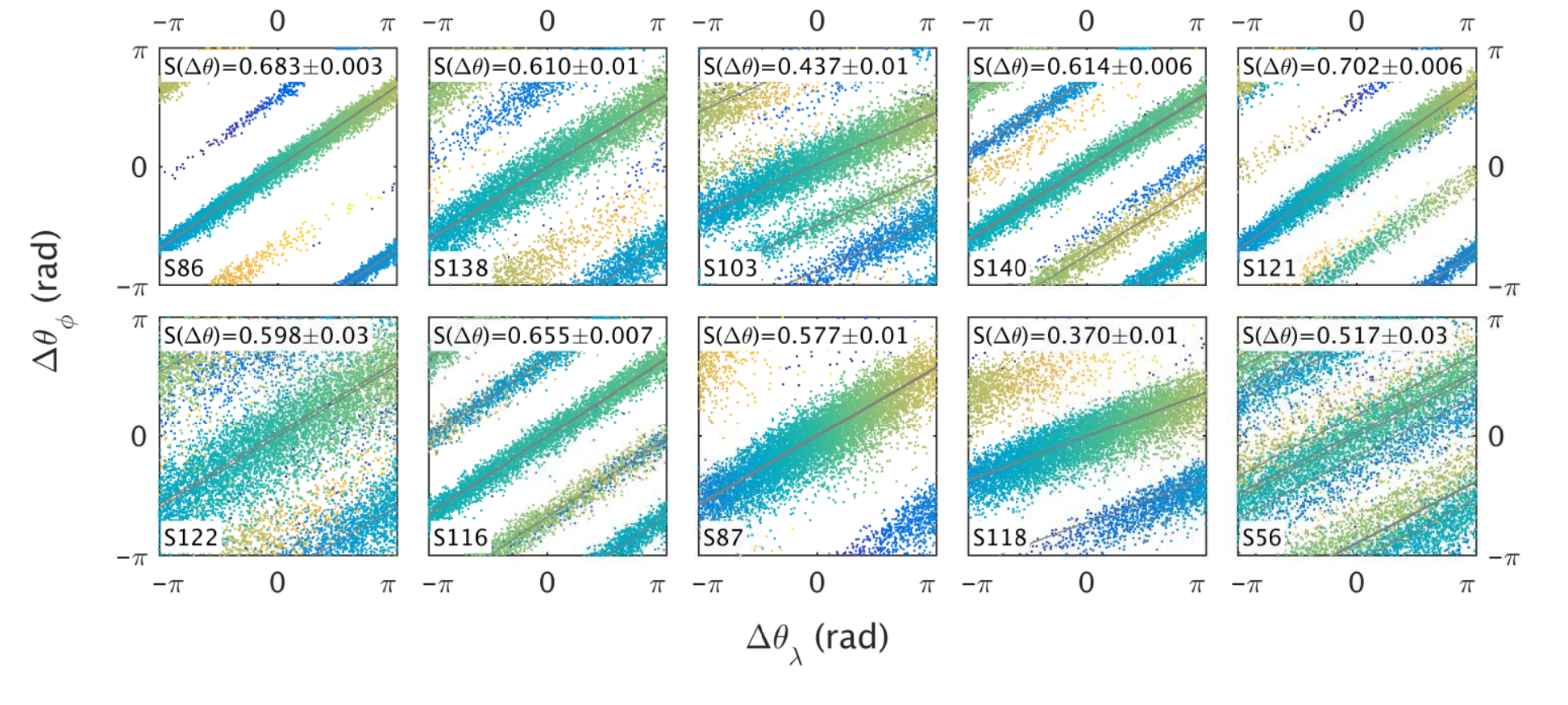}}
\vspace{-1cm}
\caption{\small Test-particle streams in the true $\lambda$ - $\phi$ frequency (top) and angle (bottom) spaces after 10 Gyr of evolution in the Kuzmin-Kutuzov potential. The colours represent the energy gradient with the most bound particles in yellow and those least bound in blue. The panels have been centred on the current position of the centre of mass of the progenitor system. The insets give the best-fitting slopes to the distributions as well as the angle-frequency misalignment $\Delta S$.}
\label{fig4:simulation_anglesfrequencies_lambdaphi}
\end{figure*}

\begin{figure*}[!htbp]
\centering
\noindent\makebox[\textwidth]{
\includegraphics[width=0.95\textwidth]{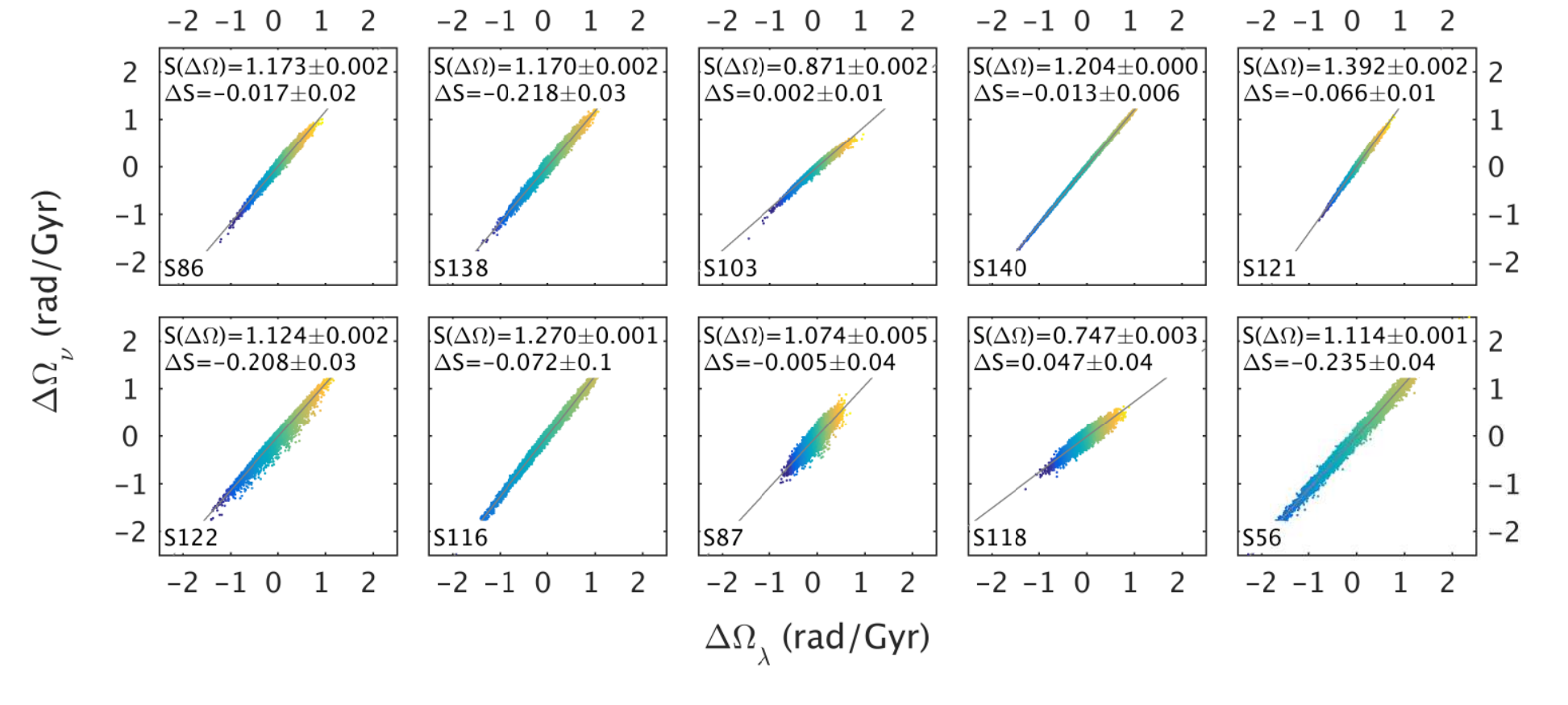}}
\noindent\makebox[\textwidth]{
\includegraphics[width=0.95\textwidth]{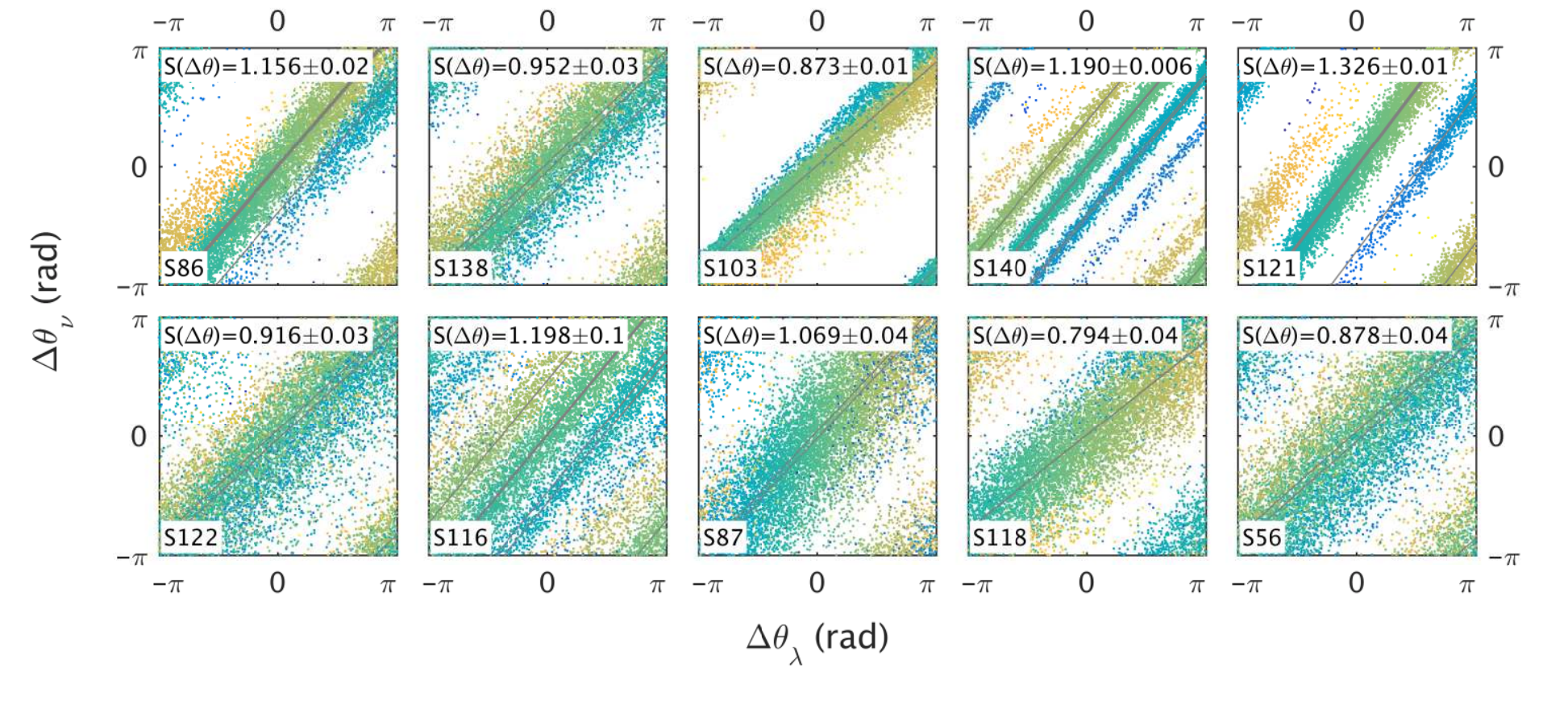}}
\vspace{-1cm}
\caption{\small Same as in Fig.~\ref{fig4:simulation_anglesfrequencies_lambdaphi}, but now for the $\nu$ - $\phi$ frequency and angle spaces.}
\label{fig4:simulation_anglesfrequencies_nuphi}
\end{figure*}

\begin{figure*}[!htbp]
\centering
\noindent\makebox[\textwidth]{
\includegraphics[width=0.5\textwidth]{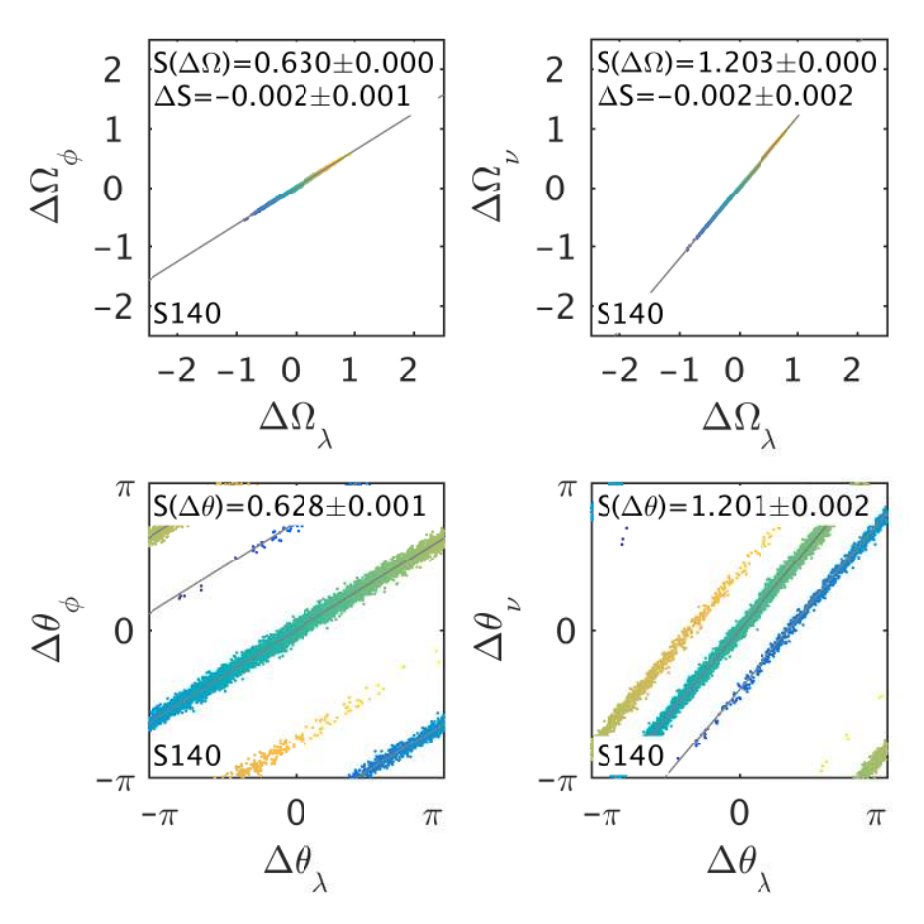}}
\vspace{-1cm}
\caption{\small Stream S140 but now with a Carina-like progenitor,  after 10 Gyr of evolution in the Kuzmin-Kutuzov potential. The colours and insets are the same as in Fig.~\ref{fig4:simulation_anglesfrequencies_lambdaphi}. With the smaller progenitor the stream is much thinner and the error on the fitting is significantly reduced, which basically removes the angle-frequency misalignment.}
\label{fig4:simulation_anglesfrequencies_carina}
\end{figure*}

\begin{table*}[!htbp]
\small
\centering
\caption{\small Overview of the fitted slopes in Figs.~\ref{fig4:simulation_anglesfrequencies_lambdaphi} and \ref{fig4:simulation_anglesfrequencies_nuphi} of our test-particle streams integrated in an axisymmetric Staeckel potential.} 
\def\arraystretch{1.2}
\begin{tabular*}{0.9\textwidth}{rllllll}
\hline
\hline
\specialcell[t]{Stream} & \specialcell[t]{$S(\Delta\theta_{\lambda,\phi})$\ \ \ } & \specialcell[t]{$S(\Omega_{\lambda,\phi})$\ \ \ } & \specialcell[t]{$\Delta S_{\lambda,\,\phi}$\ \ \ } & \specialcell[t]{$S(\Delta\theta_{\lambda,\nu})$\ \ \ } & \specialcell[t]{$S(\Omega_{\lambda,\nu})$\ \ \ } & \specialcell[t]{$\Delta S_{\lambda,\,\nu}$\ \ \ } \\
\hline
S56 & $0.517\pm0.03$ & $0.573\pm0.0006$ & $-0.056\pm0.03$ & $0.878\pm0.04$ & $1.114\pm0.001$ & $-0.235\pm0.04$ \\
S86 & $0.683\pm0.003$ & $0.656\pm0.0006$ & $0.028\pm0.004$ & $1.156\pm0.02$ & $1.173\pm0.002$ & $-0.017\pm0.02$ \\
S87 & $0.577\pm0.01$ & $0.576\pm0.002$ & $0.001\pm0.01$ & $1.069\pm0.04$ & $1.074\pm0.005$ & $-0.005\pm0.04$ \\
S103 & $0.437\pm0.01$ & $0.447\pm0.002$ & $-0.010\pm0.01$ & $0.873\pm0.01$ & $0.871\pm0.002$ & $0.002\pm0.01$ \\
S116 & $0.655\pm0.007$ & $0.669\pm0.0005$ & $-0.014\pm0.007$ & $1.198\pm0.1$ & $1.270\pm0.001$ & $-0.072\pm0.1$ \\
S118 & $0.370\pm0.01$ & $0.365\pm0.002$ & $0.005\pm0.01$ & $0.794\pm0.04$ & $0.747\pm0.003$ & $0.047\pm0.04$ \\
S121 & $0.702\pm0.006$ & $0.709\pm0.0009$ & $-0.008\pm0.006$ & $1.326\pm0.01$ & $1.392\pm0.002$ & $-0.066\pm0.01$ \\
S122 & $0.598\pm0.03$ & $0.579\pm0.002$ & $0.019\pm0.03$ & $0.916\pm0.03$ & $1.124\pm0.002$ & $-0.208\pm0.03$ \\
S138 & $0.610\pm0.01$ & $0.620\pm0.001$ & $-0.010\pm0.01$ & $0.952\pm0.03$ & $1.170\pm0.002$ & $-0.218\pm0.03$ \\
S140 & $0.614\pm0.006$ & $0.630\pm0.0004$ & $-0.016\pm0.006$ & $1.190\pm0.006$ & $1.204\pm0.0004$ & $-0.013\pm0.006$ \\
\hline\\
\end{tabular*}
\label{tab:simulation_slopes}
\end{table*}

For the Kuzmin-Kutuzov potential the Hamilton-Jacobi equation separates in prolate spheroidal coordinates ($\lambda$, $\nu$, $\phi$). This means that we can compute the true actions ($J_\lambda$, $J_\nu$, $J_\phi$), the corresponding frequencies, and the angles in a straightforward manner\footnote{For a discussion how to compute the actions in an axisymmetric potential we refer to \citet{Helmi1999, Sanders2012}.}. At large distances from the centre $\lambda$ behaves like the radial coordinate $r$ from the spherical case, while $\nu$ follows almost the longitudinal angle $\vartheta$. For this reason, we will generally compare $\Omega_\lambda$ with $\Omega_r$, and $\Omega_\nu$ with $\Omega_\vartheta$ (and the same for their corresponding angles). 

The results for each of the streams in Fig.~\ref{fig4:simulation_XY} are shown in Fig.~\ref{fig4:simulation_anglesfrequencies_lambdaphi} and Fig.~\ref{fig4:simulation_anglesfrequencies_nuphi} for both projections of frequency space (top) and angle space (bottom). In angle space, the streams are distributed along straight lines (just as in frequency space), spreading out after 10 Gyr of evolution.  
The insets in the figures show the slopes obtained from fitting
straight lines to the distributions, where the error is estimated from
bootstrapping the fits 200 times. These results are also summarised in Table~\ref{tab:simulation_slopes}.

We also give in the top panels the angle-frequency misalignment
$\Delta S = S(\Delta\theta) - S(\Delta\Omega)$, with the corresponding
error. Notice that typically $|\Delta S| \sim 0.01-0.02$, which is
larger than found by \cite{BuistHelmi2015} for the streams evolved in
a static spherical potential.  This is mostly due to the larger
progenitor used here which causes the streams to be wider, and this has an impact
both on the initial spreads and on the fitting procedure. This is
explicitly demonstrated in
Fig.~\ref{fig4:simulation_anglesfrequencies_carina}, where we show a
stream on the same orbit as S140 but now for the Carina-like
progenitor ($\sigma_\textrm{pos}=0.1$~kpc and $\sigma_\textrm{vel}=5$~kpc/Gyr) used in \cite{BuistHelmi2015} and evolved for 10 Gyr. We see
that the angle-frequency misalignment is now consistent with being
zero.

\sectionmark{Behaviour of streams in approximate potentials of varying shape}
\section{Action-angle behaviour of streams in approximate potentials of varying shape}
\sectionmark{Behaviour of streams in approximate potentials of varying shape}
\label{sec4:aabehavioursimulations}

Thus far we have discussed mostly from a theoretical point of view
what to expect in angle and frequency space in the true potential, and
why streams are on straight lines. In this section we explore what
happens when the wrong potential is used to compute the angles and
frequencies. To this end we employ the test-particle simulations of
Sec.~\ref{sec4:testparticlesaxisymmetric} run in the axisymmetric
Staeckel potential, but we will assume a spherical potential
(Sec.~\ref{sec4:wrong_sph}), a Staeckel potential but with a different
flattening parameter ($q_\textrm{K}'=1/q_\textrm{K}$,
Sec.~\ref{sec4:wrong_qk}), and a spherical potential with a more
dissimilar radial mass distribution (Sec.~\ref{sec4:wrong_nfw}). We
also explore the effect of self-gravity in Sec.~\ref{sec4:n-body-ch4}

\subsection{Spherical approximation}
\label{sec4:wrong_sph}
\begin{figure*}[!htbp]
\vspace*{-0.2cm}
\centering
\noindent\makebox[\textwidth]{
\includegraphics[width=0.95\textwidth]{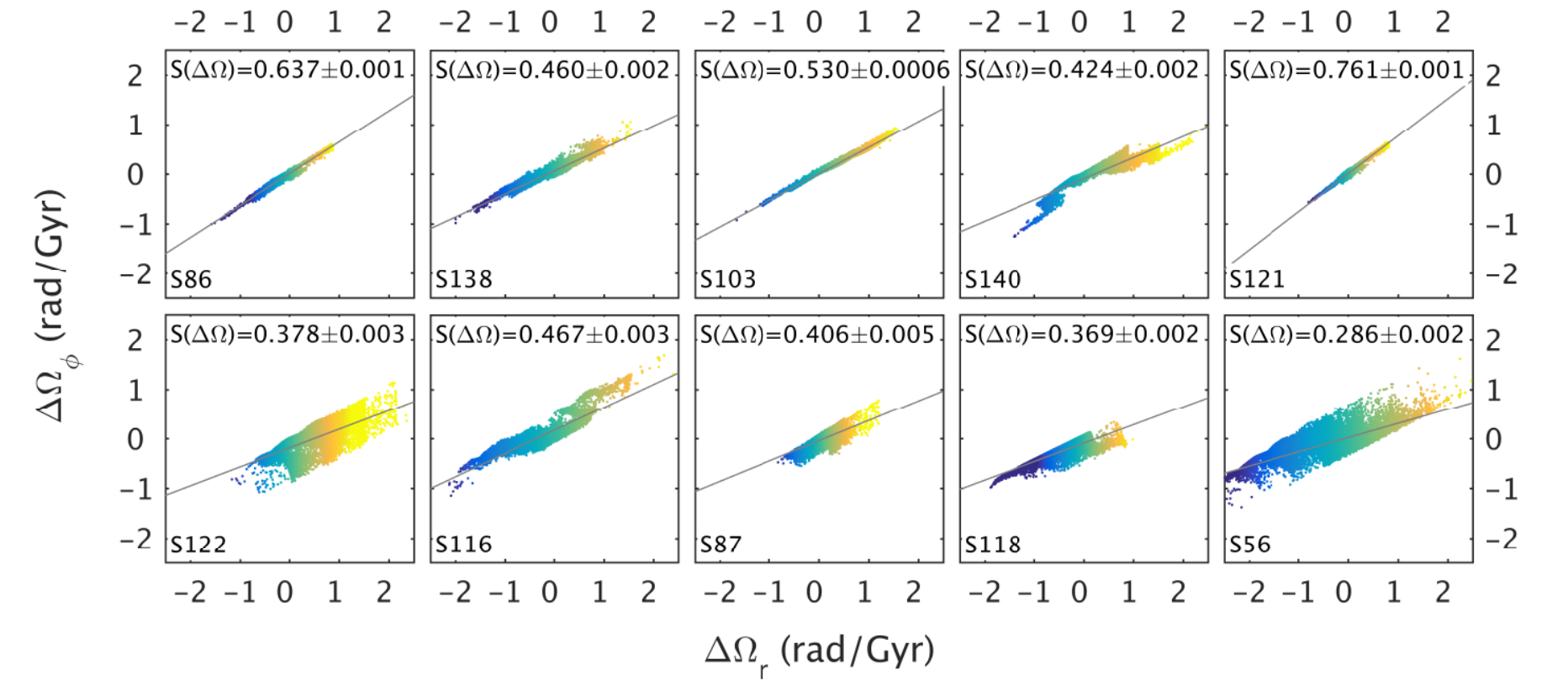}}
\noindent\makebox[\textwidth]{
\includegraphics[width=0.95\textwidth]{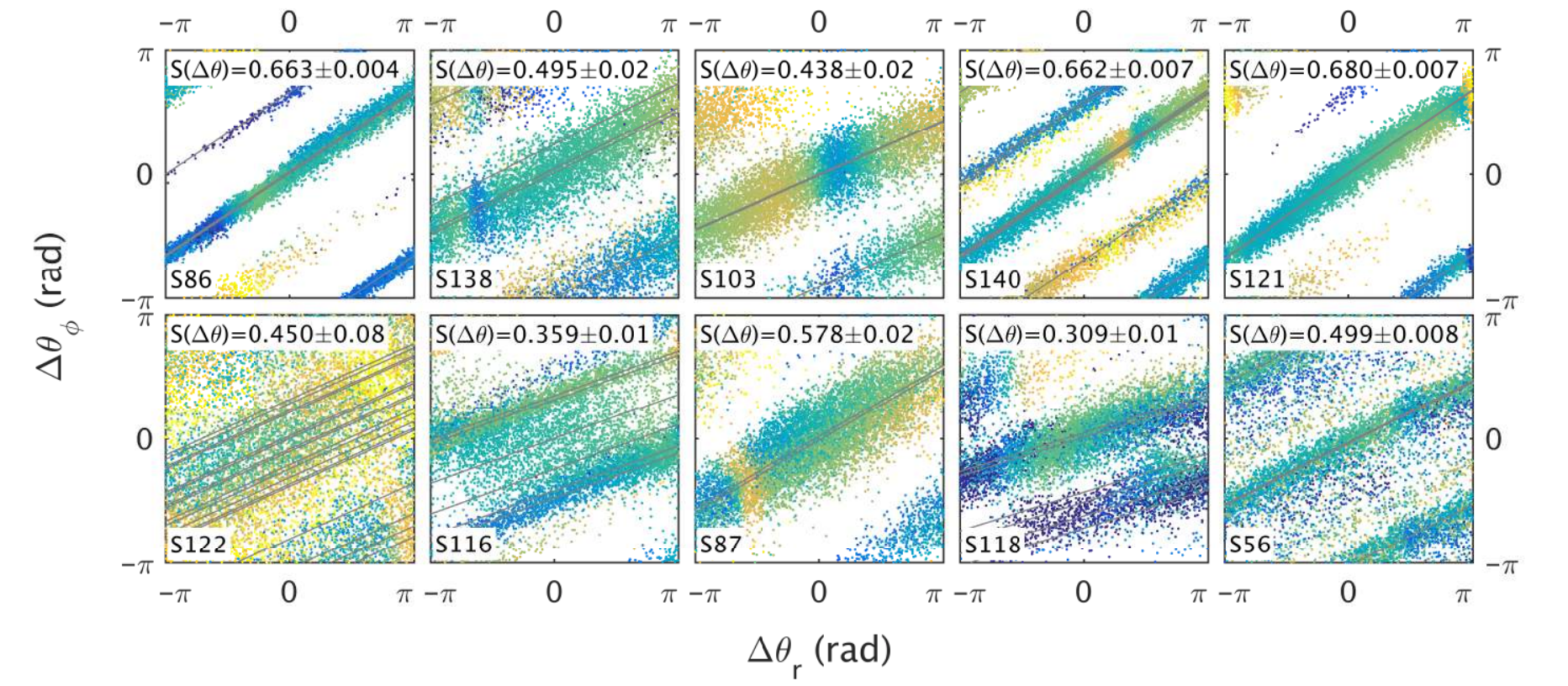}}
\noindent\makebox[\textwidth]{
\includegraphics[width=0.95\textwidth]{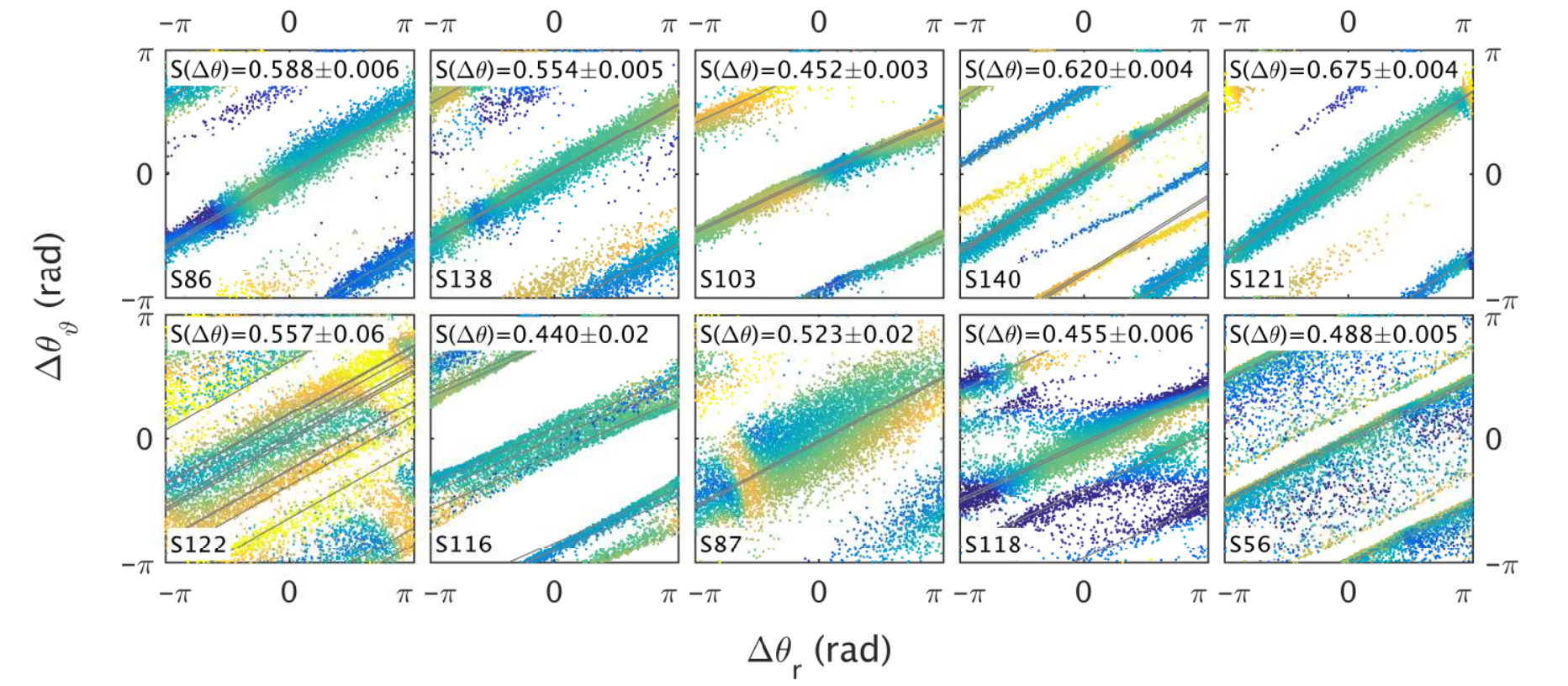}}
\vspace*{-0.8cm}
\caption{\small Test-particle streams evolved in the axisymmetric
  Kuzmin-Kutuzov potential and shown in
  Figs.~\ref{fig4:simulation_XY}-\ref{fig4:simulation_anglesfrequencies_nuphi},
  depicted now assuming a spherical approximation to the potential in
  the space of frequencies $\Delta\Omega_r$-$\Delta\Omega_\phi$ (top
  panels) and angles $\Delta\theta_r$-$\Delta\theta_\phi$ (middle
  panels) and $\Delta\theta_r$-$\Delta\theta_\vartheta$ (bottom
  panels). The
  colours represent the energy gradient in the isochrone
  potential. The panels have been centred on the progenitor
  position. The insets indicate the fitted slopes for the approximate
  potential. The errors were found by bootstrapping the fit 200
  times.}
\label{fig4:simulation_with_spherical}
\end{figure*}

We now compute the angles and frequencies for the test-particle
simulations assuming the isochrone potential. This potential
constitutes the spherical limit of the Kuzmin-Kutuzov potential, and
its characteristic parameters were defined in
Sec.~\ref{sec4:potentialsetup}. The resulting angle and frequency
distributions are shown in Fig.~\ref{fig4:simulation_with_spherical},
where we do not include the $\Omega_r$-$\Omega_\vartheta$ space as it
is redundant in the spherical case. Instead, the behaviour in the
angles $\theta_\phi$ and $\theta_\vartheta$ is relevant since these
additionally depend on the physical location of the particles (which
encodes information about the potential in which they were evolved as
discussed earlier in Sec.~\ref{sec4:aabehaviouraquarius}). The characteristic parameters of the fitted straight lines
are also listed in Table~\ref{tab:simulation_slopes_spherical}.

\begin{table*}[!htbp]
\small
\centering
\caption{\small Overview of the fitted slopes in Fig.~\ref{fig4:simulation_with_spherical} obtained assuming a spherical isochrone potential for our test-particle streams evolved in an axisymmetric Staeckel potential.} 
\def\arraystretch{1.2}
\begin{tabular*}{0.6\textwidth}{rlll}
\hline
\hline
\specialcell[t]{Stream} & \specialcell[t]{$S(\Delta\theta_{r,\phi})$\ \ \ } & \specialcell[t]{$S(\Delta\theta_{r,\vartheta})$\ \ \ } & \specialcell[t]{$S(\Delta\Omega)$\ \ \ } \\
\hline
S56 & $0.499\pm0.008$ & $0.488\pm0.005$ & $0.286\pm0.002$ \\
S86 & $0.663\pm0.004$ & $0.588\pm0.006$ & $0.637\pm0.001$ \\
S87 & $0.578\pm0.02$ & $0.523\pm0.02$ & $0.406\pm0.005$ \\
S103 & $0.438\pm0.02$ & $0.452\pm0.003$ & $0.530\pm0.0006$ \\
S116 & $0.359\pm0.01$ & $0.440\pm0.02$ & $0.467\pm0.003$ \\
S118 & $0.309\pm0.01$ & $0.455\pm0.006$ & $0.369\pm0.002$ \\
S121 & $0.680\pm0.007$ & $0.675\pm0.004$ & $0.761\pm0.001$ \\
S122 & $0.450\pm0.08$ & $0.557\pm0.06$ & $0.378\pm0.003$ \\
S138 & $0.495\pm0.02$ & $0.554\pm0.005$ & $0.460\pm0.002$ \\
S140 & $0.662\pm0.007$ & $0.620\pm0.004$ & $0.424\pm0.002$ \\
\hline\\
\end{tabular*}
\label{tab:simulation_slopes_spherical}
\end{table*}

In the top panel of Fig.~\ref{fig4:simulation_with_spherical} we show
the resulting frequency distributions. We find that quite a few
streams have become thicker and distorted in the spherical approximation
to the potential in comparison to their behaviour in the true
axisymmetric potential. Some of the streams seem quite irregular such
as S140, a characteristic that was also seen for some of the Aquarius
streams (see e.g.\ stream A107 in 
Fig.~\ref{fig4:aquarius_frequenciesAD}). As a consequence of the distortions, it is generally more difficult to
fit straight lines to the frequency distribution, and therefore also the
determination of their slopes is less reliable (see e.g.\ S122 and S138).

\begin{figure}[!htbp]
\centering
\includegraphics[width=0.5\textwidth]{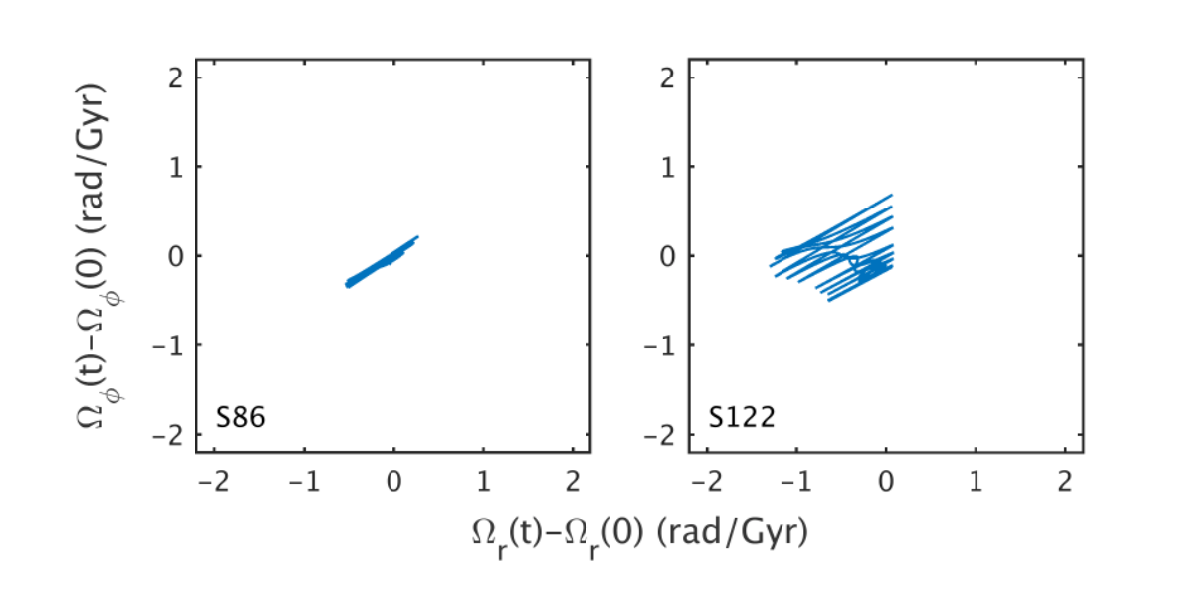}
\vspace{-0.5cm}
\caption{\small Frequencies computed in the limiting spherical potential for two progenitor orbits from the test-particle simulations in the Kuzmin-Kutuzov potential (see Fig.~\ref{fig4:simulation_XY}).}
\label{fig4:simulation_with_spherical_orbits}
\end{figure}

To understand this behaviour it is interesting to explore how 
orbits map onto frequency space if
the frequencies are computed at each time step in the approximate
spherical potential. This is 
show in Fig.~\ref{fig4:simulation_with_spherical_orbits} for two different orbits. 
The trajectories oscillate in frequency,
while if the potential had been truly spherical they would collapse into a single
point. These oscillations are therefore directly related 
to how well the isochrone potential represents the mass distribution. We take from this
that the thickening of streams is caused by the degree to which the orbits of those particles are described by a spherical potential.

The central panel of Fig.~\ref{fig4:simulation_with_spherical} 
shows the $\theta_r$-$\theta_\phi$ space. Streams such as S103 and S121
appear quite similar to their counterparts in
Fig.~\ref{fig4:simulation_anglesfrequencies_lambdaphi}, plotted in their
natural (true) angle space ($\theta_\lambda$-$\theta_\phi$). Notice that when the straight line fits
look reasonable, the slopes in this spherical angle space and
in the corresponding Staeckel angles are similar,
differing by $\sim 0.05$. Like for the Aquarius
streams, the slopes in $\theta_r$-$\theta_\phi$ span a range between
$\sim 0.4-0.7$. In all cases, the streams have become thicker. The
energy gradient along the streams seems especially discontinuous at
some locations, while for some streams
such as S140 and S121 it is somewhat better retained.

The bottom panels of Fig.~\ref{fig4:simulation_with_spherical} 
shows the angles $\theta_r$-$\theta_\vartheta$. The comparison of
$\theta_\lambda$-$\theta_\nu$ to $\theta_r$-$\theta_\vartheta$ is less
straightforward because $\Omega_\vartheta$ and $\Omega_\nu$ differ by
a factor 2 in the spherical limit. 

We conclude that using a spherical approximation to the potential
leads to an increase of the spread in frequency space, by making the
streams longer, wider and/or distorted, but the energy gradient
remains quite intact in this space. The streams in angle space often
look much more well-behaved, although they do become thicker and the
energy gradient is not preserved. This good behaviour probably
reflects that the spherical potential has the right average enclosed
mass (as argued by \citealt{BuistHelmi2015}).  We find that the fitted
slopes in angle and frequency space deviate significantly from each
other, indicating that the true potential in which the streams have
evolved has not been used for the angle and frequency computations.
We have explored possible correlations between slopes differences
  and orbital characteristics in our stream sample that could perhaps
  be used to infer the asphericity of the true potential. However, we
  found no specific trends in the differences in slopes between the
  angle and frequency spaces, nor between the angles in the spaces
  $\theta_\theta - \theta_r$) and $\theta_\phi - \theta_r$).

\subsection{Axisymmetric Staeckel potential with different flattening}
\label{sec4:wrong_qk}

\begin{figure*}[!htbp]
\centering
\noindent\makebox[\textwidth]{
\includegraphics[width=0.95\textwidth]{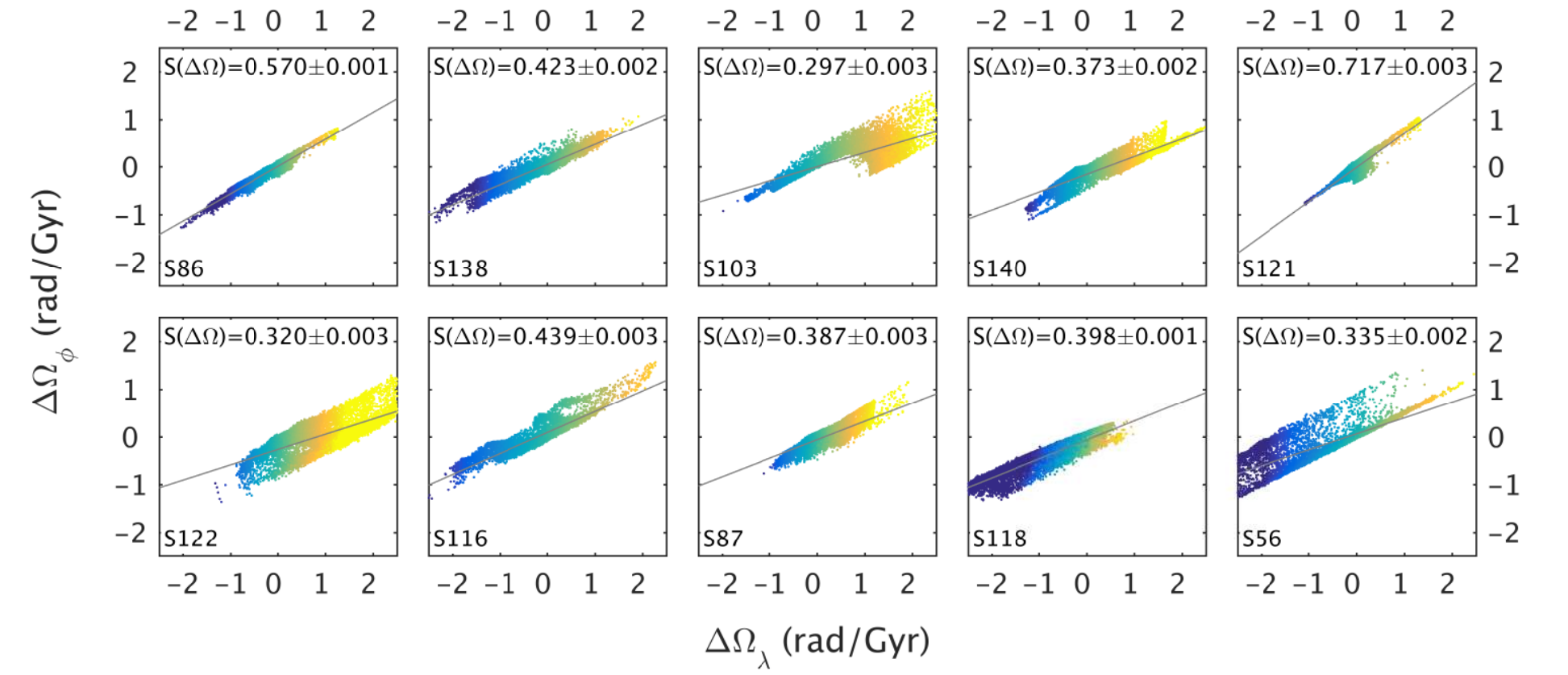}}
\noindent\makebox[\textwidth]{
\includegraphics[width=0.95\textwidth]{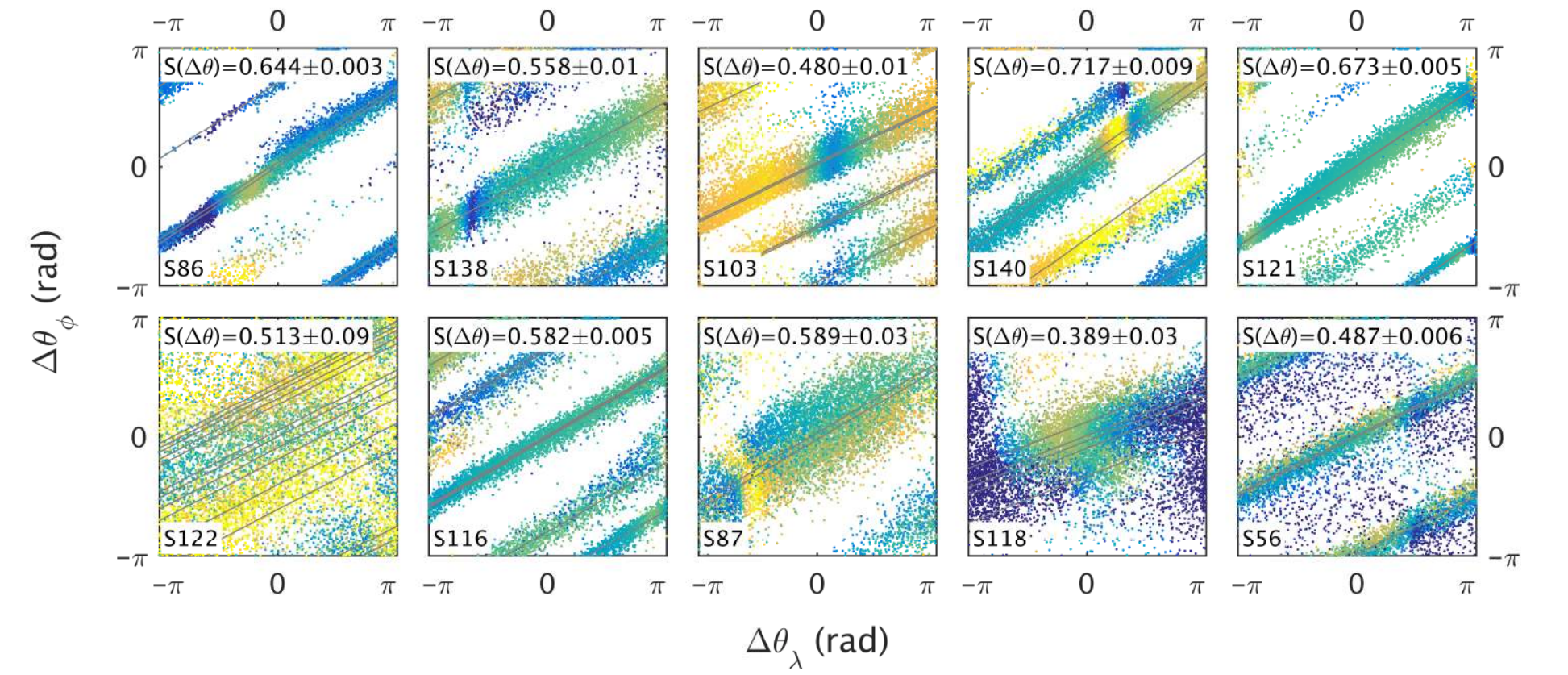}}
\vspace{-0.6cm}
\caption{\small Test-particle streams from
  Fig.~\ref{fig4:simulation_XY}-~\ref{fig4:simulation_anglesfrequencies_nuphi}
  with the frequencies and angles in the $\lambda$-$\phi$ projections computed for a Staeckel potential with flattening $q_\textrm{K}' = 1/q_\textrm{K}$. The colours represent the energy gradient with the most bound particles in yellow and the least bound particles in blue.}
\label{fig4:simulation_frequencies_invq}
\end{figure*}

\begin{figure*}[!htbp]
\centering
\noindent\makebox[\textwidth]{
\includegraphics[width=0.95\textwidth]{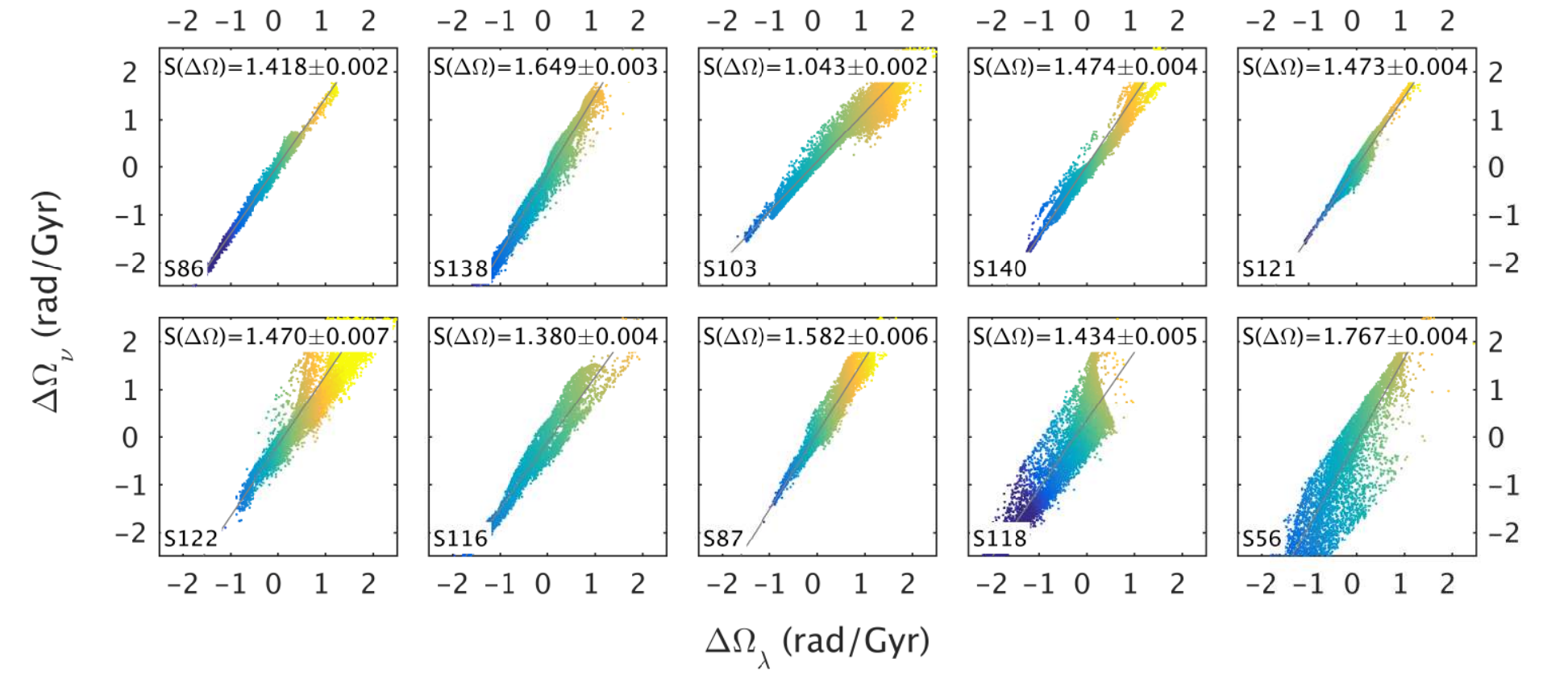}}
\noindent\makebox[\textwidth]{
\includegraphics[width=0.95\textwidth]{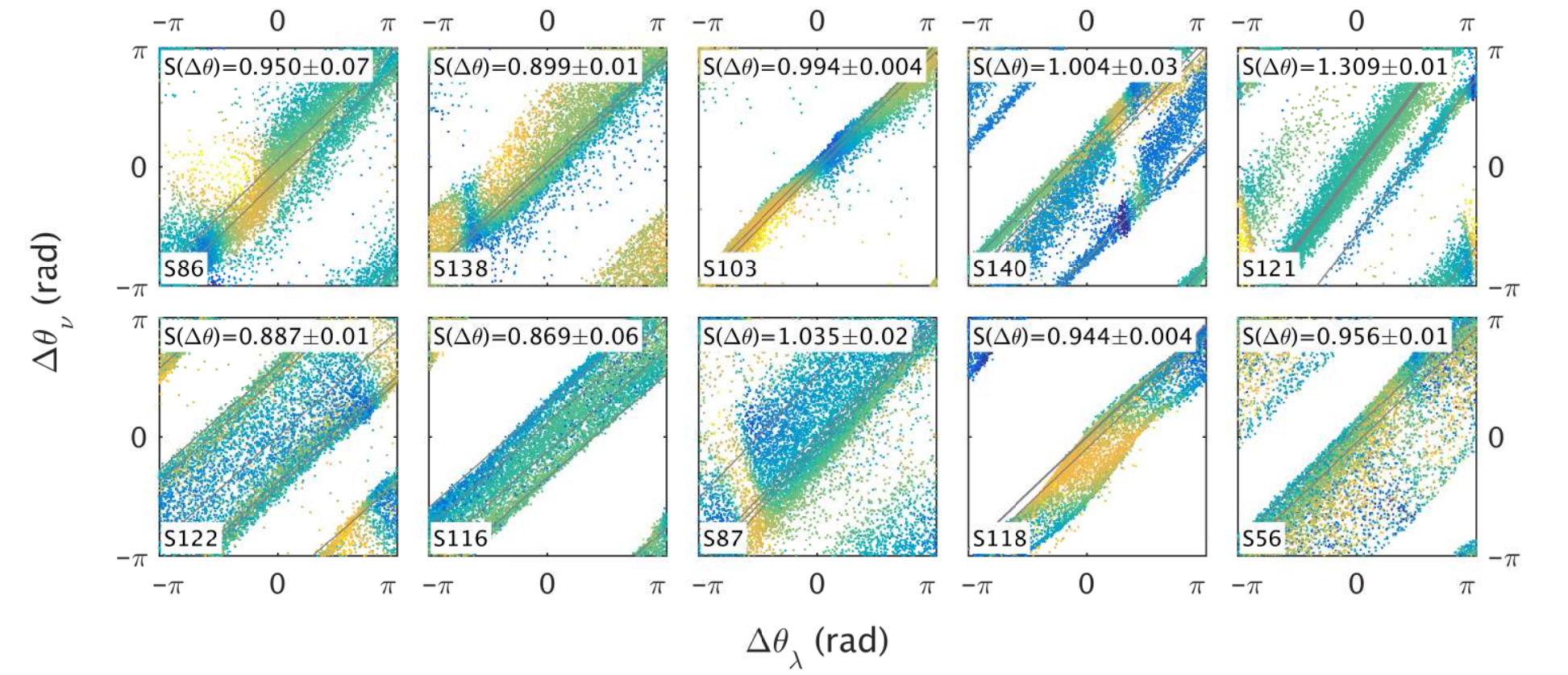}}
\vspace*{-0.6cm}
\caption{\small Same as Fig.~\ref{fig4:simulation_frequencies_invq} but now showing the $\lambda$-$\nu$ projections.}
\label{fig4:simulation_frequencies_invq2}
\end{figure*}

\begin{table*}[!htbp]
\small
\centering
\caption{\small Overview of the fitted slopes in Figs.~\ref{fig4:simulation_frequencies_invq} and \ref{fig4:simulation_frequencies_invq2} obtained for our test-particle streams assuming an axisymmetric Staeckel potential with flattening $q_\textrm{K}' = 1/q_\textrm{K}$.} 
\def\arraystretch{1.2}
\begin{tabular*}{0.6\textwidth}{rllll}
\hline
\hline
\specialcell[t]{Stream} & \specialcell[t]{$S(\Delta\theta_{\lambda,\phi})$\ \ \ } & \specialcell[t]{$S(\Delta\Omega_{\lambda,\phi})$\ \ \ } & \specialcell[t]{$S(\Delta\theta_{\lambda,\nu})$\ \ \ } & \specialcell[t]{$S(\Delta\Omega_{\lambda,\nu})$\ \ \ } \\
\hline
S56 & $0.487\pm0.006$ & $0.335\pm0.002$ & $0.956\pm0.01$ & $1.767\pm0.004$ \\
S86 & $0.644\pm0.003$ & $0.570\pm0.001$ & $0.950\pm0.07$ & $1.418\pm0.002$ \\
S87 & $0.589\pm0.03$ & $0.387\pm0.003$ & $1.035\pm0.02$ & $1.582\pm0.006$ \\
S103 & $0.480\pm0.01$ & $0.297\pm0.003$ & $0.994\pm0.004$ & $1.043\pm0.002$ \\
S116 & $0.582\pm0.005$ & $0.439\pm0.003$ & $0.869\pm0.06$ & $1.380\pm0.004$ \\
S118 & $0.389\pm0.03$ & $0.398\pm0.001$ & $0.944\pm0.004$ & $1.434\pm0.005$ \\
S121 & $0.673\pm0.005$ & $0.717\pm0.003$ & $1.309\pm0.01$ & $1.473\pm0.004$ \\
S122 & $0.513\pm0.09$ & $0.320\pm0.003$ & $0.887\pm0.01$ & $1.470\pm0.007$ \\
S138 & $0.558\pm0.01$ & $0.423\pm0.002$ & $0.899\pm0.01$ & $1.649\pm0.003$ \\
S140 & $0.717\pm0.009$ & $0.373\pm0.002$ & $1.004\pm0.03$ & $1.474\pm0.004$ \\
\hline\\
\end{tabular*}
\label{tab:simulation_slopes_invq}
\end{table*}

In this section we compute the angles and frequencies for the
Kuzmin-Kutuzov potential, but now assuming a flattening $q_\textrm{K}'
= 1/q_\textrm{K}$, which reverses the axis lengths $a_\textrm{K}$ and
$c_\textrm{K}$ compared to the true potential. For the simulations of
Sec.~\ref{sec4:testparticlesaxisymmetric} we used a prolate potential,
and inverting $q_\textrm{K}$ results in an oblate shape. 

The resulting
distributions in frequency and angle space are shown in
Figs.~\ref{fig4:simulation_frequencies_invq},
\ref{fig4:simulation_frequencies_invq2}, and listed in Table~\ref{tab:simulation_slopes_invq}, and qualitatively are very
similar to what we saw in Sec.~\ref{sec4:wrong_sph} and in previous
figures. The frequency distributions are even more broadened than when
assuming a spherical potential, and for some of the streams such as
S56 also more extended (i.e.\ beyond the boundaries of the box, which
we retained for easier comparison, and which is centred on the
progenitor's centre of mass). This broadening appears to be asymmetric
with respect to the progenitor. As in the previous section, the energy
gradient in frequency space remains almost intact while in angle space
it almost cannot be discerned.

The slopes fitted to the angle and frequency distributions differ considerably,
which we, for a large part can attribute to the difficulty of the fitting in frequency space.
In both angle spaces we see that some of the streams depict small
scale wiggles, such as streams S86, S118 and S140, which are more
apparent in $\theta_\lambda$-$\theta_\nu$ space. This is probably
because the shape of the potential is significantly different from
the true form, and from the spherical shape assumed in the previous
section. 

\subsection{Spherical approximation with incorrect radial form}
\label{sec4:wrong_nfw}

\begin{figure*}[!htbp]
\centering
\vspace*{-0.2cm}
\noindent\makebox[\textwidth]{
\includegraphics[width=0.95\textwidth]{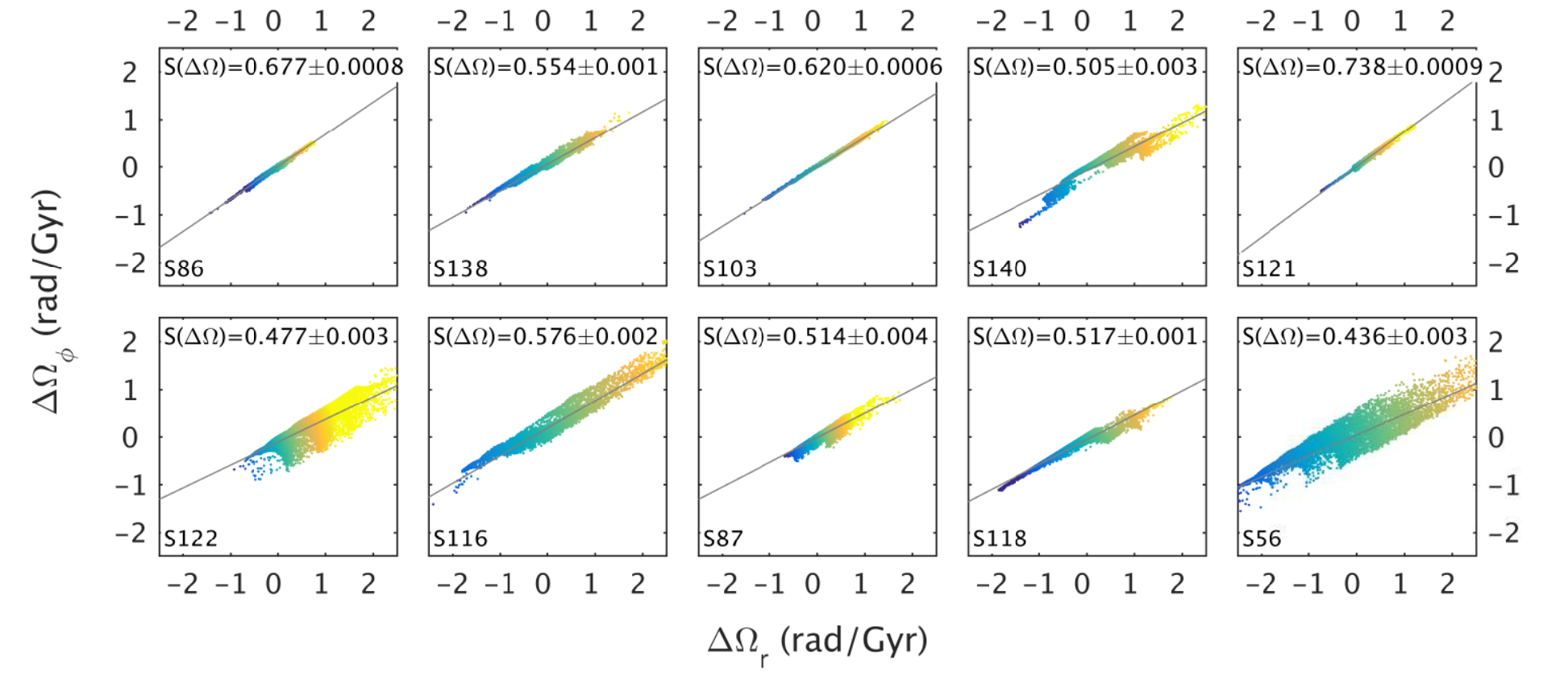}}
\noindent\makebox[\textwidth]{
\includegraphics[width=0.95\textwidth]{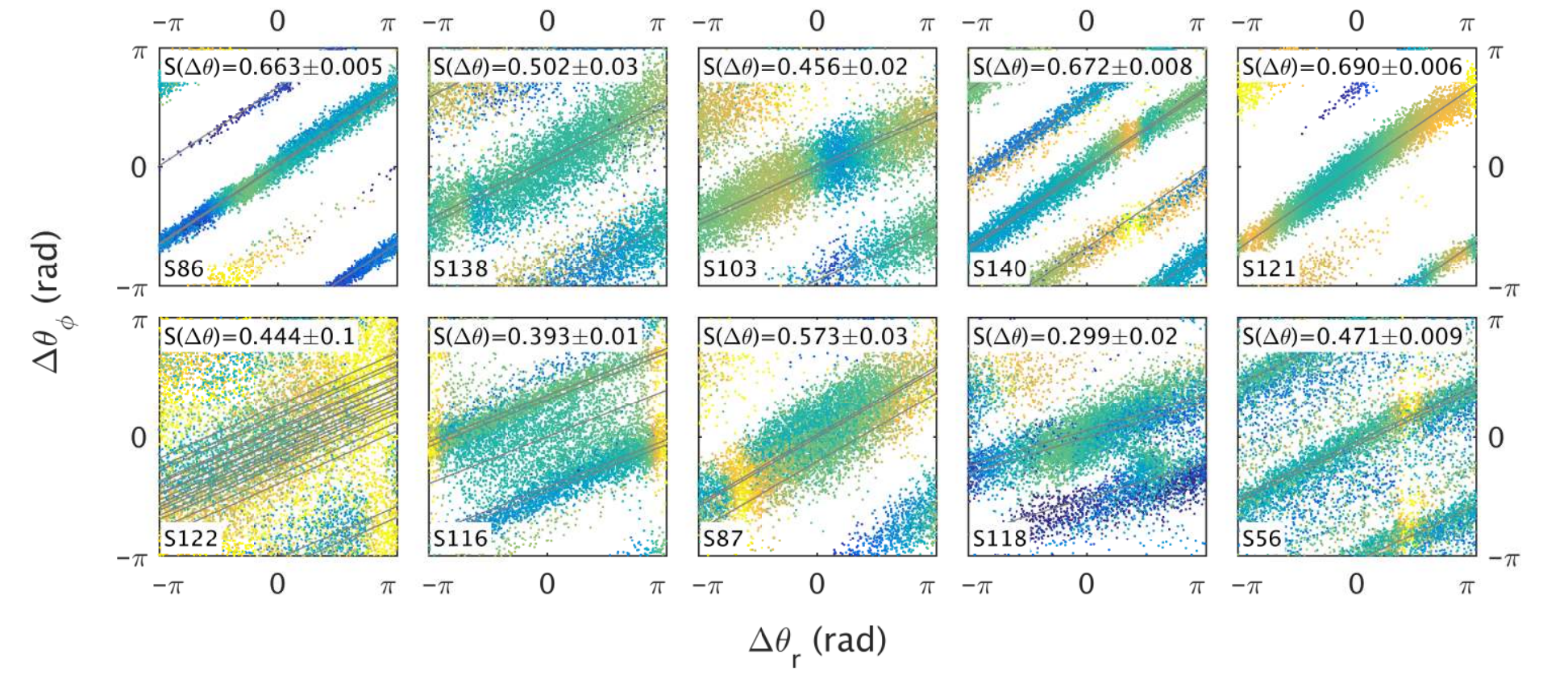}}
\noindent\makebox[\textwidth]{
\includegraphics[width=0.95\textwidth]{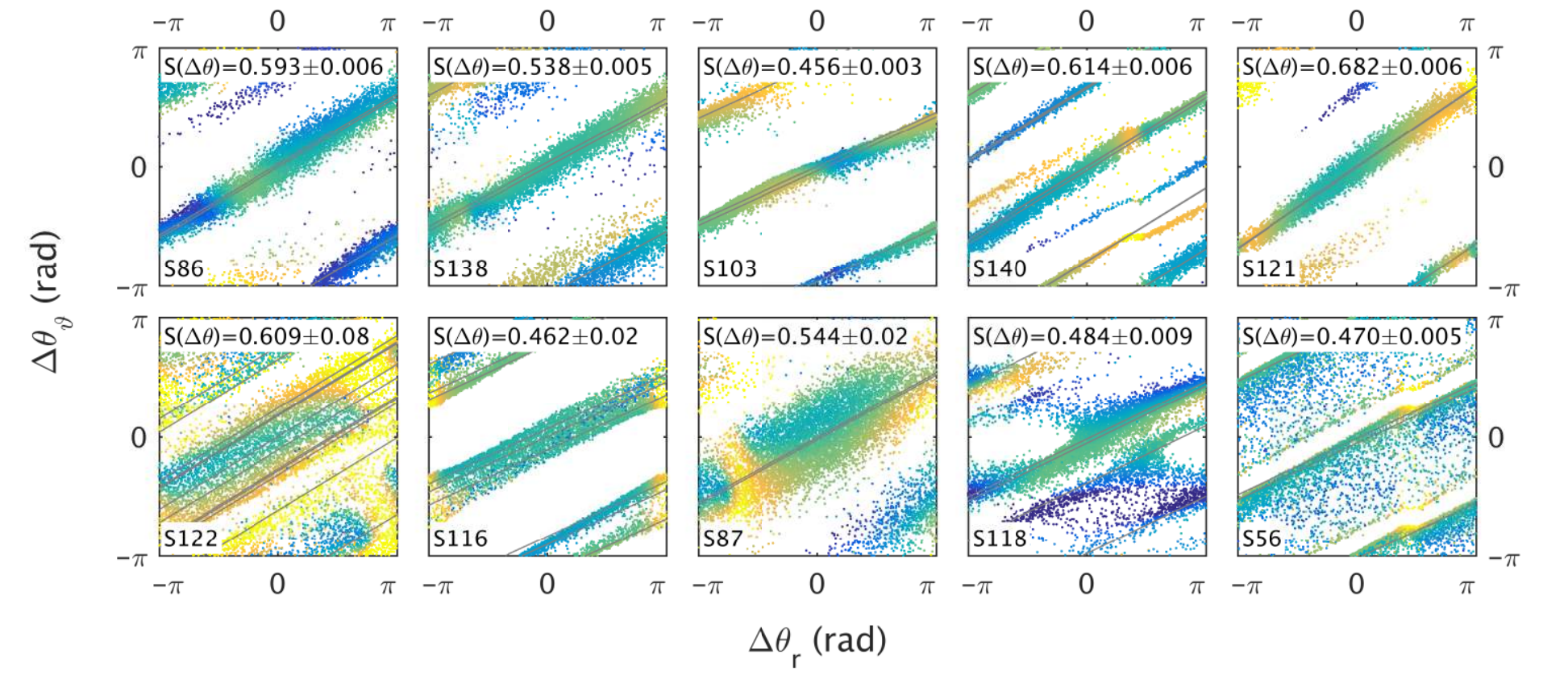}}
\vspace*{-0.8cm}
\caption{\small Test-particle streams from Fig.~\ref{fig4:simulation_XY}--\ref{fig4:simulation_anglesfrequencies_nuphi} with the frequencies $\Delta\Omega_r$-$\Delta\Omega_\phi$ (top panels) and angles $\Delta\theta_r$-$\Delta\theta_\phi$ (middle panels) and $\Delta\theta_r$-$\Delta\theta_\vartheta$ (bottom panels), computed using an NFW potential that has the same slope at a fixed radius as the limiting spherical isochrone potential (see Fig.~\ref{fig4:fit_isochrone_to_NFW}). The colours represent the energy gradient in the NFW potential. The panels have been centred on the progenitor position. The insets indicate the
  fitted slopes for the approximate potential. The errors were found by bootstrapping the fit 200 times.}
\label{fig4:simulation_wrong_fr}
\end{figure*}

\begin{table*}[!htbp]
\small
\centering
\caption{\small Overview of the fitted slopes in Fig.~\ref{fig4:simulation_wrong_fr}  obtained assuming a spherical NFW potential for our test-particle streams evolved in an axisymmetric Staeckel potential.} 
\def\arraystretch{1.2}
\begin{tabular*}{0.6\textwidth}{rlll}
\hline
\hline
\specialcell[t]{Stream} & \specialcell[t]{$S(\Delta\theta_{r,\phi})$\ \ \ } & \specialcell[t]{$S(\Delta\theta_{r,\vartheta})$\ \ \ } & \specialcell[t]{$S(\Delta\Omega)$\ \ \ } \\
\hline
S56 & $0.471\pm0.009$ & $0.470\pm0.005$ & $0.436\pm0.003$ \\
S86 & $0.663\pm0.005$ & $0.593\pm0.006$ & $0.677\pm0.0008$ \\
S87 & $0.573\pm0.03$ & $0.544\pm0.02$ & $0.514\pm0.004$ \\
S103 & $0.456\pm0.02$ & $0.456\pm0.003$ & $0.620\pm0.0006$ \\
S116 & $0.393\pm0.01$ & $0.462\pm0.02$ & $0.576\pm0.002$ \\
S118 & $0.299\pm0.02$ & $0.484\pm0.009$ & $0.517\pm0.001$ \\
S121 & $0.690\pm0.006$ & $0.682\pm0.006$ & $0.738\pm0.0009$ \\
S122 & $0.444\pm0.1$ & $0.609\pm0.08$ & $0.477\pm0.003$ \\
S138 & $0.502\pm0.03$ & $0.538\pm0.005$ & $0.554\pm0.001$ \\
S140 & $0.672\pm0.008$ & $0.614\pm0.006$ & $0.505\pm0.003$ \\
\hline\\
\end{tabular*}
\label{tab:simulation_slopes_wrong_fr}
\end{table*}

We now focus on the impact of computing the actions and angles in a
spherical potential whose radial dependence is quite different from
the Kuzmin-Kutuzov potential in the spherical limit. We explore an NFW potential that has the
same enclosed mass and mass slope at $r_\textrm{fix} = 50$ kpc (see
also Fig.~\ref{fig4:fit_isochrone_to_NFW}).

In Fig.~\ref{fig4:simulation_wrong_fr} and Table~\ref{tab:simulation_slopes_wrong_fr} we show the resulting
distribution of these angles and frequencies for the streams evolved
in the axisymmetric Kuzmin-Kutuzov potential of
Sec.~\ref{sec4:testparticlesaxisymmetric}. At first sight, many of the
streams look very similar to those in
Fig.~\ref{fig4:simulation_with_spherical}, which corresponds to the
computations assuming a spherical isochrone mass distribution.

In frequency space we see that some of the streams are longer and
thinner (see e.g.\ S140 and S118). The slopes
derived from fitting straight lines to these distributions are different from those
computed in the isochrone potential, such as for S86 and S103. In angle space the differences are
more subtle, and the typical difference between the fitted slopes
between the isochrone and NFW cases is $\sim 0.05$, which is comparable to
the estimated error in the fit, but also the incorrect shape of the potential may contribute here. The energy gradient in angle space for the
isochrone and NFW potentials is not always the same, with the isochrone
potential showing more fluctuations in the $\theta_r$
direction. Overall we find that the frequencies are the most sensitive
to the potential, but the differences are small because the slope and
the enclosed mass at $r_\textrm{fix}=50$ kpc are equal for the NFW and
isochrone profiles. The behaviour in frequency and angle space is only  slightly worse than
for the spherical isochrone potential.
\begin{figure*}[!htbp]
\centering
\noindent\makebox[\textwidth]{
\includegraphics[width=0.95\textwidth]{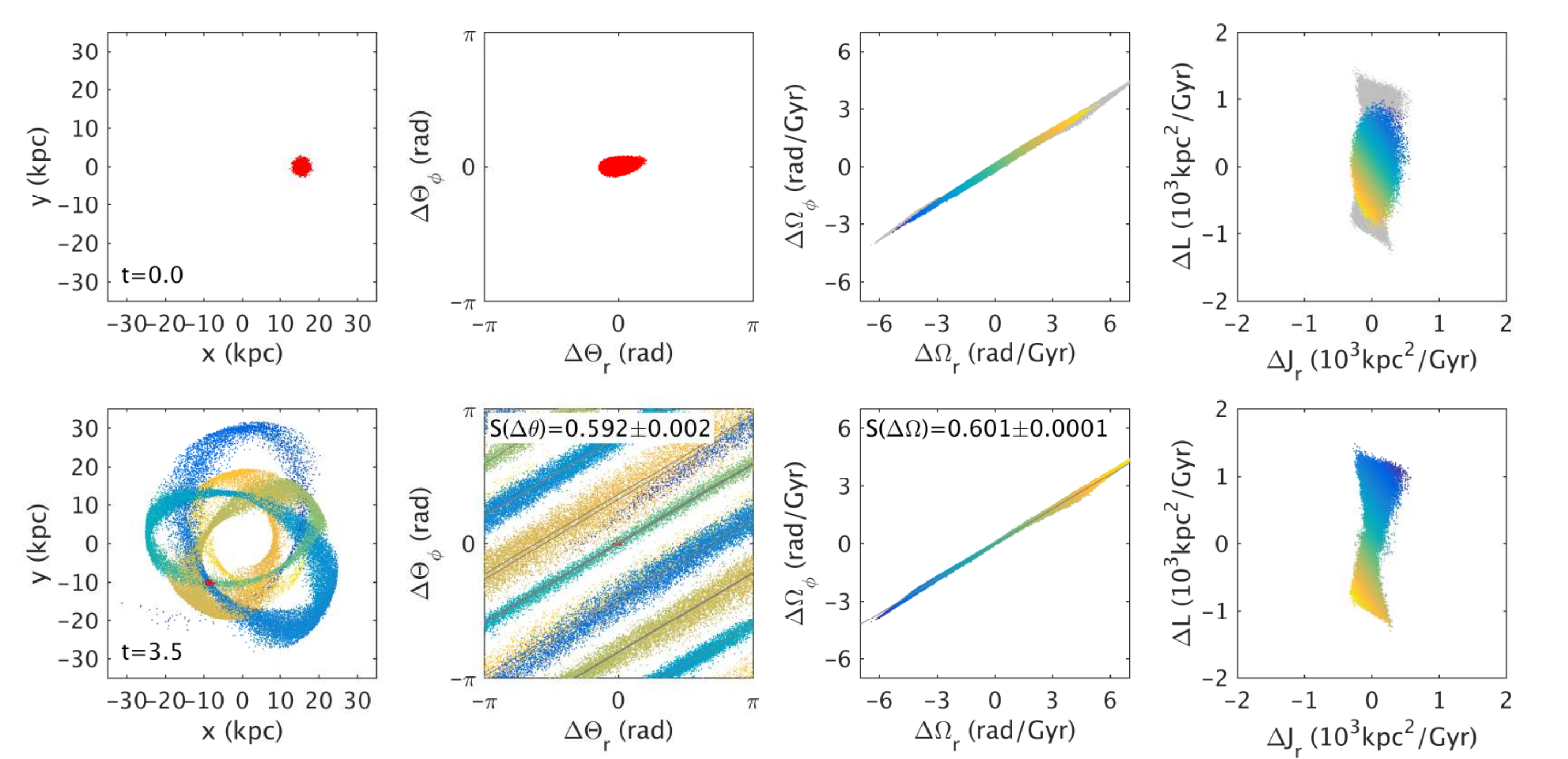}}
\caption{\small Projections of configuration, angle, frequency and
  action space at $t=0$ (top panels) and $t=3.5$ Gyr (bottom panels)
  for an N-body stream evolved in a spherical NFW potential. The first
  column shows the projection of the stream on the orbital plane of
  the progenitor. Particles still
  bound to the progenitor are plotted in red. In the third column we
  have overlaid the distribution of frequencies at $t=10$ Gyr in
  grey. In the panels on the right we also
  show the final distribution of actions at $t=10$ Gyr in grey. The actions,
  angles and frequencies have been centred on the coordinates of the centre of mass of the progenitor and the
  colours represent the energy gradient.}
\label{fig4:simulation_nbody}
\end{figure*}

\subsection{The effects of self-gravity}
\label{sec4:n-body-ch4}

In reality the progenitors of streams will initially be bound by their
own self-gravity. Particles that become unbound with time will typically be
released at specific points along the orbit (close to pericentre),
rather than continuously as modelled thus far. This results in the
leading and trailing arms being offset from each other in
configuration space \citep{Johnston1998} and in
energy-angular momentum space \citep{Gibbons2014}. The process of
disruption also causes particles to define a bow-tie structure in
action (and energy-angular momentum) space. Since one of our
goals is to understand the properties of streams in the cosmological
Aquarius N-body simulations, we attempt
here to establish what the effect of self-gravity is on the distribution of particles in frequency and angle spaces.

To this end, we simulated the evolution of a $3.7 \times 10^8$
$\textrm{M}_\odot$ progenitor for 10 Gyr on an orbit with
$r_\textrm{apo}=23.6$ kpc and $r_\textrm{peri}=10.2$ kpc in a
spherical NFW potential with $M_\textrm{s}=1.5\times10^{11}$
$\textrm{M}_\odot$ and $r_\textrm{s}=12$ kpc using an N-body code that
uses a quadrupole expansion to model the internal gravitational
potential of the system \citep{Helmi2001}. Fig.~\ref{fig4:simulation_nbody} shows snapshots for $t=0$
and $t=3.5$ Gyr. The left panels plot the projection of the stream on
the progenitor's orbital plane, where the red particles correspond to
those that are (still) bound. The stream has already spread out
significantly by $t=3.5$ Gyr because the progenitor is large and the
orbit is rather confined to the inner regions of the host, which
results in fast evolution. By this time, the progenitor has almost
completely dissolved (i.e.\ only a few particles are marked in red).

In the second panel we show the angle distribution initially and at
$3.5$ Gyr, with the particles bound to the progenitor marked in red. By
$3.5$ Gyr many wraps fill up the angle space, to which we fitted
parallel straight lines and whose slope is shown in the inset. The
frequency distribution at both times is shown in the third column
panels, which is overlaid onto the distribution at $t=10$ Gyr in
grey. We learn from this that at $t=3.5$ Gyr, the distribution has
reached its final (bow-tie like) shape, and the evolution of the
particles is simply governed by $\theta \sim \Omega t$ by this point in time.

The expected gap near the progenitor in action space forms slowly and
is only subtle at the final time \citep[see][for a more detailed
discussion of the process]{Gibbons2014}. Furthermore, we find no
significant offset in configuration space between the leading and
trailing arms in our simulation, nor any epicyclic oscillations in the
stream. This is most likely related to the fast disruption of the
progenitor as a consequence of its low density contrast with respect
to the host. This is quite different to what is seen for N-body
simulations of globular clusters whose disruption process is slower
because these are strongly bound gravitational systems
\citep{Kuepper2010,Kuepper2012}.

At $3.5$ Gyr, the distribution of angles and frequencies follow each
other very closely, as quantified by the slope of the fitted straight
lines in the bottom panels, and this is in agreement with the findings of
\citet{Sanders2013a}. The difference between test-particle 
and N-body simulations is especially seen in the bow-tie shape in
action space and frequency space, but it is almost absent in
angle space. The dynamics of streams is otherwise mostly the same. Furthermore, the
slopes of the straight lines along which the stream is distributed in angle and in frequency
space are the same, in agreement with the results by \cite{Sanders2013b}. Only for very massive progenitors ($\sim 10^{10}$
$\textrm{M}_\odot$) this picture may change because of interactions
between the stream and the progenitor \citep{Choi2009}, or because the
overall potential of the halo changes while the progenitor system is in the process of being disrupted \citep{VeraCiroHelmi2013,
  Gomez2015}. 

This analysis leads us to believe that self-gravity has a negligible impact on the results presented in earlier sections of the paper.

\section{Discussion and conclusions}
\label{sec4:discussionconclusion}

We have studied the behaviour of streams in fully cosmological N-body simulations of the formation of stellar halos from the Aquarius project \citep{Springel2008}. These stellar halos were produced by tagging dark matter particles in these otherwise dark-matter only simulations according to the GALFORM semi-analytic galaxy formation model \citep{Cooper2010, Lowing2014}. From these halos we selected a set of stream-like objects and used the `tagged' particles to build up a catalogue of stellar streams. Our interest was to understand the behaviour of these streams, evolved in a fully cosmological time-dependent framework, particularly when studied in action-angle space. 

Since the Aquarius halos' potential is not analytic, we explored as a first approximation their behaviour when a spherical NFW potential was assumed. For the best fitting NFW mass distribution we computed the streams' angles and frequencies. We found that many of the streams in the Aquarius halos show
several wraps in angle space that appear to be on relatively straight
lines, as reported in other works for streams evolved in static or
evolving smooth potentials \citep{Sanders2013a,
  BuistHelmi2015}. However in many cases these lines are not
parallel. We also found patchy features and wiggly behaviour in angle
space. In frequency space, often the structures are very broad but relatively
linear and depict some amount of irregularity. The width of these
streams and features are typically larger than what we have seen before
in simple simulations of streams evolved in a smooth spherical
potential.

To understand the nature of the various features, we proceeded to
explore how various deviations from spherical symmetry could be
affecting the behaviour of streams. We have been able to demonstrate
that, independently of the form of the host potential, if the angles
and frequencies are computed self-consistently then the streams are
expected to be along straight lines in frequency and angle space. This
is because the Hessian of the Hamiltonian generally has one eigenvalue
that is much larger than the others and this dictates to a large extent the
direction in which the streams will expand \citep{Tremaine1999}. The
exact direction depends also on the action distribution. These results
are valid provided the progenitor of the stream is relatively compact
in phase-space.

We next focused on why streams evolved in a particular potential but
whose angles and frequencies have been computed in a
different approximate potential are still on straight lines, as we
found for the Aquarius simulations. To this end we ran a set of
test-particle simulations in an axisymmetric prolate
Kuzmin-Kutuzov potential, which is of Staeckel form. We computed the
angles and frequencies in this potential, and found again the
characteristic linear appearance of streams in these spaces. Next we
assumed different forms of the potential and computed the angles and
frequencies for those cases as well. We found that even if this
procedure is not self-consistent, streams are still distributed along
relatively straight lines. However, in frequency space streams became 
typically thicker and somewhat distorted, and in angle space
 they depict wiggly behaviour. For example we found
that using a potential with the wrong flattening (spherical or oblate,
instead of prolate) has a strong effect on the size of streams in
frequency space. On the other hand, differences in the angles and frequencies
  distributions for spherical potentials of different radial form remained
subtle provided the enclosed mass was approximately correct within in
the radial extent probed by the streams orbits. 

In all cases, the energy gradient along the stream seems almost intact
in frequency space (as seen for the Aquarius streams)
but clearly distorted or broken in angle space. The straight lines
that we fit to the angle and frequency distributions differ in slope, even when the
potential is assumed to be spherical, contrary to expectations. This is the clearest
indication that the shape assumed for the angle computation is
incorrect and does not correspond to that of the potential in which
the streams were evolved. Finally, we also investigated what happens
to the actions and angles in a simulation with self-gravity. The
largest difference is that during the disruption process of the
progenitor, the action distribution of the particles that eventually form the
stream is altered, but otherwise the dynamics of streams are the same as in
test-particle simulations.

In conclusion, we have been able to reproduce and understand most of
the features seen in the approximate angles and frequencies for the
Aquarius streams, with the exception of the `noisy' and `patchy'
appearance of the streams in angle and configuration space. We believe
these can be attributed to interactions of a stream with dark matter
substructures, which are known to give rise to disturbed morphologies
\citep{Bonaca2014}. Such interactions may also introduce non-adiabatic
time-dependent effects on streams that lead to the formation of gaps
\citep{Yoon2011, Carlberg2013, Ngan2014}. 

Finally, since the angle-frequency misalignments found for the 
Aquarius streams can mostly be attributed to using the wrong 
potential, this implies that they cannot be used determine the mass 
growth history of the Aquarius dark matter halos as we had proposed 
in \citet{BuistHelmi2015}. This may be resolved with approximate 
schemes to compute the actions in a triaxial potential 
\citep[e.g.][]{Sanders2014a, Bovy2014} and by using the distortions of 
sufficiently thin streams in angle and frequency space to further 
constrain the present-day potential. 

A similar conclusion may be drawn for the determination of the
time-evolution of the Milky Way's gravitational potential, although
the challenge in this case is greater. As discussed in
\citet{BuistHelmi2015}, the ability to measure time dependence also
requires the presence of nearby thin and long streams, as only for
such streams, {\it Gaia} will be able to determine their full
phase-space coordinates precisely. Clearly the first step is to have
an accurate model for the present-day mass distribution in the Milky
Way.  Once this has been constructed and we are fortunate enough to be
able to exploit the presence of suitable streams, measuring the angle-frequency misalignment
to determine the evolution of our Galaxy's gravitational potential may
become feasible.
\\

We are grateful to Volker Springel, Simon White, and Carlos
Frenk for generously allowing us to use the Aquarius
simulations. Carlos Vera-Ciro is acknowledged for his help in the
analysis of these simulations. H.J.T.B. and A.H. gratefully acknowledge financial support from ERC-Starting Grant GALACTICA-240271 and a Vici grant. We wish to thank
Tim de Zeeuw for his helpful suggestions on an earlier draft, and the anonymous referee for their constructive comments on this manuscript.

\bibliographystyle{aa}
\bibliography{article}

\appendix

\section{Stream properties}
\label{sec4:Appendix4C}

In Table \ref{tab:stream_properties} we list the apocentre and pericentre distances of the orbits integrated in a spherical NFW potential as described in Sec.~\ref{sec4:streamsmorphologyandorbits}. We also provide the total dark matter masses of the progenitors with more than 500 dark matter particles for all the streams considered in that Section.

\begin{table*}[!htbp]
\small
\centering
\caption{\small Stream properties for halo Aq-A and Aq-D}
\def\arraystretch{1.2}
\begin{tabular*}{0.73\textwidth}{rrrr|rrrr}
\hline
\hline
\specialcell[t]{Stream\\} & \specialcell[t]{$r_\textrm{orbit,\,peri}$\ \ \ \\(kpc)} & \specialcell[t]{$r_\textrm{orbit,\,apo}$\ \ \ \\(kpc)} & \specialcell[t]{$M_\textrm{DM,\,progenitor}$\\(M${}_\odot$)} \ \ \ & \specialcell[t]{Stream\\} & \specialcell[t]{$r_\textrm{orbit,\,peri}$\ \ \ \\(kpc)} & \specialcell[t]{$r_\textrm{orbit,\,apo}$\ \ \ \\(kpc)} & \specialcell[t]{$M_\textrm{DM,\,progenitor}$\\(M${}_\odot$)}\ \ \  \\
\hline

A98 & $ 2.5$ & $35.5$ & $ 2.0\times 10^8$ & D56 & $ 2.4$ & $40.6$ & $ 7.3\times 10^8$\\
A99 & $ 6.2$ & $60.4$ & $ 1.3\times 10^7$ & D72 & $ 2.9$ & $127.8$ & $ 6.1\times 10^8$\\
A104 & $ 4.5$ & $44.5$ & $ 1.2\times 10^9$ & D73 & $ 4.8$ & $121.7$ & $ 6.7\times 10^8$\\
A106 & $ 8.1$ & $41.9$ & $ 1.5\times 10^7$ & D78 & $23.8$ & $194.5$ & $ 2.2\times 10^8$\\
A107 & $ 3.4$ & $57.6$ & $ 2.9\times 10^7$ & D82 & $15.6$ & $107.4$ & $ 3.4\times 10^8$\\
A108 & $12.5$ & $107.4$ & $ 5.1\times 10^7$ & D86 & $39.2$ & $84.6$ & $ 2.2\times 10^7$\\
A112 & $ 4.5$ & $59.8$ & $ 9.5\times 10^7$ & D87 & $16.1$ & $115.1$ & $ 9.5\times 10^7$\\
A116 & $ 3.2$ & $89.3$ & $ 1.2\times 10^8$ & D91 & $12.0$ & $48.8$ & $ 1.6\times 10^8$\\
A140 & $20.5$ & $198.1$ & $ 1.8\times 10^8$ & D98 & $ 2.9$ & $70.3$ & $ 1.5\times 10^9$\\
A151 & $35.0$ & $210.6$ & $ 1.4\times 10^8$ & D103 & $30.4$ & $55.6$ & $ 2.7\times 10^7$\\
A153 & $10.7$ & $189.1$ & $ 4.7\times 10^7$ & D116 & $16.9$ & $60.5$ & $ 5.2\times 10^7$\\
A158 & $33.6$ & $173.5$ & $ 7.1\times 10^7$ & D118 & $17.5$ & $42.5$ & $ 1.1\times 10^8$\\
A163 & $26.5$ & $54.6$ & $ 1.7\times 10^7$ & D120 & $13.7$ & $33.9$ & $ 8.1\times 10^7$\\
A164 & $15.9$ & $87.6$ & $ 6.2\times 10^8$ & D121 & $29.6$ & $177.3$ & $ 3.2\times 10^7$\\
A171 & $16.3$ & $76.3$ & $ 1.1\times 10^8$ & D122 & $ 6.3$ & $73.4$ & $ 3.4\times 10^7$\\
 & & &  & D125 & $ 3.0$ & $33.4$ & $ 1.9\times 10^8$\\
 & & &  & D129 & $ 4.3$ & $47.7$ & $ 1.2\times 10^8$\\
 & & &  & D137 & $10.7$ & $71.8$ & $ 9.1\times 10^6$\\
 & & &  & D138 & $15.7$ & $86.5$ & $ 2.5\times 10^7$\\
 & & &  & D140 & $ 7.7$ & $102.2$ & $ 3.1\times 10^7$\\ 
\hline\\
\end{tabular*}
\label{tab:stream_properties}
\end{table*}

\section{More streams}

\begin{figure*}[!htbp]
\centering
\noindent\makebox[\textwidth]{
\includegraphics[width=0.95\textwidth]{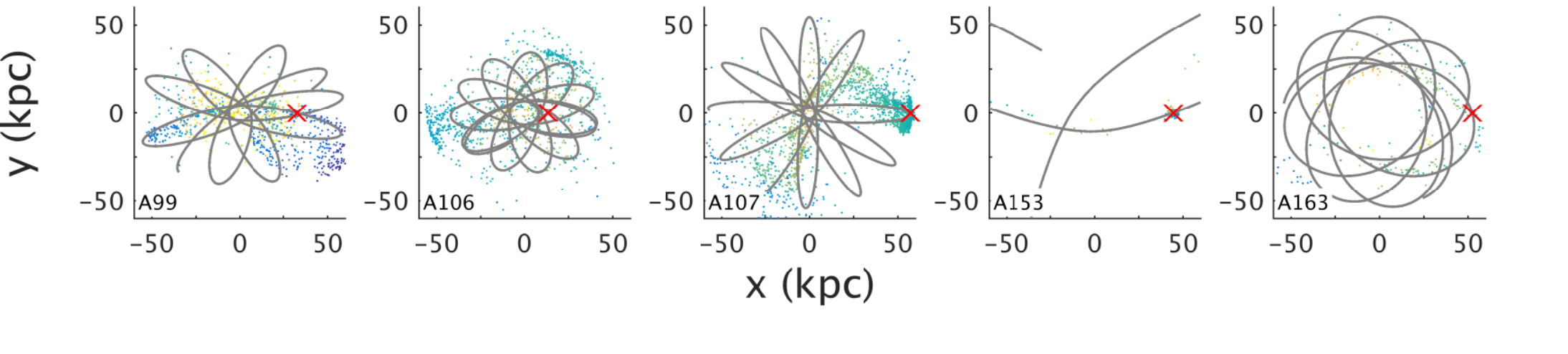}}
\noindent\makebox[\textwidth]{
\includegraphics[width=0.95\textwidth]{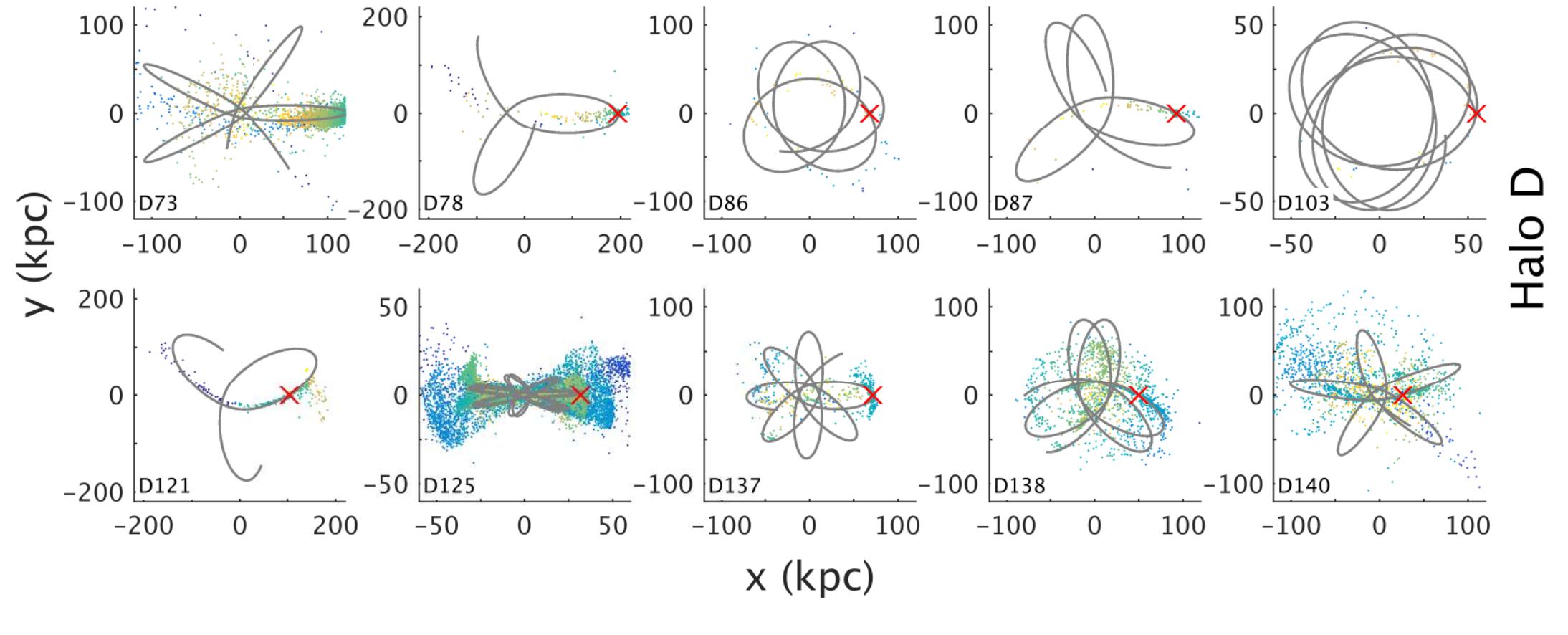}}
\caption{\small Same as Fig.~\ref{fig4:sphericalOrbitsAqADxy} but with the remaining streams streams with at least 500 dark matter particles that were not shown in that figure.}
\label{fig4:sphericalOrbitsAqADxy_rem}
\end{figure*}

\begin{figure*}
\centering
\noindent\makebox[\textwidth]{
\includegraphics[width=0.95\textwidth]{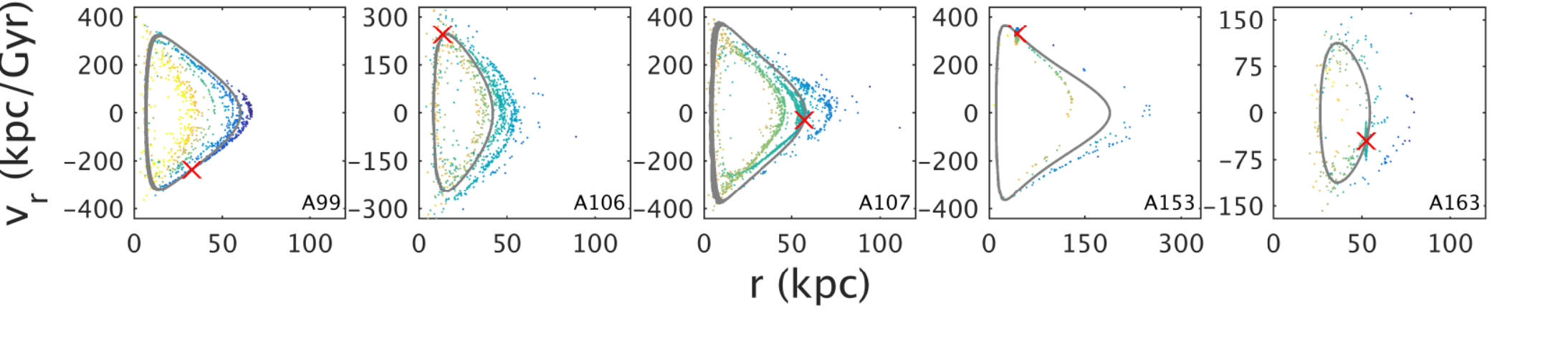}}
\noindent\makebox[\textwidth]{
\includegraphics[width=0.95\textwidth]{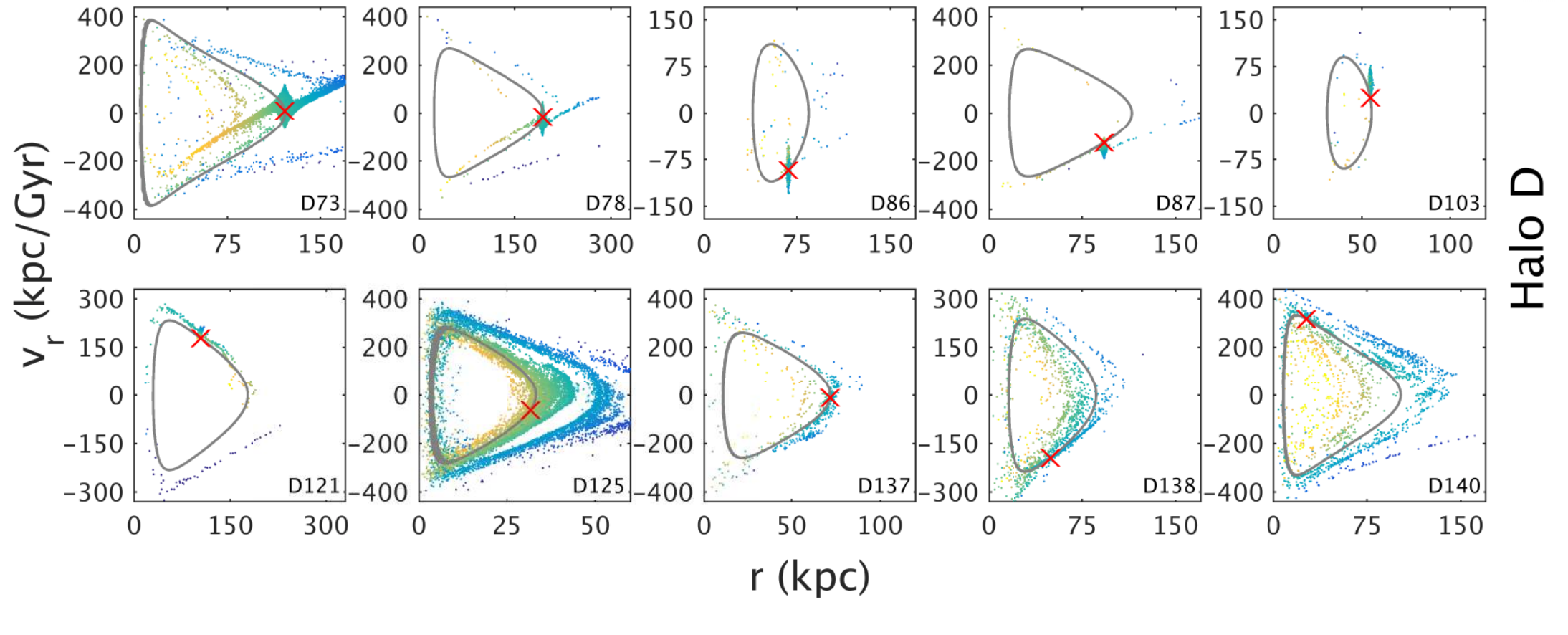}}
\caption{\small Same as Fig.~\ref{fig4:sphericalOrbitsAqADrvr} but with the remaining streams streams with at least 500 dark matter particles that were not shown in that figure.}
\label{fig4:sphericalOrbitsAqADrvr_rem}
\end{figure*}

In Section \ref{sec4:streamsinaquarius} we showed only 10 of the streams present in halo
Aq-A and Aq-D with more than 500 dark matter particles in their progenitors. In total there are 35 streams
that match our selection criteria, and the remaining 15 (5 in Aq-A and
10 in Aq-D) are shown in Fig.~\ref{fig4:sphericalOrbitsAqADxy_rem} and
\ref{fig4:sphericalOrbitsAqADrvr_rem}. As in
Figs.~\ref{fig4:sphericalOrbitsAqADxy} and
\ref{fig4:sphericalOrbitsAqADrvr} we have overplotted the orbits
integrated in the corresponding best fitting NFW spherical potential.


\section{Numerical parameters for Kuzmin-Kutuzov potential}
\label{sec4:potentialparametersnumerical}

\begin{figure*}[!htbp]
\centering
\noindent\makebox[\textwidth]{
\includegraphics[width=0.8\textwidth]{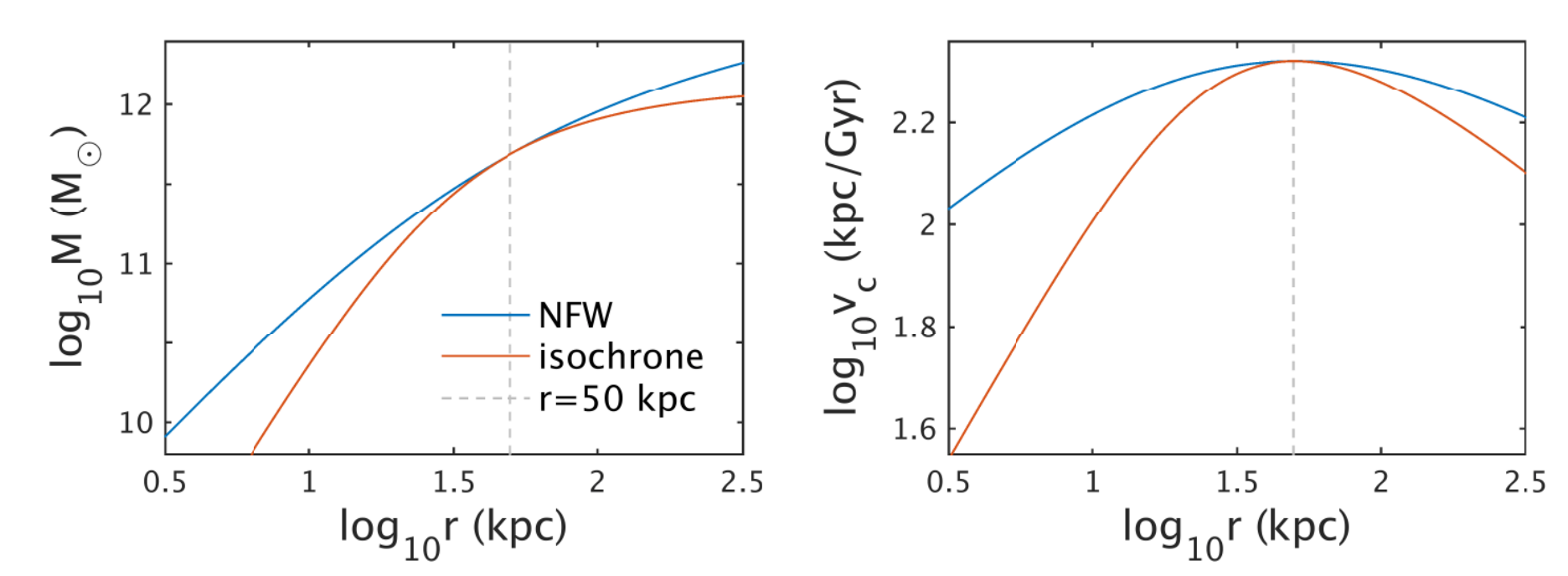}}
\caption{\small Enclosed mass and circular velocity curves for the best fitting NFW profile to halo Aq-D, and the isochrone potential that matches the enclosed mass and slope at $r_\textrm{fix} = 50$ kpc. }
\label{fig4:fit_isochrone_to_NFW}
\end{figure*}

To set the numerical values of the parameters of the axisymmetric Staeckel potential used in the experiments of Section ~\ref{sec4:testparticlesaxisymmetric} we proceed as follows. We first set up the isochrone potential by choosing a radius $r_\textrm{fix}$ at which the enclosed mass and its slope equals that of the best-fitting NFW potential for the Aquarius halos. This ensures that around $r_\textrm{fix}$ the (spherically averaged) mass distributions are similar. We choose $r_\textrm{fix}=50$ kpc because most of the streams we study are located around $50$-$100$ kpc from the halo centre. Having chosen $r_\text{fix}$, we first match the slope of the mass profile and then proceed to set the enclosed mass. The logarithmic slope $\kappa$ is
\begin{equation}
	\kappa_\textrm{NFW}(x) \equiv \frac{{\rm d}\log M_\textrm{NFW}(x)}{{\rm d} \log x}
\end{equation}
with $x = r/r_\textrm{s}$ and $r_\textrm{s}$ the scale radius of the NFW potential such that $M_\textrm{NFW}(r_\textrm{s}) = M_\textrm{s}$. The condition of equal slopes at $r_\textrm{fix}$ is
\begin{equation}
	\kappa_\textrm{NFW}(r_\textrm{fix}/r_\textrm{s}) = \kappa_\textrm{iso}(r_\textrm{fix}/r_\textrm{iso}),
\end{equation}
which we can invert to find $r_\textrm{iso}$. The scale mass $M_\textrm{iso}$ ensures the enclosed mass is equal at $r_\textrm{fix}$
\begin{equation}
	M_\textrm{s} \frac{A_\textrm{NFW}(r_\textrm{fix}/r_\textrm{s})}{A_\textrm{NFW}(1)} = M_\textrm{iso} \frac{A_\textrm{iso}(r_\textrm{fix}/r_\textrm{iso})}{A_\textrm{iso}(1)},
\end{equation}
where $A_i$ are the normalised radial mass profiles for the used potentials. In Fig.~\ref{fig4:fit_isochrone_to_NFW} we show the result of our
fitting procedure for halo Aq-D\footnote{We note
that the correspondence of $r_\textrm{fix}$ with the location of the
maximum circular velocity for the best fitting NFW potential to halo
Aq-D is a coincidence.}. The overall mass and velocity
profiles differ significantly outside $r \sim r_\textrm{fix}$.

The next step in setting up the axisymmetric Kuzmin-Kutuzov potential
is to use Eqs.~(\ref{eq4:isochrone_KuzminKutuzov}) to determine its
characteristic parameters. This requires some measurement of the shape
of the Aquarius halos, and we use the axis ratios
determined using the reduced moment of inertia tensor at $r_\textrm{fix}=50$ kpc by
\citet{VeraCiro2011}, and which closely follow the isodensity
contours. Since Aq-A and Aq-D halos are more prolate, we define
an axisymmetric equivalent of the axis ratio of the density $q_\rho$
as
\begin{equation}
q_\rho = \dfrac{2 a}{b+c},
\end{equation}
where $a$, $b$ and $c$ are the major, intermediate and minor axis
lengths. For halo Aq-D $q_\rho(50~{\rm kpc}) = 1.53$ ($q_\rho=1.39$ when the
axis ratios $b/a$ and $c/a$ are determined at 100 kpc). 

The density
profile of the Kuzmin-Kutuzov potential is
given by \citep{Dejonghe1988}:
\begin{equation}
	\rho_\textrm{K}(\lambda, \nu) = \frac{M_\textrm{K}\, c_\textrm{K}^2}{4 \pi} \frac{\lambda\nu+a_\textrm{K}^2\left(\lambda+3\sqrt{\lambda\nu}+\nu\right)}{\left(\lambda\nu\right)^{3/2}\left(\sqrt{\lambda}+\sqrt{\nu}\right)^3}.
\end{equation}
 We can use this expression to solve numerically for the value of $q_\textrm{K}$ such that the isodensity-contour at $R=50$ kpc has the desired axis ratio $q_\rho$. This results in $q_\textrm{K} = 1.87$ for halo Aq-D. Note that for this value of $q_\textrm{K}$ we find $q_{\rho,\textrm{K}} = 1.44$ at 100 kpc for the Kuzmin-Kutuzov potential, which is also not too far off from the value of $q_\rho(100~{\rm kpc})=1.39$ measured for Aq-D.

\section{Fitting algorithm}
\label{sec4:Appendix4A}

When the individual wraps of a stream are sufficiently distinct in
angle space, we can fit straight lines to these. Their slopes can then
be compared to those obtained when fitting in frequency space. In
Section \ref{sec4:aabehaviouraquarius}, we derived the angle
distributions for streams assuming a spherical potential. These
distributions are rather noisy and make the process of fitting
straight lines non-trivial and complex. Therefore, we proceed to describe now in
detail the steps that we take to measure the slopes in angle space.

\begin{figure*}[!htbp]
\centering
\noindent\makebox[\textwidth]{
\includegraphics[width=0.95\textwidth]{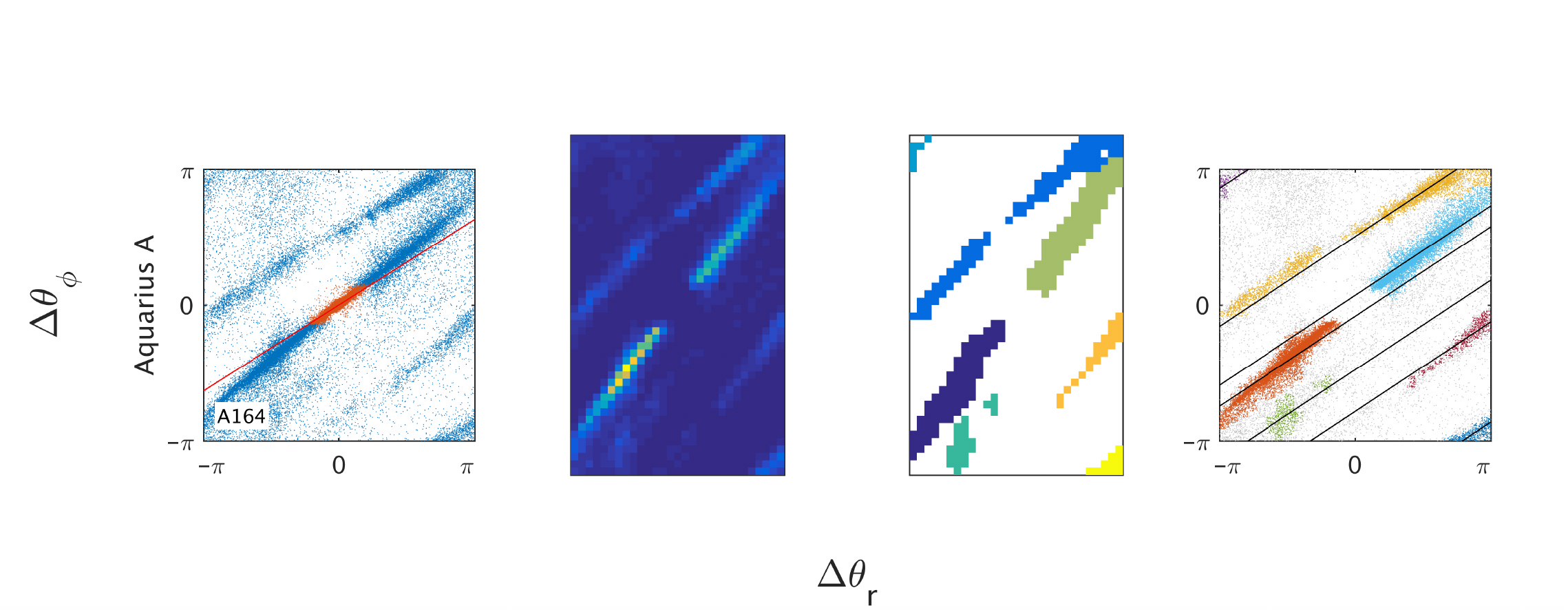}}
\noindent\makebox[\textwidth]{
\includegraphics[width=0.95\textwidth]{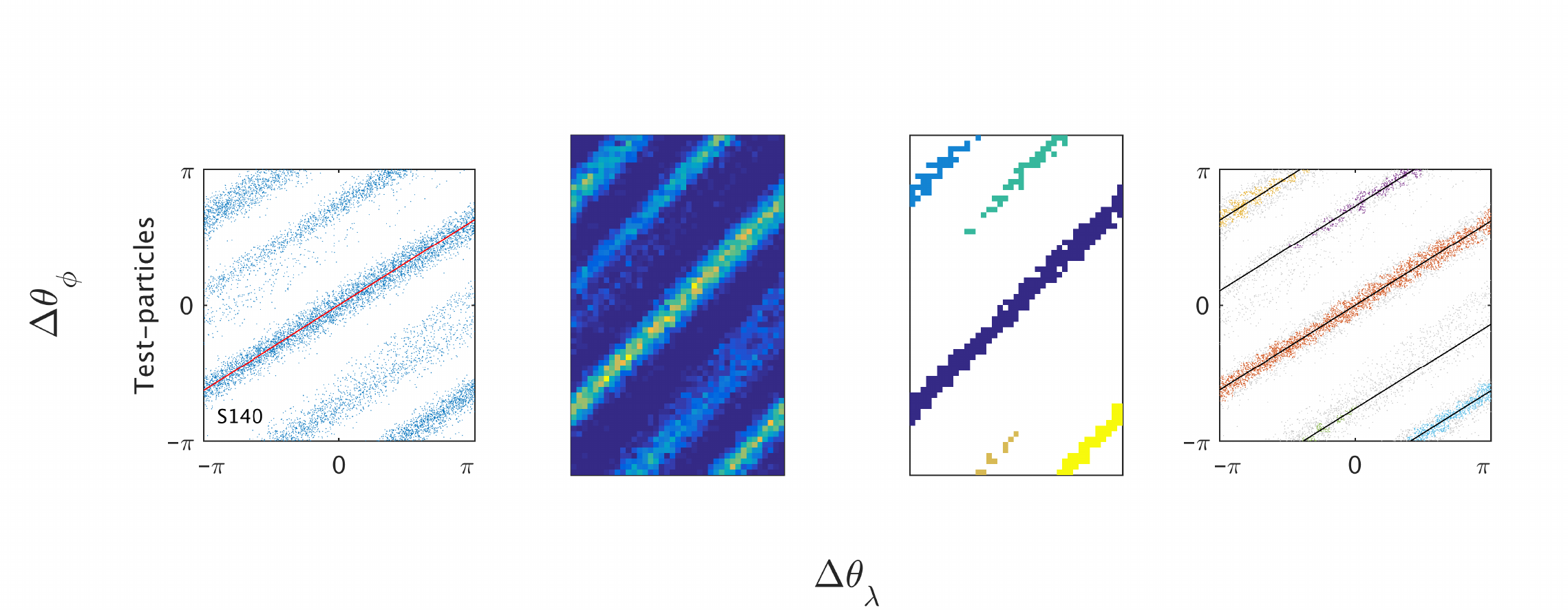}}
\caption{\small Examples of the straight lines fitting routine for
  stream A164 (top panels) and the test particle simulations'
  S140. In the left panel we show the stream in angle space with
  possible bound particles in red and the straight line obtained by
  fitting the frequency distribution. The second panel shows the
  streams in bins, where the axis have been scaled such that the
  streams are oriented at 45 deg. The third panel shows the groupings
  found by the pattern filling algorithm. The last panel shows the 
  resulting straight lines determined using parallel fitting.}
\label{fig4:fitting_example}
\end{figure*}

\begin{figure*}[!htbp]
\centering
\includegraphics[width=0.3\textwidth]{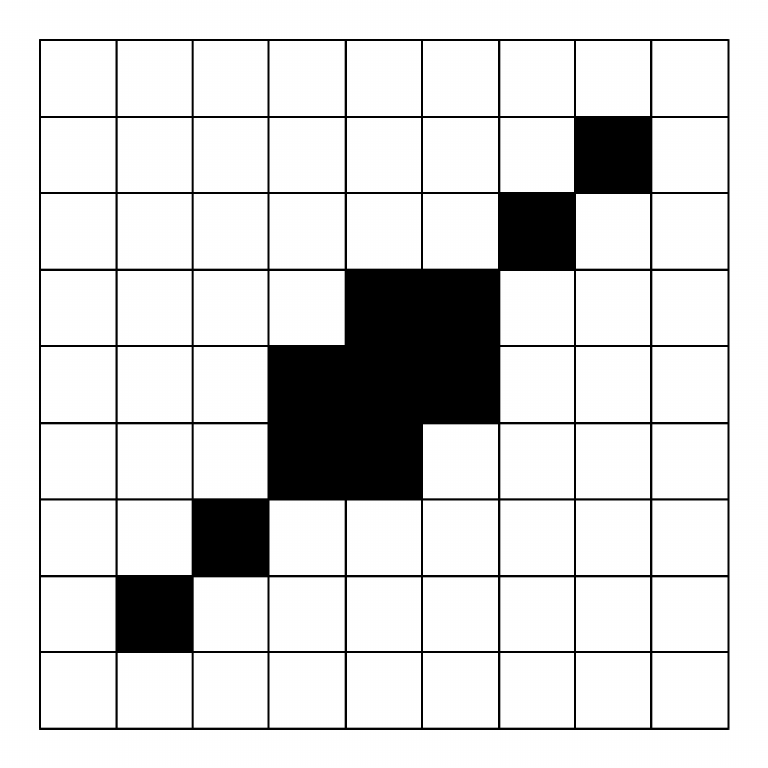}
\caption{\small Pattern used to link pixels in the angle-space image of the stream.}
\label{fig4:fitting_pattern}
\end{figure*}

In Fig.~\ref{fig4:fitting_example} we show two examples: stream A164 and the test-particle simulations stream S140 as they are being
fitted. In case of the Aquarius halos, we first use the binding
energies of the particles in the stream to find the most bound
particle and then centre angle and frequency space on this particle. If there is no
bound structure, the centre is put at the location of the highest density in
angle space. We generously remove the progenitor and particles in its
surroundings by computing the total binding energy with a much higher mass
per particle ($\sim 4$). These bound particles are marked in red in the top-left panel of
Fig.~\ref{fig4:fitting_example}.

In the next step we fit a straight line to the remaining particles
in each independent projection of frequency space using a simple least
squares method. We show this line for the $r$-$\phi$ projection of
angle space in the left panels of Fig.~\ref{fig4:fitting_example}. We
then bin the data of this angle space projection in $N$ bins
horizontally and $N/S(\Delta\Omega_\phi)$ vertically, where $N=40$ for
the test-particle streams and $N=30$ for the Aquarius streams, unless
there are fewer than 100 particles remaining. We then `clean' this image by emptying the bins that have fewer counts than the median of non-empty pixels. This is done twice to generate
enough contrast w.r.t. 'background' of particles between the streams for the Sculptor progenitors, and only once for the thinner Carina progenitors (e.g. Fig.~\ref{fig4:simulation_anglesfrequencies_carina}).

We group the particles by connecting pixels that fall within the
pattern of Fig.~\ref{fig4:fitting_pattern}, which is elongated in the
same direction as the stream, which is on an angle of 45 degrees if it
follows the frequency distribution. Some parallel streams may be `connected' at a single pixel, and the algorithm keeps them separate by not combining groups that individually had more than 25 connected pixels. This procedure results in the groups shown in the third column of Fig.~\ref{fig4:fitting_example} with different colours. It is clear from
the top row that sometimes structures are grouped together that should
not be connected, but it is difficult to completely prevent this from
happening without removing too many particles from the streams.

We then use a least-squares fit for parallel lines with different
offsets. The result is optimised by running the full fitting procedure
twice, because we can then use the slope fitted to the angles found in the first iteration when
binning angle space\footnote{What may also happen in this second iteration is that if stream wraps depict discontinuities, for example but not only, after removal of the progenitor bound particles, these parts are fitted separately. This explains why some of the fitted straight lines in the figures in the main paper  are very close to each other, such as for streams D72 and D82 in Fig.~\ref{fig4:aquarius_anglesAD_2}.}. To estimate the errors in the fitting we
bootstrap the data 200 times, which was found to be enough to get a
reasonable estimate of the errors, although this does not fully
reflect the error when the wraps in angle space overlap or if they are
not on parallel lines.

\end{document}